\documentclass[12pt]{article}
\usepackage{amsmath}
\usepackage{amssymb}
\usepackage{amsthm}
\usepackage{enumerate}
\usepackage{enumitem}
\usepackage[toc,page]{appendix}
\setlist[enumerate]{label=(\arabic*), nosep}
\usepackage{bm}
\usepackage{color}
\usepackage{amsmath}
\usepackage{graphicx}
\usepackage{enumerate}
\usepackage[authoryear]{natbib}
\usepackage{url}
\usepackage{graphicx}
\usepackage[space]{grffile}
\usepackage{latexsym}
\usepackage{textcomp}
\usepackage{longtable}
\usepackage{tabulary}
\usepackage{booktabs,array,multirow}
\usepackage{amsfonts,amsmath,amssymb}
\usepackage[table,xcdraw]{xcolor}
\usepackage{array}
\newcolumntype{"}{@{\vrule width 1pt}}
\usepackage[figuresright]{rotating}
\usepackage[colorlinks,citecolor=blue,urlcolor=blue]{hyperref}
\usepackage{amsmath}
\usepackage{amssymb}
\usepackage{bm}
\usepackage{amsmath}
\usepackage{mathrsfs}

\def\mE{\mathbb E}

\def\var{\mathrm {var}}

\def\n{\nonumber}
\def\var{\mathrm {Var}}

\def\mE{\mathbb{E}}

\def\n{\nonumber}

\def\P{{\mathbb P}}
\def\0{{\bf 0}}

\def\be{\begin{eqnarray}}
	\def\ee{\end{eqnarray}}
\def\bse{\begin{eqnarray*}}
	\def\ese{\end{eqnarray*}}
\numberwithin{equation}{section}
\newtheorem{example}{Example}[section]

\newtheorem{theorem}{Theorem}[section]
\newtheorem{lemma}{Lemma}[section]
\newtheorem{corollary}{Corollary}[section]
\newtheorem{assumption}{Assumption}[section]

\allowdisplaybreaks[4]
\def\boxit#1{\vbox{\hrule\hbox{\vrule\kern6pt  \vbox{\kern6pt#1\kern6pt}\kern6pt\vrule}\hrule}}

\allowdisplaybreaks[4]

\title{{\bf Testing Independence Between High-Dimensional Random Vectors Using Rank-Based Max-Sum Tests}}
\author{Hongfei Wang and  Binghui Liu \\
	School of Mathematics and Statistics \& KLAS,\\ Northeast Normal University\\
	Long Feng\\
	School of Statistics and Data Science, KLMDASR, LEBPS \& LPMC,\\ Nankai University
}

\begin{document}
	
	\maketitle

	%\begin{center}
	%\author{Hongfei Wang$^1$, Long Feng$^{1}$, Binghui Liu$^{1}$}\\
	%$^1$School of Mathematics and Statistics and KLAS of MOE, Northeast Normal University\\
	%\end{center}
	
	\def\spacingset#1{\renewcommand{\baselinestretch}%
		{#1}\small\normalsize} \spacingset{1}

	\begin{abstract}
		In this paper, we address the problem of testing independence between two high-dimensional random vectors. Our approach involves a series of max-sum tests based on three well-known classes of rank-based correlations. These correlation classes encompass several popular rank measures, including Spearman's $\rho$, Kendall's $\tau$, Hoeffding's D, Blum-Kiefer-Rosenblatt's R and Bergsma-Dassios-Yanagimoto's $\tau^*$.
		The key advantages of our proposed tests are threefold: (1) they do not rely on specific assumptions about the distribution of random vectors, which flexibility makes them available across various scenarios; (2) they can proficiently manage non-linear dependencies between random vectors, a critical aspect in high-dimensional contexts; (3) they have robust performance, regardless of whether the alternative hypothesis is sparse or dense.
		Notably, our proposed tests demonstrate significant advantages in various scenarios, which is suggested by extensive numerical results and an empirical application in RNA microarray analysis.
	\end{abstract}
	
	\noindent%

	\spacingset{1.45} % DON'T change the spacing!
	%%%%%%%%%%%%%%%%%%%%%%%%%%%%%%%%%%%%%%%%%%%%%%%%%%%%%%%%%%%%%%%%%%%%%%%%%%%%%%
	\section{Introduction}
	
	Testing the independence of two random vectors is a fundamental problem, which bears a significant correlation with a multitude of statistical learning problems. It encompasses a broad spectrum of aspects, including but not limited to independent component analysis, feature selection and Markov random fields  \citep{grover1985probabilistic,lu2009financial,liu2010versatile,imbens2015causal,maathuis2018handbook,fan2020statistical}, and has practical applications in many fields, such as genetic data analysis \citep{josse2016measuring} and financial data analysis \citep{guerrero1998measures}.

	In the literature, numerous measures have been proposed for determining the independence between random vectors. In the case where the dimensions of the two vectors are finite, \cite{Szekely2007} introduced the concepts of distance covariance and distance correlation, which bear similarities to the traditional notions of covariance and correlation. These concepts quantify independence by measuring the discrepancy between the population characteristic function and its empirical counterpart. Furthermore, \cite{Gretton2005} proposed the Hilbert-Schmidt Independence Criterion (HSIC), which is equivalent to the distance covariance when selecting a specific kernel \citep{Sejdinovic2013}. Rank correlation is widely recognized as an effective tool for measuring the independence of two random variables. To extend rank correlation-based tests from univariate to multivariate, projection serves as an efficient technique \citep{zhu2017,Kim2018}. In contrast to the methods based on projection, \cite{Shi2022} and \cite{DebandSen2023} defined multivariate ranks directly using measure transport theory and proposed a multivariate rank version of the distance covariance for testing independence. In addition, there are many testing methods based on spatial partitioning \citep{Gretton2010,heller2013consistent}.

	In the case of high dimensions, \cite{szekely2013distance} considered the regime where the dimensions of the two random vectors diverge but the sample size remains fixed, and showed that for each random vector with exchangeable components, the bias-corrected sample distance correlation statistic converges to a t-distribution with the degrees of freedom related to the sample size. Recently, \cite{Gaolan} suggested using the bias-corrected sample distance correlation statistic and established its asymptotic distribution when both the sample size and dimension diverge.  \cite{Zhuchangbo} pointed out that a test based on the distance/Hilbert-Schmidt covariance can only capture linear dependencies in high-dimensional data. To deal with nonlinearly dependence, they proposed some tests based on an aggregation of marginal sample distance/Hilbert?Schmidt covariances, and established the asymptotic null distributions in the situations with medium to low sample sizes.
	In addition, \cite{chakraborty2021new} proposed a t-test based on the sum of all group-wise squared sample generalized dCov, which can capture
	group-wise nonlinear dependence that cannot be detected by the usual distance covariance and HSIC in the high dimensional regime.

	Recently, \cite{Qiutao} proposed a general framework for testing the dependence of two random vectors to randomly select two subspaces consisting of components of the vectors respectively,  which can work for nonlinear dependence detection in a high-dimensional setup. \cite{zhouyeqing2024} found that under any strictly monotonic transformation of coordinates, distance correlation remains consistent but not invariant. To deal with this issue, they proposed a series of tests based on the rank correlations, including Hoefding's D, Blum-Kiefer-Rosenblatt's R and Bergsma-Dassios-Yanagimoto's $\tau^*$, and established their asymptotic null distributions. The key advantages of such rank-based tests are twofold: (1) they do not rely on specific assumptions about the distribution of random vectors, which flexibility makes them robust across various scenarios; (2) they can handle non-linear dependencies effectively, which is crucial in high-dimensional settings.
Note that the tests proposed by \cite{zhouyeqing2024} are of the sum-type and are primarily relevant to dense alternative hypotheses, while having lower power under sparse alternative hypotheses. In contrast, \cite{Cai2023} proposed a comprehensive framework centered around classifiers, which is particularly effective in detecting sparse dependency signals in high-dimensional setups.
	
	In this paper, we first propose a series of max-type and sum-type tests based on three classes of rank-based correlations that were studied in \cite{distributionfree,testsum} and \cite{drton2020}. These classes of correlations encompass many commonly used rank correlations, such as Spearman's $\rho$, Kendall's $\tau$, Hoeffding's D, Blum-Kiefer-Rosenblatt's R and Bergsma-Dassios-Yanagimoto's $\tau^*$ correlations.
	To pursue robust performance regardless of whether the alternative hypothesis is sparse or dense, we propose a series of max-sum tests by combining the max-type and sum-type test statistics, and establish the asymptotic properties of these max-sum tests. Of course, the proposed max-sum tests also possess the two advantages of the aforementioned rank-based approach. Compared with many existing tests, the proposed max-sum tests have robust and advantageous performance in various  scenarios under both sparse and dense alternative hypotheses, suggested by the numerical results as well as an empirical application.
	
	% It is worth noting that the method of theoretical proof is different from that of mutual independence test \citep{wang2024}. Here, we need to consider the dependence structures of the two random vectors.
	
	The main contributions of this paper is listed as follows:
	\begin{enumerate}
		\item %While most current methodologies utilize sum-type test procedures, these often fall short when dealing with sparse alternatives.
		This paper is the first to use max-type rank-based procedures for testing the independence between two high-dimensional random vectors. The limiting null distribution of the proposed max-type test statistics based on three classes of rank-based correlations are established. Furthermore,  the consistency and rate-optimality of these max-type tests are established.
		
		\item We have proposed some sum-type procedures for testing the independence between two high-dimensional random vectors, and establish the asymptotic normality of these test statistics based on three classes of rank-based correlations. These tests are based on the sum of the squares of the rank correlations, which generally outperform the sum-type tests proposed by \cite{zhouyeqing2024} based on the sum of the rank correlations. Indeed, a test based on the sum of the rank correlations may lack power if the sum of the rank correlations equals or closes to zero.
		
		\item We have established the asymptotical independence of the proposed sum-type and max-type test statistics based on each of the rank correlation belonging  to the above three classes. In fact, such type of study is very popular in various testing issues, such as in \cite{xu2016adaptive}, \cite{he2021}, \cite{feng2022max}, \cite{wang2023}, \cite{feng2022dependent} and \cite{yu2022jasa}. These studies typically require a certain distribution assumption, whereas such assumption is not needed in this paper.
	\end{enumerate}

	\textsc{Notation}
	Let $m$ be any positive integer such that $m\geq 1$. The symbol $\mathfrak{S}({1},\cdots,{m})$ is used to represent the set of all permutations of the integers from 1 to $m$. The notation $\mathbb{R}^m$ is used to denote the $p$-dimensional real space. The notation ${I}(\cdot)$ is used to denote the indicator function of a set. The uniform distribution in the interval $(a,b)$ is represented by $U(a,b)$, and the joint distribution of two independent variables that following $U(a,b)$ is represented by $U(a,b)^2$.
	The distribution of standard normal distribution is represented by $\Phi(x)$ for $x\in \mathbb{R}$.  The factorial of any positive integer $m$ is denoted by $m!$. The symbol $\lfloor x\rfloor$ represents the greatest integer less than or equal to any real number $x$. The symbol $\lceil x\rceil $ represents the smallest integer larger than or equal to any real number $x$. The symbol $\lambda_{\max}(\cdot)$ represents the largest eigenvalue of a matrix.
	
	The rest of this paper is organized as follows. In Section \ref{sec:statistic}, we propose a series of tests based on  three classes  of rank correlations, and establish their asymptotic properties. In Section \ref{sec:simu}, we present extensive numerical results of the proposed tests in comparison with some of their competitors, followed by an empirical application in Section \ref{empirical study}. Finally, we conclude this paper in Section \ref{sec:dis} with some discussions and relegate the technical proofs to the Supplementary Material.
	
	\section{Test statistics}\label{sec:statistic}
	
	We consider the problem of testing the independence between a $p$-dimensional random vector $\bm{X}=(X_1,\cdots,X_p)^{\top}$ and a $q$-dimensional random vector $\bm{Y}=(Y_1,\cdots,Y_q)^{\top}$. The aim is to test the hypothesis:
	\begin{align}\label{H0}
		H_0: \bm{X} \text{ is independent of } \bm{Y},
	\end{align}
	based on the sample of $n$ independent replicates of the random vector $(\bm{X}^\top,\bm Y^\top)^\top$, i.e. $\{(\bm{X}_k^\top, \bm{Y}_k^\top)^\top\}_{k=1}^n$, where $\bm{X}_k=(X_{k,1},\cdots,X_{k,p})^{\top}$ and $\bm{Y}_k=(Y_{k,1},\cdots,Y_{k,q})^{\top}$, for each $1\leq k \leq n$. In this paper, we consider the case of $p+q\rightarrow \infty$, so without losing generality, we assume that $p\rightarrow \infty$.
	
	In this section, we will consider three classes of rank-based statistics for testing the independence between two random vectors: simple linear rank-based statistics, non-degenerate rank-based U-statistics and degenerate rank-based U-statistics. Based on these classes of rank-based statistics, we propose the corresponding rank-based max-type, sum-type and max-sum tests for testing the independence between two random vectors.
	
	%Next, we will extend three classes of rank-based statistics to test the independence of two random vectors. In this case, we permit at least one dimension of $\bm{X}$ and $\bm Y$ to tend to infinity. For simplicity, we assume that $p\rightarrow \infty$, without placing any constraints on the dimension $q$.
	
	\subsection{Simple linear rank-based statistics}
	
	The first class of rank-based statistics is the simple linear rank-based statistics. Such class of statistics were previously studied by \cite{distributionfree} for testing the mutual independence between all entries of a random vector. This class includes the statistic based on Spearman's $\rho$ correlation proposed by \cite{Spearman1904}.
	
	Specifically, for each $1\leq i \leq p$ and $1\leq j \leq q$, the simple linear rank-based statistic for the index pair $(i,j)$ is $$\tilde{V}_{ij}\doteq n^{-1}\sum_{k=1}^{n}f\left\{k /(n+1)\right\}g\left\{R_{n k}^{ij} /(n+1)\right\},$$
	where $f(\cdot)$ and $g(\cdot)$ are two Lipschitz functions, and for each $1\leq k \leq n$, $R_{n k}^{ij} \doteq Q_{n k^{\prime}}^{j}$ with $1\leq k^{\prime} \leq n$ satisfying $R_{n k^{\prime}}^i=k$. Here,  $R_{n k^{\prime}}^{i}$ denotes the rank of $X_{k^{\prime},i}$ in $\left\{X_{1,i}, \ldots, X_{n,i}\right\}$ and $Q_{n k^{\prime}}^{j}$ denotes the rank of $Y_{k^{\prime},j}$ in $\left\{Y_{1,j}, \ldots, Y_{n,j}\right\}$.   In fact, $\tilde{V}_{ij}$ measures the dependence between the pair of variables $X_i$ and $Y_j$ via the agreement between the ranks of two random sequences of $X_i$ and $Y_j$, i.e. $\left\{X_{1,i}, \ldots, X_{n,i}\right\}$ and $\left\{Y_{1,j}, \ldots, Y_{n,j}\right\}$.
	
	\subsubsection{The max-type statistics}
	
	Based on $\tilde{V}_{ij}$, the max-type statistic is
	$$L_V\doteq \max _{1 \leq i\leq p, 1\leq j \leq p}\left|\tilde{V}_{i j}\right|. $$
	
	Before establishing the asymptotic null distribution of $L_V$, we need to present some notation and definitions. Let $N=pq$. Define $\left\{\upsilon_1, \cdots, \upsilon_{N}\right\}=\big\{\tilde{V}_{i j} / \sigma_{L_V}: 1 \leq i\leq p, 1\leq j \leq q\big\}$, where $\sigma_{{L_V}}^2 = \var_{H_{0}}(\tilde{V}_{ij})$. Define $\boldsymbol{\Sigma}_V=\left(\sigma_{st}^V\right)_{1 \leq s, t \leq N}$ with $\sigma_{s t}^V=\operatorname{cor}\left(\upsilon_s, \upsilon_t\right)$ for each $1 \leq s, t \leq N$. In addition, define a class of matrices  $$\mathcal{C}\doteq \left\{\boldsymbol{\Sigma}=\left(\sigma_{st}\right)_{1 \leq s, t \leq N}: \boldsymbol{\Sigma} \text{ satisfies Assumption } \ref{assump1}\right\}.$$
	\begin{assumption} \label{assump1}
		For some $\varepsilon \in(0,1)$, $\left|\sigma_{st}\right| \leq \varepsilon$ for all $1 \leq s<t \leq N$ and $N \geq 2$. For some positive sequences $\left\{\varphi_N: N \geq 1\right\}$ and $\left\{\varsigma_N: N \geq 1\right\}$ with $\varphi_N=o(1 / \log N)$ and $\varsigma_N \rightarrow 0$ as $N \rightarrow \infty$,  $\left|C_N\right| / N \rightarrow 0$ as $N \rightarrow \infty$, where $C_N=\left\{1 \leq s \leq N: \left|B_{N, s}\right| \geq N^{\varsigma}\right\}$ and $B_{N, s}=\left\{1 \leq t \leq N:\left|\sigma_{ st}\right| \geq \varphi_N\right\}$ for each $1 \leq s \leq N$.
	\end{assumption}
	
	Based on these definitions, we present an assumption need to be imposed.
	\begin{assumption}\label{assump2}
		(1) Let $c_{n,k}=n^{-1}f\left\{k /(n+1)\right\}$ for each $1\leq k \leq n$, and  $\bar{c}_{n} = \sum_{k=1}^{n} c_{n, k}$. For constants $C_1,C_2> 0$,
		$$
		\max _{1 \leq k \leq n}\left|c_{n, k}-\bar{c}_{n}\right|^2 \leq  \frac{C_1}{n} \sum_{k=1}^{n}\left(c_{n, k}-\bar{c}_{n}\right)^2,~
		\left|\sum_{k=1}^{n}\left(c_{n,k}-\bar{c}_{n}\right)^3\right|^2 \leq  \frac{C_2}{n}\left\{\sum_{k=1}^{n}\left(c_{n, k}-\bar{c}_{n}\right)^2\right\}^3.
		$$
		(2) $g(\cdot)$ is differentiable with bounded Lipschitz constant.
		%\item $\boldsymbol{\Sigma}_V\in \mathcal{C}$.  \label{cond3}%latter assumptions for asymptotic independence
	\end{assumption}
	\begin{theorem}\label{th:LV}
		Suppose that $\boldsymbol{\Sigma}_V\in \mathcal{C}$ and Assumption \ref{assump2} holds. Further suppose that $\mE_{H_0}(\tilde{V}_{ij})=0$. Then, under $H_0$, for any $y \in \mathbb{R}$,
		$$
		\left|\mathbb{P}\left(L_V^2 / \sigma_{L_V}^2-2 \log N+\log \log N \leq y\right)-\exp \left\{-\pi^{-1 / 2} \exp (-y / 2)\right\}\right|=o(1),
		$$
		as $n,p \rightarrow \infty$, where $N=o\left(n^{\epsilon}\right)$ for some $\epsilon>0$.
	\end{theorem}
	
	According to Theorem \ref{th:LV}, the max-type test based on $L_V$ rejects the null hypothesis in \eqref{H0} when $L_V^2 / \sigma_{L_V}^2-2 \log N+\log \log N\geq q_{\alpha}$, where $q_\alpha=-\log \pi-2\log\log (1-\alpha)^{-1}$ and $\alpha$ is the significance level.
	
	Next, we turn to the power analysis of the max-type test based on $L_V$. For any $c>0$, define
	$$
	\mathcal{U}(c) \doteq \left\{M=\left(m_{ij}\right)_{1 \leq i \leq p, 1\leq j\leq q}: \max _{1 \leq i \leq p, 1\leq j\leq q} |m_{ij}| \geq c(\log N/n)^{1 / 2}\right\} .
	$$
	For each matrix in $\mathcal{U}(c)$, at least one element has a magnitude greater than $c(\log N/n)^{1 / 2}$. Define $\hat{\mathbf{V}}=(\hat{V}_{i j})_{1 \leq i \leq p, 1\leq j\leq q}$, where
	$
	\hat{V}_{i j}=n^{-1/2}\sigma_{L_V}^{-1}\tilde{V}_{i j}
	$
	for each $1 \leq i\leq p$ and each $1\leq j \leq q$. Let $\mathbf{V} =(V_{i j})_{1\leq i\leq p,1\leq j\leq q}= \mE(\hat{\mathbf{V}})$.
	Consider the following local alternative hypothesis
	$$
	H_{\mathrm{a}}^V(c)\doteq \left\{F(\cdot,\cdot): \mathbf{V} \in \mathcal{U}(c) \text{ if } (\bm{X},\bm{Y})\sim F(\cdot,\cdot)\right\},
	$$
	where $F(\bm{x},\bm{y})$ with $\bm{x}\in \mathbb{R}^p$ and $\bm{y}\in \mathbb{R}^q$ is the joint distribution function of $(\boldsymbol{X},\boldsymbol{Y})$.
	
	\begin{theorem}\label{th:LVH1}
		Suppose that $\mE_{H_0}(\tilde{V}_{ij})=0$, $n\sigma_{L_{V}}^2 =a_{1}\{1+o(1)\}$ and $\max\{|f(0)|, |g(0)|\}\leq a_2$ for some positive constants $a_1$ and $a_{2}$, and the Lipschitz constants of $f(\cdot)$ and $g(\cdot)$ are bounded. Then,   if $(\bm{X},\bm{Y})\sim F(\cdot,\cdot)$,
		$$
		\inf_{F(\cdot,\cdot) \in H_{\mathrm{a}}^{V}\left(A_1\right)} \mathbb{P}\left(L_V^2 / \sigma_{L_V}^2-2 \log N+\log \log N\geq q_{\alpha}\right)=1-o(1),
		$$
		as $n, p\rightarrow \infty$, where $A_1$ is a large scalar that depends only on $a_1$ and $a_{2}$ as well as the Lipschitz constants.
	\end{theorem}

	We will further establish the rate-optimality of the above test. First, consider the following alternative hypothesis
	$$
	H_a^{R}(c) \doteq \left\{F(\cdot,\cdot): \mathbf{R}\in \mathcal{U}(c) \text{ if } (\bm{X},\bm{Y})\sim F(\cdot,\cdot)\right\},
	$$
	where $\mathbf{R}=(R_{i j})_{1\leq i\leq p,1\leq j\leq q}=\mE(\bm{X}\bm{Y}^{\top})$. Let $\mathcal{T}_\alpha$ denote the class of measurable tests for testing the independence between $\bm X$ and $\bm{Y}$ with the significance level $\alpha$, i.e. $\mathcal{T}_\alpha =\left\{T_\alpha: \mathbb{P}_{H_0}\left(T_\alpha=1\right) \leq \alpha\right\}$.
	
	\begin{theorem}\label{R optimal}
		Suppose that $0<c_0<1$ and $\beta$ is a positive constant satisfying $\alpha+\beta<1$. Then, under the regime $\log N =o(n)$, for sufficiently large $n$ and $p$,
		$$
		\inf_{T_\alpha \in \mathcal{T}_\alpha} \sup _{F(\cdot,\cdot) \in H_a^{R}\left(c_0\right)} \mathbb{P}\left(T_\alpha=0\right) \geq 1-\alpha-\beta.
		$$
	\end{theorem}
	
	Theorem \ref{R optimal} indicates that for some constant $c_0 < 1$, any measurable  test  $T_\alpha$ in $\mathcal{T}_\alpha$ is unable to distinguish between the null hypothesis and the alternative hypothesis when $\max_{1\leq i\leq p,1\leq j\leq q} |R_{ij}| \leq c_0(\log N/n)^{1/2}$. Based on Theorems \ref{th:LVH1} and \ref{R optimal}, we can establish the rate-optimality of the max-type test based on $L_V$ as follows.
	
	\begin{assumption}\label{cond H1}
		Suppose that $(\bm{X}^\top,\bm{Y}^\top)^\top$ follows Gaussian distribution, and for sufficiently large $n$ and $p$, $c V_{i j} \leq R_{i j} \leq C V_{i j}$ for each $1 \leq i\leq p$ and $1\leq  j \leq q$, where $c$ and $C$ are two positive constants.
	\end{assumption}
	
	\begin{theorem}\label{th:LVH1 optimal}
		Suppose that the assumptions in Theorems \ref{th:LV}-\ref{th:LVH1} hold. Further suppose that Assumption \ref{cond H1} holds. Then, the max-type test based on $L_V$ is rate-optimal, i.e. there exist two constants $B_1$ and $B_2$ with $B_1<B_2$, such that:
		\begin{itemize}
			\item[(i)] $\sup _{F(\cdot,\cdot)\in H_{\mathrm{a}}^{V}\left(B_2\right) } \mathbb{P}\left\{{I}\left(L_V^2 / \sigma_{L_V}^2-2 \log N+\log \log N\geq q_{\alpha}\right)=0\right\}=o(1)$,
			\item[(ii)] for any $\beta>0$ satisfying $\alpha+\beta<1$,
			$
			\inf _{T_\alpha \in \mathcal{T}_\alpha} \sup _{F(\cdot,\cdot)\in H_{\mathrm{a}}^V\left(B_1\right)} \mathbb{P}\left(T_\alpha=0\right) \geq 1-\alpha-\beta,
			$
			for sufficiently large $n$ and $p$.
		\end{itemize}
	\end{theorem}
	%Theorem \ref{th:LVH1 optimal} suggests that the rate of $ (\log N/n)^{1/2}$ cannot be further reduced.

	\subsubsection{The sum-type statistics}\label{SV}
	The sum-type statistic based on $\tilde{V}_{ij}$ is $$S_V \doteq \underset{1 \leq i\leq p,
		1\leq j \leq p}{\sum}\left\{\tilde{V}_{i j}^2-\mE_{H_0}\left(\tilde{V}_{i j}^2\right)\right\}.$$
	To establish the asymptotic null distribution of $S_V$, we impose the following assumptions.
	\begin{assumption}\label{assmp3}
		$\{X_i\}_{i=1}^p$ and $\{Y_j\}_{j=1}^q$ are $m$-dependent random sequences. Specifically, there exists two integers $0< m_p < p$ and $0< m_q < q$, such that for any $k\geq 1$, $X_{k+m_p+1}$ is independent of $X_k$ and $Y_{k+m_q+1}$ is independent of $Y_k$.
	\end{assumption}
	
	Note that the m-dependent structure in Assumption \ref{assmp3} is a commonly used assumption in genomic data analysis \citep{farcomeni2005multiple,li2022simultaneous} and financial time series analysis \citep{MOON2013143}.
	
	\begin{assumption}\label{cond5}
		For some $\delta>0$ and $-1\leq \gamma <1,$ $\lim_{p \rightarrow \infty}m_p^{1+(1-\gamma)(1+2/\delta)}/p= 0$ and
		$\max\left\{m_p^{\gamma},m_p,m_p^{1+\gamma}p^{\frac{-\gamma}{1+(1-\gamma)(1+2/\delta)}}\right\}m_q pq/\var\left(n\sum_{1 \leq i\leq p,
			1\leq j \leq p}\tilde{V}_{ij}^2\right)=O(1).$
	\end{assumption}
	%Assumption \ref{assmp3} allows for weak dependence between the components of random vectors $\bm X$, the same applies to $\bm Y$, which is often assumed in studies of time series and genome associations.}

Based on these assumptions, we establish the asymptotic null distribution of $S_V$ as follows.

\begin{theorem}\label{th:SV}
	Suppose that Assumptions \ref{assump2} and \ref{assmp3}-\ref{cond5} hold. Further suppose that $f$ is bounded on the interval $(0, 1)$ and $\mE_{H_{0}}(\tilde{V}_{st})=0$. Then, under $H_0$, $S_{V} /\sigma_{S_{V}} \rightarrow \Phi(x)$ in distribution as $n, p\rightarrow \infty$, where $\sigma^2_{S_{V}}\doteq \var\left(\sum_{1\leq i \leq p, 1\leq j \leq q}\tilde{V}_{ij}^2\right).$
\end{theorem}
The sum-type test based on $S_V$ can be executed by using a permutation method. Specifically, we consider $B$ random permutations of $\{1,\cdots,n\}$, denoted by $\bm{\pi}^b=(\pi^{b}_{1},\ldots,\pi^{b}_{n}) \in \mathfrak{S}({1},\cdots,{n})$, $1\leq b\leq B$. For each $1\leq b\leq B$, we obtain $n$ samples of $\bm X$, denoted as $\bm X_{\pi^{b}_1},\ldots,\bm X_{ \pi^{b}_n}$, where for each $1\leq i\leq n$, $\bm X_{\pi_i^{b}}$ is independent of $\bm Y_i$.
For each $1\leq b\leq B$, let $S_V^b$ denote the permutation version of $S_V$ based on the samples of $\{(\bm X_{\pi^{b}_i},\bm Y_i)\}_{i=1}^n$ obtained from the $b$th permutation. Then, we use the sample variance of $\{S_{V}^{1},\ldots,S_V^{B}\}$ as the estimate of $\sigma^2_{S_{V}}$, denoted as $\hat{\sigma}^2_{S_{V}}$.

According to Theorem \ref{th:SV}, the sum-type test based on $S_{V}$ rejects the null hypothesis in \eqref{H0} when $S_{V} /\hat{\sigma}_{S_{V}}\geq z_\alpha$, where $z_\alpha$ is the upper $\alpha$-quantile of the standard normal distribution.

\subsubsection{The max-sum statistics}\label{CV}

To combine the advantages of the max-type and sum-type statistics mentioned above, in the following we will propose a max-sum statistic via Fisher's combination method to combine these two statistics:
\begin{align}\label{TCV}
	T_{C}^{V}\doteq -2\log P_{S_V}-2\log P_{L_V},
\end{align}
where $P_{S_V}$ and $P_{L_V}$ are the p-values of the tests based $S_V$ and $L_V$, respectively, i.e. $P_{S_V}=1-\Phi(S_V/\hat{\sigma}_{S_V})$ and $P_{L_V}=1-G(L_V^2 / \sigma_{L_V}^2-2 \log N+\log \log N)$. Here, $G(y)=\exp \left\{-\pi^{-1 / 2} \exp (-y / 2)\right\}$ with $y\in \mathbb{R}$ is the type-I extreme value distribution function.

Before establishing the asymptotic properties of $T_{C}^{V}$, we need to investigate the asymptotic independence between $S_V$ and $L_V$.
\begin{theorem}\label{th:CV}
	Suppose that the assumptions in Theorems \ref{th:LV} and \ref{th:SV} hold, and for some $C>0$, $\max_{1 \leq s \leq N} \sum_{t=1}^N (\sigma_{st}^{V})^2 \leq(\log N)^C$ for all $N \geq 3$. Then, under $H_0$, $S_{V} /{\sigma}_{S_{V}}$ is asymptotically independent of $L_V^2 / \sigma_{L_V}^2-2 \log N+\log \log N$ as $n$, $p\rightarrow\infty$.
\end{theorem}

Note that it is not contradictory to  simultaneously assume that $\boldsymbol{\Sigma}_V\in \mathcal{C}$ and Assumption \ref{assmp3} holds. In fact,
for each $1\leq s \leq N$, let $\tilde{\upsilon}_s=\tilde{V}_{i_sj_s}$ and $\mathcal{E}_{s}=\{\tilde{\upsilon}_t=\tilde{V}_{i_tj_t}: i_s-m_p\leq i_t\leq i_s+m_p,j_s-m_q\leq j_t\leq j_s+m_q\}$. It can be seen that $\tilde{\upsilon}_s$ is independent of $\{\tilde{\upsilon}_1,\ldots,\tilde{\upsilon}_N\}\setminus \mathcal{E}_{s}$, as $\{X_i\}_{i=1}^p$ and $\{Y_j\}_{j=1}^q$ are $m$-dependent sequences.
Since the cardinality of $\mathcal{E}_{s}$ is $(2m_p+1)(2m_q+1)$, the corresponding $|B_{N, s}|$ in Assumption \ref{assump1} is equal to $(2m_p+1)(2m_q+1)$. In situation where $m_pm_q=o(N^{\varsigma})$, we have that $\boldsymbol{\Sigma}_V\in \mathcal{C}$. For example, when $\varsigma=(\log N)^{-k}$ with $k \in (0,1)$ and $m_pm_q/\exp^{(\log N)^{1-k}} \to 0$, it is straightforward to derive that $\boldsymbol{\Sigma}_V\in \mathcal{C}$.

Then, the asymptotic null distribution of $T_C^V$ is provided as follows.

\begin{corollary}
	Suppose that the assumptions in Theorem \ref{th:CV} hold. Then, under $H_0$, $T_C^V$ converges to $\chi_4^2$ in distribution as $n,p\rightarrow \infty$, where $\chi_4^2$ is the chi-square random variable with 4 degrees of freedom.
\end{corollary}

According to this corollary, the max-sum test based on $T_{C}^{V}$ rejects the null hypothesis in \eqref{H0} when $T_C^V>w_\alpha$, where $w_\alpha$ is the upper $\alpha$-quantile of the distribution of $\chi_4^2$.

Next, we present the power analysis of the max-sum test based on $T_C^V$. From the  construction of $T_C^V$, it can be seen that
\begin{align}\label{TCVpowerinequality}
	\mathbb{P}_{H_{1}}\{T_C^V>w_{\alpha}\}\geq &\max\{\mathbb{P}_{H_{1}}\left( -2\log P_{L_{V}}>w_{\alpha}/2\right)+\mathbb{P}_{H_{1}}\left(-2\log  P_{S_{V}}<w_{\alpha}/2\right),\nonumber\\
	&\quad\quad\quad\mathbb{P}_{H_{1}}\left( -2\log P_{S_{V}}>w_{\alpha}/2\right)+\mathbb{P}_{H_{1}}\left(-2\log  P_{L_{V}}<w_{\alpha}/2\right)\}.
\end{align}
Hence, if the power of the test based on $S_V$ or $L_V$ approaches 1, it follows that the power of the test based on $T_C^V$ also approaches 1.

\begin{corollary}\label{corLC}
	Suppose that the assumptions in Theorems \ref{th:LVH1} and \ref{th:CV} hold. Then, for some large scalar $A_2$, as $n,p \rightarrow \infty,$
	$$\inf _{F(\bm{X},\bm{Y}) \in H_{\mathrm{a}}^{V}\left(A_2\right)}\mathbb{P}(T_C^V>w_\alpha)=1-o(1).$$
	
\end{corollary}
Corollary \ref{corLC} indicates that the power of the max-sum test based on $T_C^V$ approaches one under the local alternative $H_{\mathrm{a}}^{V}\left(A_2\right)$.

Recall that the statistic based on Spearman's $\rho$ correlation proposed by \cite{Spearman1904} is an example of the class of simple linear rank-based statistics. Below, we will specifically demonstrate the max-sum test based on this correlation.
\begin{example}\label{EX1}
	For each $1\leq i\leq p$ and $1\leq j\leq q$, the rank-based statistic based on Spearman's $\rho$ correlation is
	\begin{align}
		\tilde{\rho}_{ij}\doteq \frac{12}{n(n^2-1)}\sum_{k=1}^{n}\Big(k-\frac{n+1}{2}\Big)\Big(R_{nk}^{ij}-\frac{n+1}{2}\Big).
	\end{align}
	Obviously, we have $\mathbb{E}_{H_0}(\tilde{\rho}_{ij})=0$. Based on $\tilde{\rho}_{ij}$, define $L_\rho=\max_{1\leq i\leq p,1\leq j\leq q}|\tilde{\rho}_{ij}|$, $S_\rho={\sum}_{1\leq i\leq p,1\leq j\leq q}\{\tilde{\rho}_{ij}^2-\mE_{H_0}(\tilde{\rho}_{ij}^2)\}$ and $T_{C}^{\rho}=-2\log P_{S_\rho}-2\log P_{L_\rho}$,
	where $P_{S_\rho}=1-\Phi(S_\rho/\hat{\sigma}_{S_\rho})$, $P_{L_\rho}=1-G(L_\rho^2 / \sigma_{L_\rho}^2-2 \log N+\log \log N)$ and $\sigma_{L_\rho}^2=\var_{H_0}(\tilde{\rho}_{ij})=(n-1)^{-1}$ \citep{hoeiffding1948n1}.
	Here, $\hat{\sigma}_{S_\rho}$ is obtained in a permutation way.
	Hence, the test based on $T_{C}^{\rho} $ rejects the null hypothesis in \eqref{H0} when $T_{C}^{\rho}>w_{\alpha}$.
\end{example}

\subsection{Non-degenerate  rank-based U-statistics}

The second class of rank-based statistics is the non-degenerate rank-based U-statistics. This type of statistics were previously studied by \cite{distributionfree} and \cite{testsum} for testing mutual independence, which include the test statistic based on Kendall's $\tau$ correlation proposed by \cite{kendall1938new}.

Specifically, for each $1\leq i \leq p$ and $1\leq j \leq q$, the non-degenerate rank-based U-statistic for the index pair $(i,j)$ is
\begin{align}\label{Uij}
	\tilde{U}_{i j}\doteq \frac{(n-m)!m!}{n!} \sum_{1 \leq k_1 < k_2, \cdots, < k_m \leq n} h\left\{\left(X_{k_1, i}, Y_{k_1, j}\right), \cdots,\left(X_{k_m, i}, Y_{k_m, j}\right)\right\},
\end{align}
where $h: \mathbb{R}^{2m}\rightarrow \mathbb{R}$ is a kernel function.  We assume that $h$ is symmetric,  because if $h$ is asymmetric, we can find a symmetric kernel $h'$ equivalent to it, i.e. the value of the U-statistic based on $h'$ is equal to that based on $h$ \citep{wang2024}.  We assume that $h$ is rank-based, i.e., $h$ has the property that  $h\left\{\left(X_{k_1, i}, Y_{k_1, j}\right), \cdots,\left(X_{k_m, i}, Y_{k_m, j}\right)\right\}\equiv h\left\{\left(R_{n k_1}^{i*}, Q_{n k_1}^{j*}\right), \cdots,\left(R_{n k_m}^{i*}, Q_{n k_m}^{j*}\right)\right\}$, where for each $1\leq s\leq m$, $R_{n k_s}^{i*}$ denotes the rank of $X_{k_s, i}$ in $\{X_{k_1, i},\cdots,X_{k_m, i}\}$ and $Q_{n k_s}^{j*}$ denotes the rank of $Y_{k_s, i}$ in $\{Y_{k_1, j},\cdots,Y_{k_m, j}\}$  \citep{testsum}.

%Let $(\bm{X},\bm{Y})\sim F(\cdot,\cdot)$ and $(X_i,Y_j)\sim F_{ij}(\cdot,\cdot)$ for each $1\leq i \leq p$ and $1\leq j \leq q$. Under $H_0$, $h$ is said to be non-degenerate if it is non-degenerate under $F_{ij}(\cdot,\cdot)$ for each $1\leq i \leq p$ and $1\leq j \leq q$.
Furthermore, we assume that $h$ is bounded and non-degenerate. The definition of non-degeneracy is as follows. Let $\{\bm{Z},\bm{Z}_1, \ldots, \bm{Z}_m\}$ denote $m+1$ mutually independent random vectors with the same distribution $\mathbb{P}_{\bm{Z}}$. For each $1\leq \ell \leq m$, let $\bm{z}_{\ell}$ denote any point in the value space of $\bm{Z}_{\ell}$, and let
\begin{align}
	\mathbb{E}(h)\doteq & \mathbb{E} h\left(\bm{Z}_1, \ldots, \bm{Z}_m\right),\\
	h_{\ell}\left(\bm{z}_1 \ldots, \bm{z}_{\ell} ; \mathbb{P}_{\bm{Z}}\right)\doteq & \mathbb{E} h\left(\bm{z}_1 \ldots, \bm{z}_{\ell}, \bm{Z}_{\ell+1}, \ldots, \bm{Z}_m\right),\\
	h^{(\ell)}\left(\bm{z}_1, \ldots, \bm{z}_{\ell} ; \mathbb{P}_{\bm{Z}}\right)
	\doteq &h_{\ell}\left(\bm{z}_1, \ldots, \bm{z}_{\ell} ; \mathbb{P}_{\bm{Z}}\right)-\mathbb{E}(h)\nonumber\\
	&-\sum_{k=1}^{\ell-1} \sum_{1 \leq i_1<\cdots<i_k \leq \ell} h^{(k)}\left(\bm{z}_{i_1}, \ldots, \bm{z}_{i_k} ; \mathbb{P}_{\bm{Z}}\right).\label{2.8}
\end{align}
The kernel $h$ is said to be non-degenerate under $\mathbb{P}_{\bm{Z}}$, if the variance of $h_1(\cdot)$, i.e. $\textrm{Var}(h_1)=\mathbb{E}\left\{h_1\left(\bm{Z}_1; \mathbb{P}_{\bm{Z}}\right)- \mathbb{E} h_1\left(\bm{Z}_1; \mathbb{P}_{\bm{Z}}\right)\right\}^2$ with $\mathbb{E} h_1\left(\bm{Z}_1; \mathbb{P}_{\bm{Z}}\right)=\mathbb{E}(h)$ is non-zero. On the other hand, the kernel $h$ is said to be degenerate with order $d>1$, if $\textrm{Var}(h_{d-1})=0$ and $\textrm{Var}(h_{d})\neq 0$ under $\mathbb{P}_{\bm{Z}}$.

\subsubsection{The max-type statistics}

Based on $\tilde{U}_{ij}$, the max-type statistic is
$$L_U \doteq \max_{1 \leq i\leq p,
	1\leq j \leq q}|\tilde{U}_{i j}|.
$$

Define $\big\{u_1,\ldots,u_N\big\}=\big\{\tilde{U}_{i j}/\sigma_{L_U}: 1\leq i\leq p,1\leq j\leq q\big\}$, where $\sigma_{L_U}^2=\var(\tilde{U}_{ij})$.
Define $\mathbf{\Sigma}_{U}=(\sigma^{U}_{st})_{1\leq s,t\leq N}$ with $\sigma_{s t}^U=\operatorname{cor}\left(u_s, u_t\right)$ for each $1\leq s,t\leq N$. To establish the asymptotic null distribution of $L_U$, we need to impose the following assumption.
\begin{assumption}\label{nondeh}
	$h$ is rank-based, symmetric and bounded. Furthermore, $h$ is mean-zero and non-degenerate under $U(0,1)^2$.
\end{assumption}
%\begin{assumption}\label{condi:sigmaU}
%$\boldsymbol{\Sigma}_V\in \mathcal{C}$. For some $C>0$, $\max _{1 \leq s \leq N} \sum_{t=1}^N (\sigma_{st}^{U})^2 \leq(\log N)^C$ for all $N \geq 3$.
%\end{assumption}

\begin{theorem}\label{th:LU}
	Suppose that $\boldsymbol{\Sigma}_U\in \mathcal{C}$ and Assumption \ref{nondeh} holds. Then, under $H_0$, for any $y \in \mathbb{R}$,
	$$
	\left|\mathbb{P}\left(L_U^2 / \sigma_{L_U}^2-2 \log N+\log \log N \leq y\right)-\exp \left\{-\pi^{-1 / 2} \exp (-y / 2)\right\}\right|=o_y(1),
	$$
	as $n,p \rightarrow \infty$, where $N=o\big(n^\epsilon\big)$ for some $\epsilon>0$.
\end{theorem}
According to Theorem \ref{th:LU}, the max-type test based on $L_U$ rejects the null hypothesis in \eqref{H0} when $L_U^2 / \sigma_{L_U}^2-2 \log N+\log \log N\geq q_{\alpha}$.

Next, we start the power analysis of the max-type test based on $L_U$. Define $\hat{\mathbf{U}}=(\hat{U}_{i j})_{1 \leq i\leq p, 1\leq j \leq q}$ with $\hat{U}_{i j}=n^{-1/2}\sigma_{L_U}^{-1}\tilde{U}_{i j}$ for each $1 \leq i\leq p$ and $1\leq j \leq q$.  Let $\mathbf{U} =(U_{i j})_{1\leq i\leq p,1\leq j\leq q}= \mE(\hat{\mathbf{U}})$. Consider the following alternative hypothesis
$$
H_{\mathrm{a}}^U(c) \doteq \left\{F(\cdot,\cdot): \mathbf{U}\in \mathcal{U}(c) \text{ if } (\bm{X},\bm{Y})\sim F(\cdot,\cdot)\right\}.
$$

\begin{theorem}\label{th:LUH1}
	Suppose that Assumption \ref{nondeh} holds and $|h(\cdot)| \leq a_3$ for some $a_3>0$. Further suppose that
	$$
	m^2 \operatorname{var}\left[\mE\left\{h\left(\bm{Z}_{1}, \ldots,\bm{Z}_{m}\right) \mid\bm{Z}_{1}\right\}\right]=a_{4}\{1+o(1)\}
	$$
	for some $a_4>0$, where $\bm{Z}_{1},\ldots,\bm{Z}_{m}$ are $m$ independent $2$-dimensional random vectors following $U(0,1)^2$. Then, if $(\bm{X},\bm{Y})\sim F(\cdot,\cdot)$,
	$$
	\inf _{F(\cdot,\cdot) \in H_{\mathrm{a}}^U\left(A_3\right)} \mathbb{P}\left(L_U^2 / \sigma_{L_U}^2-2 \log N+\log \log N\geq q_{\alpha}\right)=1-o(1),
	$$
	as $n,p\rightarrow \infty,$ where $A_3$ is a large scalar depending only on $a_3, a_4$ and $m$.
\end{theorem}

Furthermore, we present the rate-optimality of the max-type test based on $L_U$.
\begin{assumption}\label{assumpLUrate}
	Suppose that $(\bm{X}^\top,\bm{Y}^\top)^\top$ follows Gaussian distribution, and for sufficiently large $n$ and $p$, $c U_{i j} \leq R_{i j} \leq C U_{i j}$ for all $1 \leq i\leq p$ and $1\leq  j \leq q$, where $c$ and $C$ are two positive constants.
\end{assumption}

\begin{theorem}\label{th:LUH1 optimal}
	Suppose that the assumptions in Theorems \ref{th:LU}-\ref{th:LUH1} hold. Further suppose that Assumption \ref{assumpLUrate} holds. Then, the max-type test based on $L_U$ is rate-optimal, i.e. there exist two constants $B_3$ and $B_4$ with $B_3<B_4$, such that:
	\begin{itemize}
		\item[(i)] $\sup _{F(\cdot,\cdot) \in H_{\mathrm{a}}^U\left(B_4\right)} \mathbb{P}\left\{{I}(L_U^2 / \sigma_{L_U}^2-2 \log N+\log \log N\geq q_{\alpha})=0\right\}=o(1)$,
		\item[(ii)] for any $\beta>0$ satisfying $\alpha+\beta<1$,
		$\inf _{T_\alpha \in \mathcal{T}_\alpha} \sup _{F(\cdot,\cdot) \in H_{\mathrm{a}}^U\left(B_3\right)} \mathbb{P}\left( T_\alpha=0 \right) \geq 1-\alpha-\beta$,
		for sufficiently large $n$ and $p$.
	\end{itemize}
\end{theorem}

\subsubsection{The sum-type statistics}\label{SU}
The sum-type statistic based on $\tilde{U}_{i j}$ is
$$S_U \doteq \underset{1 \leq i\leq p,
	1\leq j \leq q}{\sum}\left\{\tilde{U}_{i j}^2-\mE_{H_0}\left(\tilde{U}_{i j}^2\right)\right\}.
$$

Below, we will establish the asymptotic null distribution of $S_U$ based on the following assumption together with Assumption \ref{assmp3} in Section \ref{SV}.
\begin{assumption}\label{condi:sigmaSU}
	For some $\delta>0$ and  $-1\leq \gamma <1,$  $\lim _{p \rightarrow \infty}m_p^{1+(1-\gamma)(1+2/\delta)}/p= 0$ and
	$\max\left\{m_p^{\gamma},m_p,m_p^{1+\gamma}p^{\frac{-\gamma}{1+(1-\gamma)(1+2/\delta)}}\right\}m_q pq/\var\left(n\sum_{1 \leq i\leq p,
		1\leq j \leq q}\tilde{U}_{ij}^2\right)=O(1).$
\end{assumption}
\begin{theorem}\label{th:SU}
	Suppose that Assumptions \ref{assmp3}, \ref{nondeh} and \ref{condi:sigmaSU} hold. Then, under $H_0$, $S_{U} /\sigma_{S_{U}} \rightarrow \Phi(x)$ in distribution as $n, p \rightarrow \infty$, where $\sigma^2_{S_{U}}\doteq\var\left(\sum_{1\leq i\leq p,1\leq j\leq q}\tilde{U}_{ij}^2\right).$
\end{theorem}
According to Theorem \ref{th:SU}, the sum-type test based on $S_U$ rejects the null hypothesis in \eqref{H0} when $S_{U} /\hat{\sigma}_{S_{U}}>z_\alpha$.
Similar to the sum-type test based on $S_V$ in Section \ref{SV}, we utilize the permutation technique to obtain an estimate of $\sigma^2_{S_{U}}$, denote as $\hat{\sigma}^2_{S_{U}}$.

\subsubsection{The max-sum statistics}
First, the asymptotic independence between $L_U$ and $S_U$ is established as follows.
\begin{theorem}\label{th:CU}
	Suppose that the assumptions in Theorems \ref{th:LU} and \ref{th:SU} hold, and for some $C>0$, $\max _{1 \leq s \leq N} \sum_{t=1}^N (\sigma_{st}^{U})^2 \leq(\log N)^C$ for all $N \geq 3$. Then, under $H_0$, $S_{U} /{\sigma}_{S_{U}}$ is asymptotically independent of $L_U^2 / \sigma_{L_U}^2-2 \log N+\log \log N$ as $n$, $p\rightarrow\infty$.
\end{theorem}

Then, we propose the max-sum statistic:
\begin{align}\label{TCU}
	T_{C}^{U}\doteq -2\log P_{S_U}-2\log P_{L_U},
\end{align}
where $	P_{S_U}=1-\Phi(S_U/\hat{\sigma}_{S_U})$ and $P_{L_U}=1-G(L_U^2 / \sigma_{L_U}^2-2 \log N+\log \log N)$. Based on the asymptotic independence in Theorem \ref{th:CU},  the asymptotic null distribution of $T_C^U$ is obtained.

\begin{corollary}
	Suppose that the assumptions in Theorem \ref{th:CU} hold. Then, under $H_0$, $T_C^U$ converges to $\chi_4^2$ in distribution as $n,p\rightarrow \infty$.
\end{corollary}

According to this corollary, the max-sum test based on $T_C^U$ rejects the null hypothesis in \eqref{H0} when $T_C^U>w_\alpha$.

Similar to Corollary \ref{corLC} in Section \ref{CV}, we present the power analysis of the max-sum test based on $T_C^U$.

\begin{corollary}
	Suppose that the assumptions in Theorems \ref{th:LUH1} and \ref{th:CU} hold. Then, for some sufficiently large scalar $A_4$, as $n,p \rightarrow \infty,$
	$$\inf _{F(\cdot,\cdot) \in H_{\mathrm{a}}^{U}\left(A_4\right)}\mathbb{P}(T_C^U>w_\alpha)=1-o(1).$$
	
\end{corollary}
Therefore, the power of the test based on $T_C^U$ approaches one as $n,p\rightarrow \infty$, under the sparse local alternative $H_{\mathrm{a}}^{U}\left(A_4\right)$.

The statistic based on Kendall's $\tau$ correlation proposed by \cite{kendall1938new} is an example of the class of non-degenerate rank-based U-statistics. Based on this correlation, the max-sum test is demonstrated as follows.

\begin{example}
	For each $1\leq i\leq p$ and $1\leq j\leq q$, the rank-based statistic based on Kendall's $\tau$ correlation is
	\begin{align}
		\tilde{\tau}_{ij}\doteq \frac{2}{n(n-1)}
		\underset{1\leq k_{1}<k_{2}\leq n}{\sum} h_{\tau}\left\{\left(X_{k_1, i}, Y_{k_1, j}\right),\left(X_{k_2, i}, Y_{k_2, j}\right)\right\},
	\end{align}
	where $h_{\tau}\left(\bm x_{1},\bm x_{2}\right)=\operatorname{sign}\left\{\left(x_{11}-x_{21}\right)\left(x_{12}-x_{22}\right)\right\}$ for any $\bm x_{1}=(x_{11},x_{12})$ and $\bm x_{2}=(x_{21},x_{22})$. Here, if $x\neq 0$, then $\operatorname{sign}(x)=x/|x|$, otherwise $\operatorname{sign}(x)=0$. Obviously, we have $\mathbb{E}_{H_0}(\tilde{\tau}_{ij})=0$. Define
	$L_\tau=\max_{1\leq i\leq p,1\leq j\leq q}|\tilde{\tau}_{ij}|$, $S_\tau=\underset{1\leq i\leq p,1\leq j\leq q}{\sum}\tilde{\tau}_{ij}^2-\mE_{H_0}(\tilde{\tau}_{ij}^2)$ and $T_{C}^{\tau}=-2\log P_{S_\tau}-2\log P_{L_\tau}$,
	where $P_{S_\tau}=1-\Phi(S_\tau/\hat{\sigma}_{S_\tau})$, $P_{L_\tau}=1-G(L_\tau^2 / \sigma_{L_\tau}^2-2 \log N+\log \log N)$ and $\sigma_{L_\tau}^2=\var_{H_0}(\tilde{\tau}_{ij})=\mE_{H_0}(\tilde{\tau}_{ij}^2)=2(2n + 5)/\{9n(n -1)\}$.  Similarly, $\hat{\sigma}_{S_\tau}$ is obtained in a permutation way. Hence, the max-sum test based on $T_{C}^{\tau}$ rejects the null hypothesis in \eqref{H0} when $T_{C}^{\tau}>w_{\alpha}$.
\end{example}

\subsection{Degenerate rank-based  U-statistics}

In this subsection, we consider the third class of rank-based statistics, i.e. the degenerate rank-based U-statistics, previously studied by \cite{testsum} and \cite{drton2020}. The motivation for proposing this class of statistics is to test the non-monotonic dependence between variables. Test methods based on the first two classes of statistics, such as the statistics based on Spearman's $\rho$ and Kendall's $\tau$ correlations, often cannot effectively detect such dependence \citep{Hoeffding1948}.    This class of statistics includes several commonly used rank-based statistics, such as those based on Hoeffding's D \citep{Hoeffding1948}, Blum-Kiefer-Rosenblatt's R \citep{Blum1961} and Bergsma-Dassios-Yanagimoto's $\tau^{*}$ \citep{Bergsma2014} correlations.

Note that this class of statistics is still based on $\tilde{U}_{i j}$ defined in \eqref{Uij}, but the kernel $h$ here is degenerate with order 2.

\subsubsection{The max-type statistics}

The max-type statistic based on $\tilde{U}_{i j}$ with a degenerate kernel $h$ is written as
$$L_Q \doteq \max _{1 \leq i\leq p,
	1\leq j \leq q}\tilde{U}_{i j}.
$$
Here, to distinguish it from max-type statistic $L_U$ based on a non-degenerate kernel, we use the subscript `$Q$' to represent the statistic based on a degenerate kernel.

To establish the asymptotic null distribution of $L_Q$, we impose the following two assumptions.
\begin{assumption}\label{assumptuihua}
	(1) Assume that the kernel $h$ is rank-based, symmetric and bounded. In situation where $\{\bm{Z}_1, \ldots, \bm{Z}_m\}$ are mutually independent random vectors following $U(0,1)^2$, $\mathbb{E}\left\{h\left(\bm{Z}_1,\dots, \bm{Z}_m\right)\right\}=0$ and $\mathbb{E}\left\{h_1\left(\bm{Z}_1 \right)\right\}^2=0$.\\  (2) In situation where $\{\bm{Z}_1, \ldots, \bm{Z}_m\}$ are mutually independent random vectors following $U(0,1)^2$,  $h_2\left(\cdot, \cdot \right)$ has uniformly bounded eigenfunctions, i.e. for each $\bm{z}_1=(z_{11},z_{12})$ and $\bm{z}_2=(z_{21},z_{22})$,
	$$
	h_2\left(\bm{z}_1, \bm{z}_2 \right)=\sum_{v=1}^{\infty} \lambda_v \phi_v\left(\bm{z}_1\right) \phi_v\left(\bm{z}_2\right),
	$$
	where $\left\{\lambda_v\right\}$ and $\left\{\phi_v\right\}$ are the eigenvalues and eigenfunctions satisfying the following integral equation:
	$$
	\mathbb{E} h_2\left(\bm{z}_1, \bm{Z}_2\right) \phi\left(\bm{Z}_2\right)=\lambda \phi\left(\bm{z}_1\right) \text { for all } \bm{z}_1,
	$$
	with $\lambda_1 \geq \lambda_2 \geq \cdots \geq 0$, $ \Lambda=\sum_{v=1}^{\infty} \lambda_v \in(0, \infty)$ and $\sup _v\left\|\phi_v\right\|_{\infty}<\infty$.
\end{assumption}
Before presenting Assumption \ref{condi:sigmaQ}, we introduce some additional notation and definitions.
Define $\boldsymbol{S}_{k, i j}=\{F_{i}(X_{k,i}),G_{j}(Y_{k,j})\},$ for all $1\leq k\leq n,$ $1\leq i\leq p$ and $1\leq j\leq q,$
where $F_{i}(\cdot)$ and $G_{j}(\cdot)$ are the cumulative distribution functions of $X_{k,i}$ and $Y_{k,j},$ respectively.
If $h$ is rank-based,
$h\left(\boldsymbol{S}_{1, ij}, \cdots, \boldsymbol{S}_{m, i j}\right)=
h\left(\boldsymbol{R}_{1, ij},\cdots,\boldsymbol{R}_{m, i j}\right),$
where $\boldsymbol{R}_{k, i j}=(R_{k,i}^{1},R_{k,j}^{2})$, $R_{k,i}^{1}$ is the rank of $F_i(X_{k,i})$ in $\{F_i(X_{1,i}),\cdots, F_i(X_{m,i})\}$ and $R_{k,j}^{2}$ is the rank of $G_j(Y_{k,j})$ in $\{G_j(Y_{1,j}),\cdots, G_j(Y_{m,j})\}$. In this case, the equation still holds after replacing $R_{k,i}^{1}$ with the rank of $X_{k,i}$ in $\{X_{1,i},\cdots, X_{m,i}\}$, because $F_i(\cdot)$ is an increasing function. The same applies when replacing $R_{k,j}^{2}$ with the rank of $Y_{k,j}$ in $\{Y_{1,j},\cdots, Y_{m,j}\}$.
Hence, for each $1\leq i\leq p$ and $1\leq j\leq q,$ $h\left(\boldsymbol{S}_{1, ij}, \cdots, \boldsymbol{S}_{m, i j}\right)=h\left(\bm{Z}_{1, ij}, \cdots, \bm{Z}_{m, i j}\right),$
where $\bm{Z}_{k,ij}=(X_{k,i},Y_{k,j})$. To simplify the notation, we sort $N$ pairs of $(i, j)$ with $1\leq i\leq p$ and $1\leq j\leq q$ as $(i_1,j_1),\ldots,(i_N,j_N)$. Define $Q_{v, ks}=\phi_v\left(\boldsymbol{S}_{k, i_s j_s}\right)$ for all $v\geq 1,$ $1\leq s \leq N$ and $1\leq k \leq n$. Define an absolute constant $\theta,$ such that
$$
\theta<\sup \bigg\{\gamma \in[0,1 / 3): \sum_{v>\left\lfloor n^{(1-3 \gamma) / 5}\right\rfloor} \lambda_v=O\left(n^{-\gamma}\right)\bigg\},
$$
if infinitely many eigenvalues $\lambda_v$ are nonzero, and $\theta=1 / 3$ otherwise. Define $\omega_{vs}=n^{-1 / 2} \sum_{k=1}^{n} Q_{v,ks}$ for all $1 \leq v \leq M$ and $1\leq s\leq N$, where if infinitely many eigenvalues are nonzero, $M=\lfloor n^{(1-3 \theta) / 5}\rfloor$, and if there are only finitely many nonzero eigenvalues, $M$ is the number of nonzero eigenvalues. For each $1\leq s\leq N$, define $\boldsymbol{\omega}_{s}=\left(\omega_{ 1s}, \cdots, \omega_{Ms}\right)$ and $\mathbf{\Xi}_{s}=\boldsymbol{\Sigma}_{s} \boldsymbol{\Sigma}_{s}^{\top}$, where $\boldsymbol{\Sigma}_{s}$ is the covariance matrix between $\boldsymbol{\omega}_{s}$ and $\boldsymbol{\omega}_{l}$ with $l\neq s$.

Based on the above notation and definitions, we present the following assumption that needs to be imposed.

\begin{assumption}\label{condi:sigmaQ}
	There exists a constant $\delta \in(0,1)$ satisfying $\lambda_{\max }\left(\mathbf{\Xi}_s\right) \leq \delta$ for all $1 \leq s\leq N$. Suppose that $\left\{\varphi_N: N \geq 1\right\}$ and $\left\{\varsigma_N : N \geq 1\right\}$ are positive constant sequences with $\varphi_N=o(1 / \log N)$ and $\varsigma_N \rightarrow 0$ as $N \rightarrow \infty$.  For each $1 \leq s \leq N$, define $C_N=\left\{1 \leq s \leq N: \left|B_{N, s}\right| \geq N^{\varsigma}\right\}$ and $B_{N, s}=\left\{1 \leq t \leq N: \lambda_{\max}(\mathbf{\Xi}_{st}\mathbf{\Xi}_{st}^{\top}) \geq \varphi_N^{2+2 c}\right\}$ for some constant $c>0$. Suppose that $\left|C_N\right| / N\rightarrow 0$ as $N \rightarrow \infty$.
\end{assumption}

Now, we are ready to present the asymptotic null distribution of $L_Q$.

\begin{theorem}\label{th:LQ}
	Suppose that Assumptions \ref{assumptuihua}-\ref{condi:sigmaQ} hold. Then, under $H_0$, for any absolute constant $y \in \mathbb{R}$,
	$$
	\begin{aligned}
		& \mathbb{P}\left\{\frac{n-1}{\lambda_1\binom{m}{2}} L_Q-2 \log N-\left(\mu_1-2\right) \log \log N+\frac{\Lambda}{\lambda_1} \leq y\right\} \\
		= & \exp \Big\{-\frac{\kappa}{\Gamma\left(\mu_1 / 2\right)} \exp \big(-y/2\big)\Big\}+o(1),
	\end{aligned}
	$$
	as $n,p \rightarrow \infty$, when $\log N=o\left(n^\theta\right)$. Here, $\kappa=\prod_{v=\mu_1+1}^{\infty}\left(1-\lambda_v / \lambda_1\right)^{-1 / 2}$, $\mu_1$ is the multiplicity of the largest eigenvalue $\lambda_1$ in the sequence $\{\lambda_1, \lambda_2, \cdots,\}$ and $\Gamma(z)=\int_0^{\infty} x^{z-1} e^{-x} d x$ is the Gamma function.
\end{theorem}

According to Theorem \ref{th:LQ}, the max-type test based on $L_Q$ rejects the null hypothesis in \eqref{H0}, when  $(n-1)\lambda_1^{-1}\tbinom{m}{2}^{-1} L_Q- 2 \log N-(\mu_1-2)\log \log N+\Lambda/\lambda_1 \geq l_\alpha $. Here, $$l_\alpha^h=\log \frac{ \kappa^2}{\left\{\Gamma\left(\mu_1 / 2\right)\right\}^2}-2 \log \log (1-\alpha)^{-1}$$ is the upper $\alpha$-quantile of the distribution function $L^h(y)=\exp\{- \exp(-y/2)\cdot\kappa/\Gamma(\mu_1/2)\}$, where we use subscript `$h$' to distinguish the parameters $\kappa$ and $\mu_1$ determined by $h$.

Next, we present the power analysis of the max-type test based on $L_Q$. For any kernel $h$, constants $\gamma>0$ and $p,q \in \mathbb{Z}^{+}$, define a family of distribution functions as follows:
\begin{align}
	\mathcal{D}(\gamma, p,q ; h)\doteq &\Big\{F(\cdot,\cdot): \var\left\{h^{(1)}\left(\bm{Z}_{k,ij}\right)\right\} \leq \gamma \mathbb{E}h(\bm{Z}_{1,ij},\cdots,\bm{Z}_{m,ij})\nonumber\\
	& ~~\text { for all } 1\leq k\leq n, 1 \leq i\leq p \text{ and } 1\leq j \leq q, \text{ if } (\bm{X},\bm{Y})\sim F(\cdot,\cdot)\Big\}.
\end{align}
If $(\bm{X},\bm{Y})\sim F(\cdot,\cdot)\in \mathcal{D}(\gamma, p,q ; h)$, then $\var\left\{h^{(1)}\left(\bm{Z}_{k,ij}\right)\right\} \leq \gamma \mathbb{E}h(\bm{Z}_{1,ij},\cdots,\bm{Z}_{m,ij})$ for each $1\leq k\leq n$, $1\leq i\leq p$ and $1\leq j\leq q$; further, if $\bm{X}$ is independent of $\bm{Y}$ and $h(\cdot)$ satisfies Assumption \ref{assumptuihua}, then
$$
\var\left\{h^{(1)}\left(\bm{Z}_{k,ij}\right)\right\} = \mathbb{E}h(\bm{Z}_{1,ij},\cdots,\bm{Z}_{m,ij})=0.
$$
Define
$$
\mathcal{V}(c) \doteq \left\{M=\left(m_{ij}\right)_{1 \leq i \leq p, 1\leq j\leq q}: \max _{1 \leq i \leq p, 1\leq j\leq q} |m_{ij}|  \geq c(\log N/n)\right\}.
$$

The following theorem establishes the property of $L_Q$ under the local alternative $\mathcal{V}(c)$.

\begin{theorem}\label{th:LQH1}
	Suppose that Assumption \ref{assumptuihua} holds. Then, for any $\gamma>0$, there exists some sufficiently large $A_5$ depending on $\gamma$, such that
	$$
	\inf _{F(\cdot,\cdot)\in \mathcal{D}(\gamma, p,q ; h),\breve{\mathbf{U}} \in \mathcal{V}(A_5)} \mathbb{P}\left\{\frac{n-1}{\lambda_1\binom{m}{2}}L_Q- 2 \log N-(\mu_1-2)\log \log N+\frac{\Lambda}{\lambda_1} \geq l_\alpha^h \right\}=1-o(1),
	$$
	as $n,p \rightarrow \infty$, where $\breve{\mathbf{U}}=(\breve{U}_{ij})_{1 \leq i \leq p, 1\leq j\leq q}=\mE(\tilde{U})$.
\end{theorem}

Furthermore, we present the rate-optimality of the max-type test based on $L_Q$.

\begin{assumption}\label{LQoptimal}
	Suppose that $(\bm{X}^\top,\bm{Y}^\top)^\top$ follows Gaussian distribution, and for sufficiently large $n$ and $p$, $c \breve{U}_{i j} \leqslant R^2_{i j} \leqslant C \breve{U}_{i j}$ for all $1 \leq i\leq p$ and $1\leq  j \leq q$, where $c$ and $C$ are two positive constants.  Further suppose that there exists an absolute constant $\gamma>0$, such that the distribution of $(\bm{X}^\top,\bm{Y}^\top)^\top$ belongs to $\mathcal{D}(\gamma,p,q;h)$.
\end{assumption}

Then, we establish the rate-optimality result of $L_Q$ in the following theorem.
\begin{theorem}\label{th:LQH1 optimal}
	Suppose that the assumptions in Theorems \ref{th:LQ} and \ref{th:LQH1} hold. Further suppose that Assumption \ref{LQoptimal} holds. Then, the max-type test based on $L_Q$ is rate-optimal, i.e. there exist two constants $B_5$ and $B_6$ with $B_5<B_6$, such that:
	\begin{itemize}
		\item[(i)] $\sup_{\breve{\mathbf{U}} \in \mathcal{V}(B_6)}   \mathbb{P}\left[{I}\left\{\frac{n-1}{\lambda_1\binom{m}{2}}L_Q- 2 \log N-(\mu_1-2)\log \log N+\frac{\Lambda}{\lambda_1} \geq  l_\alpha^h\right\}=0\right]=o(1)$,
		\item[(ii)] for any $\beta>0$ satisfying $\alpha+\beta<1$,
		$\inf_{T_\alpha \in \mathcal{T}_\alpha} \sup _{\breve{\mathbf{U}} \in \mathcal{V}(B_5)} \mathbb{P}\left(T_\alpha =0\right) \geqslant 1-\alpha-\beta,$
		for sufficiently large $n$ and $p$.
	\end{itemize}
\end{theorem}

%\begin{remark}
%	When $(\bm X,\bm Y)$ follows Gaussian distribution, according to Lemma 4.1 in \cite{drton2020}, we can obtain that there exists an absolute constant $\gamma >0$ such that $p+q$-dimensional Gaussian distribution is in $\mathcal{D}(\gamma,p,q;h_{D})$, $\mathcal{D}(\gamma,p,q;h_{R})$ and $\mathcal{D}(\gamma,p,q;h_{\tau^{*}})$. Moreover, note the fact in \cite{drton2020} that for any $1\leq i\leq p$, $1\leq j\leq q$,
%	$$\tilde{D}_{ij},\tilde{R}_{ij},\tilde{\tau}^{*}_{ij}\asymp r_{ij}^2 \text{ as } r_{ij}\rightarrow 0.$$ Hence, we can easily obtain the rate-optimality of the tests $L_D$, $L_R$ and $L_{\tau^{*}}$.
%\end{remark}

\subsubsection{The sum-type statistics}

Similar to $L_Q$, the sum-type statistic based on $\tilde{U}_{i j}$ with a degenerate kernel $h$ is written as
$$
S_Q \doteq \underset{1 \leq i\leq p,
	1\leq j \leq q}{\sum}\left\{\tilde{U}_{i j}^2-\mE_{H_0}\left(\tilde{U}_{i j}^2\right)\right\}.
$$
\begin{assumption}\label{condi:sigmaSQ}
	For some $\delta>0$ and  $-1\leq \gamma <1,$  $\lim _{p \rightarrow \infty}m_p^{1+(1-\gamma)(1+2/\delta)}/p= 0$ and
	$\max\left\{m_p^{\gamma},m_p,m_p^{1+\gamma}p^{\frac{-\gamma}{1+(1-\gamma)(1+2/\delta)}}\right\}m_q pqn^{-2}/\var\Big(n\sum_{1 \leq i\leq p,
		1\leq j \leq q}\tilde{U}_{ij}^2\Big)=O(1).$
\end{assumption}
\begin{theorem}\label{th:SQ}
	Suppose that Assumptions \ref{assmp3}, \ref{assumptuihua} and \ref{condi:sigmaSQ} hold. Then, under $H_0$, $S_{Q} /\sigma_{S_{Q}} \rightarrow \Phi(x)$ in distribution as $n, p \rightarrow \infty$, where $\sigma^2_{S_{Q}}\doteq \var\left(\sum_{1 \leq i\leq p, 	1\leq j \leq q}\tilde{U}_{ij}^2\right).$
\end{theorem}

According to Theorem \ref{th:SQ}, the sum-type test based on $S_Q$ rejects the null hypothesis in \eqref{H0}, when $S_{Q} /\hat{\sigma}_{S_{Q}}>z_\alpha$. Similar to Sections \ref{SV} and \ref{SU}, we employ the permutation method to obtain an estimate of $\sigma^2_{S_{Q}}$, denoted as $\hat{\sigma}^2_{S_{Q}}$.

\subsubsection{The max-sum statistics}

Then, we establish the asymptotic independence between $L_Q$ and $S_Q$ as follows.

\begin{theorem}\label{th:CQ}
	Suppose that the assumptions in Theorems \ref{th:LQ} and \ref{th:SQ} hold, and for some $C>0$, $\max _{1 \leq s \leq N} \sum_{t=1}^N \lambda_{\max}(\mathbf{\Xi}_{st}\mathbf{\Xi}_{st}^{\top}) \leq(\log N)^C$. Then, under $H_0$, $S_{Q} /{\sigma}_{S_{Q}}$ is asymptotically independent of $(n-1)\lambda_1^{-1}\binom{m}{2}^{-1}L_Q-2 \log N-\left(\mu_1-2\right) \log \log N+\Lambda/\lambda_1$ as $n$, $p\rightarrow\infty$.
\end{theorem}

Based on $S_Q$ and $L_Q$, we construct the max-sum statistic
\begin{align}\label{TCQ}
	T_{C}^{Q}\doteq -2\log P_{S_Q}-2\log P_{L_Q},
\end{align} where $	P_{S_Q}=1-\Phi(S_Q/\hat{\sigma}_{S_Q})$, $P_{L_Q}=1-L^h\{(n-1)\lambda_1^{-1}\tbinom{m}{2}^{-1} L_Q- 2 \log N-(\mu_1-2)\log \log N+\Lambda/\lambda_1 \}$.
Due to the asymptotic independence in Theorem \ref{th:CQ}, we establish the asymptotic null distribution of $T_C^Q$.

\begin{corollary}
	Suppose that the assumptions in Theorem \ref{th:CQ} hold. Then, under $H_0$, $T_C^Q$ converges to $\chi_4^2$ in distribution as $n,p\rightarrow \infty$.
\end{corollary}
Hence, the max-sum test based on $T_{C}^{Q}$ rejects the null hypothesis in \eqref{H0}, when $T_C^Q>w_\alpha$.

Below, we will present that the local power of the test based on $T_C^Q$ approaches one as both $n$ and $p$ tend towards infinity.

\begin{corollary}
	Suppose that the assumptions in Theorems \ref{th:LQH1} and \ref{th:CQ} hold. Then, there exists some sufficiently large $A_6$ depending on $\gamma$, such that as $n,p \rightarrow \infty,$
	$$\inf _{F(\bm X,\bm Y)\in \mathcal{D}(\gamma, p,q ; h) ,\breve{U}\in \mathcal{V}(A_6)}\mathbb{P}(T_C^Q>w_\alpha)=1-o(1).$$
\end{corollary}

Three examples of the tests based on degenerate rank-based U-statistics are presented as follows.

\begin{example}
	For each $1\leq i\leq p$ and $1\leq j\leq q$, the rank-based statistic based on Hoeffding's D correlation is
	\begin{align}
		\tilde{D}_{ij}=\frac{5!(n-5)!}{n!}
		\underset{1\leq k_{1}<\cdots<k_{5}\leq n}{\sum}	 h_{D}\left\{\left(X_{k_1, i},Y_{k_1, j}\right),\cdots,\left(X_{k_5, i}, Y_{k_5, j}\right)\right\},
	\end{align}
	where $h_{D}\left(\bm x_{{1}},\bm x_{{2}},\cdots,\bm x_{{5}}\right)$ is defined as
	\begin{align*}
		\frac{1}{16} \sum_{\substack{(\pi_{1},\cdots,\pi_{5}) \in \mathfrak{S}({1},\cdots,{5})}} & \left[\left\{{I}\left(x_{\pi_{1}1} \leq x_{\pi_{5}1}\right)-{I}\left(x_{\pi_{2}1} \leq x_{\pi_{5}1}\right)\right\}\left\{{I}\left(x_{\pi_{3}1} \leq x_{\pi_{5}1}\right)-{I}\left(x_{\pi_{4}1} \leq x_{\pi_{5}1}\right)\right\}\right]
		\\
		&\left[\left\{{I}\left(x_{\pi_{1}2} \leq x_{\pi_{5}2}\right)-{I}\left(x_{\pi_{2}2} \leq X _{\pi_{5}2}\right)\right\}\left\{{I}\left(x_{\pi_{3}2} \leq x_{\pi_{5}2}\right)-{I}\left(x_{\pi_{4}2} \leq x_{\pi_{5}2}\right)\right\}\right],
	\end{align*}
	with $\bm x_{k}=(x_{k1},x_{k2})$ for each $1\leq k\leq 5$. Obviously, we have $\mathbb{E}_{H_0}(\tilde{D}_{ij})=0$. Define
	$L_D=\max_{1\leq i\leq p,1\leq j\leq q}\tilde{D}_{ij}$, $S_D=\sum_{1\leq i\leq p,1\leq j\leq q}\{\tilde{D}_{ij}^2-\mE_{H_0}(\tilde{D}_{ij}^2)\}$ and $T_{C}^{D}=-2\log P_{S_D}-2\log P_{L_D}$, where $\mathbb{E}_{H_0}(\tilde{D}_{ij}^2)=2\left(n^{2}+5 n-32\right)/\left\{9 n(n-1)(n-3)(n-4)\right\},$ $P_{S_D}=1-\Phi(S_D/\hat{\sigma}_{S_D})$, $P_{L_D}=1-L^D\{\pi^4(n-1)L_{D}/30- 2 \log N+\log \log N+\pi^4/36 \}$, $L^D(y)=\exp \left\{-\kappa_{D}/\sqrt{\pi} \exp \left(-y/2\right)\right\}$ and $\kappa_{D}\approx2.467$. Here, $\hat{\sigma}_{S_D}$ is obtained in a permutation way. Hence, the max-sum test based on $T_{C}^{D}$ rejects the null hypothesis in \eqref{H0} when  $T_{C}^{D}>w_{\alpha}$.
\end{example}

\begin{example}
	For each $1\leq i\leq p$ and $1\leq j\leq q$, the rank-based statistic based on Blum-Kiefer-Rosenblatt's $R$ correlation is
	\begin{align}
		\tilde{R}_{ij}\doteq \frac{(n-6)!6!}{n!}
		\underset{1\leq k_{1}<\cdots<k_{6}\leq n}{\sum} h_{R} \left\{\left(X_{k_1, i}, Y_{k_1, j}\right),\cdots,\left(X_{k_6, i}, Y_{k_6, j}\right)\right\},
	\end{align}
	where $h_{R}\left(\bm x_1,\bm x_{2},\cdots,\bm x_{6}\right)$ is defined as
	\begin{align*}
		\frac{1}{32} \sum_{\substack{(\pi_{1},\cdots,\pi_{6}) \in \mathfrak{S}(1,\cdots,6)}}
		&\left[\left\{{I}\left(x_{\pi_{1}s} \leq x_{\pi_{5}s}\right)-{I}\left(x_{\pi_{2}s} \leq x_{\pi_{5}s}\right)\right\}\left\{{I}\left(x_{\pi_{3}s} \leq x_{\pi_{5}s}\right)-{I}\left(x_{\pi_{4}s} \leq x_{\pi_{5}s}\right)\right\}\right]
		\\
		&\left[\left\{{I}\left(x_{\pi_{1}t} \leq x_{\pi_{6}t}\right)-{I}\left(x_{\pi_{2}t} \leq X _{\pi_{6}t}\right)\right\}\left\{{I}\left(x_{\pi_{3}t} \leq x_{\pi_{6}t}\right)-{I}\left(x_{\pi_{4}t} \leq x_{\pi_{6}t}\right)\right\}\right]
	\end{align*}
	with $\bm x_{k}=(x_{k1},x_{k2})$ for each $1\leq k \leq 6$. Obviously, we have $\mathbb{E}_{H_0}(\tilde{R}_{ij})=0$. Define $L_R=\max_{1\leq i\leq p,1\leq j\leq q}\tilde{R}_{ij}$, $S_R=\underset{1\leq i\leq p,1\leq j\leq q}{\sum}\tilde{R}_{ij}^2-\mE_{H_0}(\tilde{R}_{ij}^2)$ and $T_{C}^{R}=-2\log P_{S_R}-2\log P_{L_R}$,
	where $P_{S_R}=1-\Phi(S_R/\hat{\sigma}_{S_R})$, $P_{L_R}=1-L^R\{\pi^4(n-1)L_{R}/90- 2 \log N+\log \log N+\pi^4/36 \}$, $\mathbb{E}_{H_0}(\tilde{R}_{ij}^2)=2(n^3-3n^2-6n+10)/\left\{n(n-1)(n-2)(n-3)(n-4)\right\},$
	$L^R(y)=\exp \left\{-\kappa_{R}/\sqrt{\pi} \exp \left(-y/2\right)\right\}$ and $\kappa_{R}\approx2.467$. $\hat{\sigma}_{S_R}$ is obtained in a permutation way. Hence, the max-sum test based on $T_{C}^{R}$ rejects the null hypothesis in \eqref{H0} when $T_{C}^{R}>w_{\alpha}$.
\end{example}

\begin{example}\label{EX5}
	For each $1\leq i\leq p$ and $1\leq j\leq q$, the rank-based statistic based on Bergsma-Dassios-Yanagimoto $\tau^*$ correlation is
	\begin{align}
		\tilde{\tau}^{*}_{ij}= \frac{(n-1)!4!}{n!}
		\underset{1\leq k_{1}<k_{2}<k_3<k_4\leq n}{\sum} h_{\tau^{*}}\left\{\left(X_{k_1, i}, Y_{k_1, j}\right),\cdots,\left(X_{k_4, i}, Y_{k_4, j}\right)\right\},
	\end{align}
	where $h_{\tau^*}\left(\bm x_1,\bm x_2,\cdots,\bm x_4\right)$ is defined as
	\begin{align*} 		
		\frac{1}{4!} \sum_{\substack{(\pi_{1},\cdots,\pi_{4}) \in \mathfrak{S}(1,\cdots,4)}}
		&\left\{\mathbb{I}\left(x_{\pi_{1}1}, x_{\pi_{3}1}<x_{\pi_{2}1}, x_{\pi_{4}1}\right)+\mathbb{I}\left(x_{\pi_{2}1}, x_{\pi_{4}1}<x_{\pi_{1}1}, x_{\pi_{3}1}\right)\right.\\
		&~\left.-\mathbb{I}\left(x_{\pi_{1}1}, x_{\pi_{4}1}<x_{\pi_{2}1}, x_{\pi_{3}1}\right)-\mathbb{I}\left(x_{\pi_{2}1}, x_{\pi_{3}1}<x_{\pi_{1}1}, x_{\pi_{4}1}\right)\right\} \\
		&\left\{\mathbb{I}\left(x_{\pi_{1}2}, x_{\pi_{3}2}<x_{\pi_{2}2}, x_{\pi_{4}2}\right)+\mathbb{I}\left(x_{\pi_{2}2}, x_{\pi_{4}2}<x_{\pi_{1}2}, x_{\pi_{3}2}\right)\right.\\
		&~\left.-\mathbb{I}\left(x_{\pi_{1}2}, x_{\pi_{4}2}<x_{\pi_{2}2}, x_{\pi_{3}2}\right)-\mathbb{I}\left(x_{\pi_{2}2}, x_{\pi_{3}2}<x_{\pi_{1}2}, x_{\pi_{4}2}\right)\right\}
	\end{align*}
	with $\bm x_{k}=(x_{k1},x_{k2})$ for each $1\leq k\leq 4$. Here,
	\begin{align*}
		\mathbb{I}\left(x_{\pi_{1}1}, x_{\pi_{3}1}<x_{\pi_{2}1}, x_{\pi_{4}1}\right)\doteq{I}\left(x_{\pi_{1}1}<x_{\pi_{2}1}\right) {I}\left(x_{\pi_{1}1}< x_{\pi_{4}1}\right) {I}\left(x_{\pi_{3}1}<x_{\pi_{2}1}\right) {I}\left(x_{\pi_{3}1}<x_{\pi_{4}1}\right),
	\end{align*}
	and other such symbols are defined in the same way. Obviously, we have $\mathbb{E}_{H_0}(\tilde{\tau}^{*}_{ij})=0$. Define $L_{\tau^{*}}=\max_{1\leq i\leq p,1\leq j\leq q}\tilde{\tau}^{*}_{ij}$, $S_{\tau^{*}}=\sum_{1\leq i\leq p,1\leq j\leq q}(\tilde{\tau}^{*}_{ij})^2-\mE_{H_0}\{(\tilde{\tau}^{*}_{ij})^2\}$ and $T_{C}^{\tau^{*}}=-2\log P_{S_{\tau^{*}}}-2\log P_{L_{\tau^{*}}}$,
	where $P_{S_{\tau^{*}}}=1-\Phi(S_{\tau^{*}}/\hat{\sigma}_{S_{\tau^{*}}})$, $P_{L_{\tau^{*}}}=1-L^{\tau^*}\{\pi^4(n-1)L_{\tau^{*}}/36- 2 \log N+\log \log N+\pi^4/36 \}$, $\mathbb{E}_{H_0}\{(\tilde{\tau}^{*}_{ij})^2\}=8(3 n^{2}+5 n-18)/\{75n(n-1)(n-2)(n-3)\}$, $L^{\tau^*}(y)=\exp \left\{-\kappa_{\tau^{*}}/\sqrt{\pi} \exp \left(-y/2\right)\right\}$ and $\kappa_{\tau^{*}}\approx2.467$.  $\hat{\sigma}_{S_{\tau^*}}$ is obtained in a permutation way. Hence,  the max-sum test based on $T_{C}^{\tau^{*}}$ rejects the null hypothesis in \eqref{H0} when $T_{C}^{\tau^{*}}>w_{\alpha}$.
\end{example}

It is worthy noting that compared with the problem of testing the mutual independence between a large number of random variables studied in \cite{wang2024}, establishing the theoretical properties of the proposed three classes of rank tests in this paper is much more difficult. This is due to the dependence structure within each of the two random vectors needs to be properly treated, which significantly differentiates the theoretical framework from that used for testing mutual independence in \cite{wang2024}. In addition, compared with the sum-type tests proposed in \cite{zhouyeqing2024}, based on the sum of the estimated rank correlations, establishing the theoretical properties of the proposed sum-type tests is much more complex, due to the squaring operation.

\section{Numerical results}\label{sec:simu}

We now present some numerical results to illustrate the finite sample performance of five representative examples of the proposed classes of rank-based testing methods.
The sum-type, max-type and max-sum tests proposed in Examples \ref{EX1}-\ref{EX5} are abbreviated as $S_\rho$, $L_{\rho}$, $T_C^\rho$, $S_\tau$, $L_{\tau}$, $T_C^\tau$, $S_D$, $L_{D}$, $T_C^D$, $S_R$, $L_{R}$, $T_C^R$, $S_{\tau^*}$, $L_{\tau^*}$ and ${T_C^{\tau^{*}}}$, respectively. We will compare these methods with some existing ones, including four tests proposed by \cite{Zhuchangbo}, i.e. the studentized distance covariance based test, the studentized marginal distance covariance based test, the studentized Hilbert-Schmidt covariance based test with Gaussian kernel and the studentized Hilbert-Schmidt covariance based test with Laplacian kernel, abbreviated as $T_{dCov}$, $T_{mdCov}$, $T_{hCov_1}$, $T_{hCov_2}$, respectively. The rescaled test based on the bias-corrected distance correlation proposed by \cite{Gaolan} is also included, which is abbreviated as $T_n$. In addition, the sum-type tests based on Hoeffding's D, Blum-Kiefer-Rosenblatt's R and Bergsma-Dassios-Yanagimoto's $\tau^*$ correlations proposed by \cite{zhouyeqing2024} are included, where the test statistics are based on the sum of estimated correlations, rather than the sum of squared estimated correlations used in the proposed sum-type tests. These three tests are abbreviated as $D$, $R$ and $\tau^*$, respectively.

From experience, the size of some of the proposed tests in Examples \ref{EX1}-\ref{EX5} is a little larger than the nominal level, which is likely due to their slow convergence speed. Hence, we have made adjustments to these methods to speed up the convergence. Specifically, for tests $S_D$, $S_R$ and $S_{\tau^*}$, we reject $H_0$ when ${S}_D/\left\{\hat{\sigma}_{S_D} \left(1+2 / n^{1 / 2}\right)\right\}>z_\alpha$, ${S}_{R}/\left\{\hat{\sigma}_{S_{R}} \left(1+2 / n^{1 / 2}\right)\right\}>z_\alpha$ and ${S}_{\tau^*}/\left\{\hat{\sigma}_{S_{\tau^*}} \left(1+2 / n^{1 / 2}\right)\right\}>z_\alpha$, respectively. To distinguish from the tests before adjustment, we have written the adjusted tests as ${S}_D^{\prime}$, ${S}_R^{\prime}$ and ${S}_{\tau^*}^{\prime}$, respectively. In addition, for test ${L}_D$, we reject $H_0$ when $\left\{\pi^4(n-1){L}_D/30-2 \log N+\log \log N+\pi^4/36\right\} /\{1+2 / \log (n p)\}>l_\alpha^D$. Similarly, the adjusted $L_D$ test is written as ${L}_D^{\prime}$. Based on all these adjusted tests, the corresponding max-sum tests  are written as ${T_{C}^D}^{\prime}$, ${T_{C}^R}^{\prime}$ and ${T_{C}^{\tau^*}}^{\prime}$, respectively.

To investigate the size performance of the above tests, we consider the following data generating process. Set $n=100$, $p=q\in \{50,150\}$ and $m_p=m_q=m=3$.
Let $\{(\bm{X}_k^\top, \bm{Y}_k^\top)^\top\}_{k=1}^n$ denote the sample of $n$ independent replicates of $(\bm{X}^\top,\bm Y^\top)^\top$, where $\bm{X}_k=(X_{k,1},\cdots,X_{k,p})^{\top}$ and $\bm{Y}_k=(Y_{k,1},\cdots,Y_{k,q})^{\top}$, for each $1\leq k \leq n$.  Let $\mathbf{E}_1=(\epsilon_{k,t}^1)_{1\leq k\leq n,1\leq t\leq(p+m)}$
and $\mathbf{E}_2=(\epsilon_{k,t}^2)_{1\leq k\leq n,1\leq t\leq(q+m)}$.
For each $1\leq k\leq n$, $1\leq i\leq p$ and $1\leq j\leq q$, let ${X}_{k,i}=\sigma_{x,1}\epsilon_{k,i+1}^1+\cdots+\sigma_{x,m}\epsilon_{k,i+m}^1$
and ${Y}_{k,j}=\sigma_{y,1}\epsilon_{k,j+1}^2+\cdots+\sigma_{y,m}\epsilon_{k,j+m}^2$,
where $\sigma_{x,1},\ldots,\sigma_{x,m}$ are independent and identically distributed (iid) from $U(1,2)$ and $\sigma_{y,1},\ldots,\sigma_{y,m}$ are iid from $U(0.5,1,5)$.
The elements of $\mathbf{E}_1$ and $\mathbf{E}_2$ are iid from one of the following two distributions:
\begin{enumerate}
	\item[(i)] normal distribution: $N(0,1)$,\label{norm}
	\item[(ii)] chi-square distribution: $(\chi_{1}^2-1)/\sqrt{2}$\label{chisq}.
\end{enumerate}

Next, we investigate the power performance of these tests. $\sigma_{x,1},\ldots,\sigma_{x,m}$, $\sigma_{y,1},\ldots,\sigma_{y,m}$, $\mathbf{E}_1$ and $\mathbf{E}_2$ are generated in the same way as above. For each $1\leq k\leq n$ and $1\leq i,j\leq p=q$, let $Z_{k,i}=\cos(\sigma_{x,1}\epsilon_{k,i+1}^{1}+\cdots+\sigma_{x,m}\epsilon_{k,i+m}^{1})$ and $Z^*_{k,j}=\sin(\sigma_{y,1}\epsilon_{k,j+1}^{2}+\cdots+\sigma_{y,m}\epsilon_{k,j+m}^{2})$ and then $X_{k,i}$ and ${Y}_{k,j}$ are generated from one of the following four settings:
\begin{enumerate}[label=(\alph*)]
	\item non-sparse case 1: ${X}_{k,i}=0.2\cos(Z_{k,i})$ and $Y_{k,j}=0.2\sin(Z_{k,j})$;\label{ns1}
	
	\item non-sparse case 2: $X_{k,i}=Z_{k,i}$ and ${Y}_{k,j}=0.01\log(|X_{k,j}|^3)$;\label{ns2}
	
	\item sparse case 1:  $X_{k,i}=Z_{k,i}$, $Y_{k,j}=0.2X_{k,j}-0.4X_{k,j+1}+0.6X_{k,j-1}+Z^*_{k,j}$ if $1\leq j\leq 3$, and $Y_{k,j}=Z^*_{k,j}$ if $4\leq j\leq q$.	\label{sl1}
	
	\item sparse case 2: $X_{k,i}=Z_{k,i}$, $Y_{k,j}=0.5\exp(X_{k,i})+Z^*_{k,j}$ if $1\leq j\leq 3$, and $Y_{k,j}=Z^*_{k,j}$ if $4\leq j\leq q$.
	\label{sn2}
\end{enumerate}

Table \ref{t1} summarizes the empirical size results of the involved tests under the null hypothesis with distributions (i)-(ii), where all results, as well as all subsequent results in this section, are based on 1,000 replications and $B=50$ permutations when implementing the proposed sum-type tests. The results suggest that all these involved tests can control the empirical size, where some max-type tests, such as $L_\rho$, are somewhat conservative.

\begin{table}[]
	\centering
	\caption{The empirical size of the involved tests under the null hypothesis with distributions (i)-(ii) at 5\% level.}\label{t1}
	\vspace{0.35cm}
	\renewcommand{\arraystretch}{0.98}
	\tabcolsep 3pt
	\scalebox{0.8}{
		\begin{tabular}{c|cc|cc}
			\hline
			\multicolumn{1}{c|}{}
			& \multicolumn{2}{c|}{normal distribution}                       & \multicolumn{2}{c}{chi-square distribution}                \\ \hline
			& \multicolumn{1}{c|}{n=100, p=50} & \multicolumn{1}{c|}{n=100, p=150}& \multicolumn{1}{c|}{n=100, p=50} &n=100, p=150 \\ \hline
			%		\multicolumn{1}{c|}{$T_1$}          & \multicolumn{1}{c|}{0.045} & \multicolumn{1}{c|}{0.040}  & \multicolumn{1}{c|}{0.046} & 0.048       \\ \hline
			%		\multicolumn{1}{c|}{$T_2$}          & \multicolumn{1}{c|}{0.045} & \multicolumn{1}{c|}{0.040}  & \multicolumn{1}{c|}{0.046} & 0.048       \\ \hline
			\multicolumn{1}{c|}{$T_n$}          & \multicolumn{1}{c|}{0.048} & \multicolumn{1}{c|}{0.041} & \multicolumn{1}{c|}{0.048} & 0.052       \\ \hline
			%		\multicolumn{1}{c|}{$dCov$}         & \multicolumn{1}{c|}{0.027} & \multicolumn{1}{c|}{0.054} & \multicolumn{1}{c|}{0.042} & 0.056       \\ \hline
			%		\multicolumn{1}{c|}{$mdCov$}        & \multicolumn{1}{c|}{0.039} & \multicolumn{1}{c|}{0.055} & \multicolumn{1}{c|}{0.051} & 0.049       \\ \hline
			\multicolumn{1}{c|}{$T_{dCov}$}     & \multicolumn{1}{c|}{0.045} & \multicolumn{1}{c|}{0.040}  & \multicolumn{1}{c|}{0.046} & 0.048       \\ \hline
			\multicolumn{1}{c|}{$T_{mdCov}$}    & \multicolumn{1}{c|}{0.041} & \multicolumn{1}{c|}{0.043} & \multicolumn{1}{c|}{0.051} & 0.053       \\ \hline
			%		\multicolumn{1}{c|}{$hCov_1$}       & \multicolumn{1}{c|}{0.026} & \multicolumn{1}{c|}{0.052} & \multicolumn{1}{c|}{0.047} & 0.051       \\ \hline
			%		\multicolumn{1}{c|}{$mhCov_1$}      & \multicolumn{1}{c|}{0.040}  & \multicolumn{1}{c|}{0.05}  & \multicolumn{1}{c|}{0.053} & 0.052       \\ \hline
			\multicolumn{1}{c|}{$T_{hCov_1}$}   & \multicolumn{1}{c|}{0.045} & \multicolumn{1}{c|}{0.040}  & \multicolumn{1}{c|}{0.046} & 0.047       \\ \hline
			%		\multicolumn{1}{c|}{$T_{mhCov_1}$}  & \multicolumn{1}{c|}{0.050}  & \multicolumn{1}{c|}{0.05}  & \multicolumn{1}{c|}{0.053} & 0.05        \\ \hline
			%		\multicolumn{1}{c|}{$hCov_2$}       & \multicolumn{1}{c|}{0.028} & \multicolumn{1}{c|}{0.051} & \multicolumn{1}{c|}{0.05}  & 0.054       \\ \hline
			%		\multicolumn{1}{c|}{$mhCov_2$}      & \multicolumn{1}{c|}{0.051} & \multicolumn{1}{c|}{0.043} & \multicolumn{1}{c|}{0.059} & 0.051       \\ \hline
			
			\multicolumn{1}{c|}{$T_{hCov_2}$}   & \multicolumn{1}{c|}{0.040}  & \multicolumn{1}{c|}{0.042} & \multicolumn{1}{c|}{0.054} & 0.052       \\ \hline
			%		\multicolumn{1}{c|}{$T_{mhCov_2}$}  & \multicolumn{1}{c|}{0.055} & \multicolumn{1}{c|}{0.043} & \multicolumn{1}{c|}{0.066} & 0.053       \\ \hline
			\multicolumn{1}{c|}{$D$}      & \multicolumn{1}{c|}{0.060} & \multicolumn{1}{c|}{0.052} & \multicolumn{1}{c|}{ 0.063 } & 0.058        \\ \hline
			\multicolumn{1}{c|}{$R$}      & \multicolumn{1}{c|}{0.056} & \multicolumn{1}{c|}{0.046} & \multicolumn{1}{c|}{0.055 } &   0.057     \\ \hline
			\multicolumn{1}{c|}{$\tau^*$}      & \multicolumn{1}{c|}{0.057} & \multicolumn{1}{c|}{0.048} & \multicolumn{1}{c|}{0.056} & 0.055        \\ \hline
			\multicolumn{1}{c|}{$S_{\rho}$}     & \multicolumn{1}{c|}{0.054} & \multicolumn{1}{c|}{0.057} & \multicolumn{1}{c|}{0.057} & 0.053       \\ \hline
			\multicolumn{1}{c|}{$L_{\rho}$}     & \multicolumn{1}{c|}{0.026} & \multicolumn{1}{c|}{0.012} & \multicolumn{1}{c|}{0.018} & 0.018       \\ \hline
			%		\multicolumn{1}{c|}{$T_C^\rho$}       & \multicolumn{1}{c|}{0.033} & \multicolumn{1}{c|}{0.035} & \multicolumn{1}{c|}{0.031} & 0.035       \\ \hline
			\multicolumn{1}{c|}{$T_C^\rho$}       & \multicolumn{1}{c|}{0.041} & \multicolumn{1}{c|}{0.029} & \multicolumn{1}{c|}{0.042} & 0.031       \\ \hline
			\multicolumn{1}{c|}{$S_\tau$}       & \multicolumn{1}{c|}{0.054} & \multicolumn{1}{c|}{0.056} & \multicolumn{1}{c|}{0.056} & 0.054       \\ \hline
			\multicolumn{1}{c|}{$L_\tau$}       & \multicolumn{1}{c|}{0.039} & \multicolumn{1}{c|}{0.035} & \multicolumn{1}{c|}{0.029} & 0.035       \\ \hline
			%		\multicolumn{1}{c|}{$T_C^\tau$}       & \multicolumn{1}{c|}{0.043} & \multicolumn{1}{c|}{0.042} & \multicolumn{1}{c|}{0.035} & 0.041       \\ \hline
			\multicolumn{1}{c|}{$T_C^\tau$}       & \multicolumn{1}{c|}{0.050}  & \multicolumn{1}{c|}{0.047} & \multicolumn{1}{c|}{0.044} & 0.040        \\ \hline
			\multicolumn{1}{c|}{$S_D^{\prime}$}          & \multicolumn{1}{c|}{0.033} & \multicolumn{1}{c|}{0.028} & \multicolumn{1}{c|}{0.033} & 0.035       \\ \hline
			\multicolumn{1}{c|}{$L_D^{\prime}$}          & \multicolumn{1}{c|}{0.040}  & \multicolumn{1}{c|}{0.044} & \multicolumn{1}{c|}{0.025} & 0.036       \\ \hline
			%		\multicolumn{1}{c|}{$T_C^D$}          & \multicolumn{1}{c|}{0.032} & \multicolumn{1}{c|}{0.04}  & \multicolumn{1}{c|}{0.026} & 0.032       \\ \hline
			\multicolumn{1}{c|}{${T_C^D}^{\prime}$}          & \multicolumn{1}{c|}{0.054} & \multicolumn{1}{c|}{0.049} & \multicolumn{1}{c|}{0.042} & 0.047       \\ \hline
			\multicolumn{1}{c|}{$S_R^{\prime}$}          & \multicolumn{1}{c|}{0.030}  & \multicolumn{1}{c|}{0.031} & \multicolumn{1}{c|}{0.036} & 0.040       \\ \hline
			\multicolumn{1}{c|}{$L_R$}          & \multicolumn{1}{c|}{0.040}  & \multicolumn{1}{c|}{0.025} & \multicolumn{1}{c|}{0.026} & 0.034       \\ \hline
			%		\multicolumn{1}{c|}{$T_C^R$}          & \multicolumn{1}{c|}{0.033} & \multicolumn{1}{c|}{0.027} & \multicolumn{1}{c|}{0.027} & 0.036       \\ \hline
			\multicolumn{1}{c|}{${T_C^R}^{\prime}$}          & \multicolumn{1}{c|}{0.054} & \multicolumn{1}{c|}{0.031} & \multicolumn{1}{c|}{0.042} & 0.045       \\ \hline
			\multicolumn{1}{c|}{$S_{\tau^{*}}^{\prime}$} & \multicolumn{1}{c|}{0.031} & \multicolumn{1}{c|}{0.030}  & \multicolumn{1}{c|}{0.035} & 0.040        \\ \hline
			\multicolumn{1}{c|}{$L_{\tau^{*}}$} & \multicolumn{1}{c|}{0.053} & \multicolumn{1}{c|}{0.039} & \multicolumn{1}{c|}{0.031} & 0.036       \\ \hline
			%		\multicolumn{1}{c|}{${T_C^{\tau^{*}}}$} & \multicolumn{1}{c|}{0.037} & \multicolumn{1}{c|}{0.032} & \multicolumn{1}{c|}{0.03}  & 0.038       \\ \hline
			\multicolumn{1}{c|}{${T_C^{\tau^{*}}}^{\prime}$} & \multicolumn{1}{c|}{0.062} & \multicolumn{1}{c|}{0.039} & \multicolumn{1}{c|}{0.051} & 0.054       \\ \hline
	\end{tabular}}
\end{table}

Table \ref{t2} summarizes the empirical power results of the involved tests under the alternative hypothesis with non-sparse settings \ref{ns1}-\ref{ns2} and distributions (i)-(ii). The table suggests that under non-sparse alternatives, the proposed tests based on Hoeffding's D, Blum-Kiefer-Rosenblatt's R and Bergsma-Dassios-Yanagimoto $\tau^*$ correlations as well as $T_{mdCov}$ outperform the remaining ones in terms of empirical power.

\begin{table}[]
	\centering
	\caption{The empirical power of the tests under the alternative hypothesis with non-sparse settings \ref{ns1}-\ref{ns2} and distributions (i)-(ii) at 5\% level.}\label{t2}
	\vspace{0.35cm}
	\renewcommand{\arraystretch}{0.98}
	\tabcolsep 3pt
	\scalebox{0.68}{
		\begin{tabular}{c|cc|cc|cc|cc}
			\hline
			& \multicolumn{4}{c|}{setting \ref{ns1}}   & \multicolumn{4}{c}{setting \ref{ns2}} \\ \hline
			\multicolumn{1}{c|}{} & \multicolumn{2}{c|}{normal distribution}                       & \multicolumn{2}{c|}{chi-square distribution}               & \multicolumn{2}{c|}{normal distribution}                      & \multicolumn{2}{c}{chi-square distribution}                 \\ \hline
			& \multicolumn{1}{c|}{n=100, p=50} & n=100, p=150 & \multicolumn{1}{c|}{n=100, p=50} & n=100, p=150 & \multicolumn{1}{c|}{n=100, p=50} & n=100, p=150 & \multicolumn{1}{c|}{n=100, p=50} & n=100, p=150 \\ \hline
			$T_n$                 & \multicolumn{1}{c|}{0.27}       & 0.10        & \multicolumn{1}{c|}{0.57}       & 1.00        & \multicolumn{1}{c|}{0.08}       & 0.07        & \multicolumn{1}{c|}{0.20}       & 0.28        \\ \hline
			%		$dCov$                & \multicolumn{1}{c|}{0.27}       & 0.10        & \multicolumn{1}{c|}{0.56}       & 1.00        & \multicolumn{1}{c|}{0.06}       & 0.06        & \multicolumn{1}{c|}{0.17}       & 0.23        \\ \hline
			%		$mdCov$               & \multicolumn{1}{c|}{1.00}       & 1.00        & \multicolumn{1}{c|}{1.00}       & 1.00        & \multicolumn{1}{c|}{1.00}       & 1.00        & \multicolumn{1}{c|}{1.00}       & 1.00        \\ \hline
			$T_{dCov}$            & \multicolumn{1}{c|}{0.27}       & 0.10        & \multicolumn{1}{c|}{0.56}       & 1.00        & \multicolumn{1}{c|}{0.08}       & 0.06        & \multicolumn{1}{c|}{0.19}       & 0.28        \\ \hline
			$T_{mdCov}$           & \multicolumn{1}{c|}{1.00}       & 1.00        & \multicolumn{1}{c|}{1.00}       & 1.00        & \multicolumn{1}{c|}{1.00}       & 1.00        & \multicolumn{1}{c|}{1.00}       & 1.00        \\ \hline
			%		$hCov_1$              & \multicolumn{1}{c|}{0.27}       & 0.10        & \multicolumn{1}{c|}{0.56}       & 1.00        & \multicolumn{1}{c|}{0.07}       & 0.05        & \multicolumn{1}{c|}{0.16}       & 0.26        \\ \hline
			%		$mhCov_1$             & \multicolumn{1}{c|}{1.00}       & 1.00        & \multicolumn{1}{c|}{1.00}       & 1.00        & \multicolumn{1}{c|}{1.00}       & 1.00        & \multicolumn{1}{c|}{1.00}       & 1.00        \\ \hline
			$T_{hCov_1}$          & \multicolumn{1}{c|}{0.27}       & 0.10        & \multicolumn{1}{c|}{0.56}       & 1.00        & \multicolumn{1}{c|}{0.08}       & 0.06        & \multicolumn{1}{c|}{0.19}       & 0.28        \\ \hline
			%		$T_{mhCov_1}$         & \multicolumn{1}{c|}{1.00}       & 1.00        & \multicolumn{1}{c|}{1.00}       & 1.00        & \multicolumn{1}{c|}{1.00}       & 1.00        & \multicolumn{1}{c|}{1.00}       & 1.00        \\ \hline
			%		$hCov_2$              & \multicolumn{1}{c|}{0.35}       & 0.11        & \multicolumn{1}{c|}{0.64}       & 1.00        & \multicolumn{1}{c|}{0.09}       & 0.06        & \multicolumn{1}{c|}{0.23}       & 0.27        \\ \hline
			%		$mhCov_2$             & \multicolumn{1}{c|}{1.00}       & 1.00        & \multicolumn{1}{c|}{1.00}       & 1.00        & \multicolumn{1}{c|}{1.00}       & 1.00        & \multicolumn{1}{c|}{1.00}       & 1.00        \\ \hline
			$T_{hCov_2}$          & \multicolumn{1}{c|}{0.36}       & 0.11        & \multicolumn{1}{c|}{0.65}       & 1.00        & \multicolumn{1}{c|}{0.10}       & 0.07        & \multicolumn{1}{c|}{0.25}       & 0.31        \\ \hline
			%		$T_{mhCov_2}$         & \multicolumn{1}{c|}{1.00}       & 1.00        & \multicolumn{1}{c|}{1.00}       & 1.00        & \multicolumn{1}{c|}{1.00}       & 1.00        & \multicolumn{1}{c|}{1.00}       & 1.00        \\ \hline
			$D$            & \multicolumn{1}{c|}{1.00 }       &  1.00        & \multicolumn{1}{c|}{1.00 }       &  1.00       & \multicolumn{1}{c|}{1.00 }       & 1.00         & \multicolumn{1}{c|}{1.00 }       &  1.00        \\ \hline
			$R$            & \multicolumn{1}{c|}{1.00 }       &  1.00        & \multicolumn{1}{c|}{1.00 }       & 1.00        & \multicolumn{1}{c|}{1.00 }       &    1.00      & \multicolumn{1}{c|}{1.00 }       &  1.00        \\ \hline
			$\tau^*$            & \multicolumn{1}{c|}{1.00 }       &  1.00        & \multicolumn{1}{c|}{1.00 }       & 1.00        & \multicolumn{1}{c|}{1.00 }       &1.00          & \multicolumn{1}{c|}{1.00 }       &       1.00   \\ \hline
			$S_{\rho}$            & \multicolumn{1}{c|}{0.88}       & 0.32        & \multicolumn{1}{c|}{0.85}       & 1.00        & \multicolumn{1}{c|}{0.17}       & 0.22        & \multicolumn{1}{c|}{0.71}       & 0.98        \\ \hline
			$L_{\rho}$            & \multicolumn{1}{c|}{0.82}       & 0.23        & \multicolumn{1}{c|}{0.35}       & 1.00        & \multicolumn{1}{c|}{0.10}       & 0.14        & \multicolumn{1}{c|}{0.55}       & 0.74        \\ \hline
			%		$T_C^\rho$              & \multicolumn{1}{c|}{0.93}       & 0.35        & \multicolumn{1}{c|}{0.81}       & 1.00        & \multicolumn{1}{c|}{0.16}       & 0.21        & \multicolumn{1}{c|}{0.73}       & 0.98        \\ \hline
			$T_C^\rho$              & \multicolumn{1}{c|}{0.96}       & 0.41        & \multicolumn{1}{c|}{0.85}       & 1.00        & \multicolumn{1}{c|}{0.19}       & 0.24        & \multicolumn{1}{c|}{0.82}       & 0.99        \\ \hline
			$S_\tau$              & \multicolumn{1}{c|}{1.00}       & 0.79        & \multicolumn{1}{c|}{0.97}       & 1.00        & \multicolumn{1}{c|}{0.44}       & 0.58        & \multicolumn{1}{c|}{0.88}       & 1.00        \\ \hline
			$L_\tau$              & \multicolumn{1}{c|}{1.00}       & 0.96        & \multicolumn{1}{c|}{0.88}       & 1.00        & \multicolumn{1}{c|}{0.54}       & 0.83        & \multicolumn{1}{c|}{0.94}       & 1.00        \\ \hline
			%		$T_C^\tau$              & \multicolumn{1}{c|}{1.00}       & 0.96        & \multicolumn{1}{c|}{0.97}       & 1.00        & \multicolumn{1}{c|}{0.59}       & 0.85        & \multicolumn{1}{c|}{0.96}       & 1.00        \\ \hline
			$T_C^\tau$              & \multicolumn{1}{c|}{1.00}       & 0.99        & \multicolumn{1}{c|}{0.99}       & 1.00        & \multicolumn{1}{c|}{0.68}       & 0.91        & \multicolumn{1}{c|}{0.98}       & 1.00        \\ \hline
			$S_D^{\prime}$                 & \multicolumn{1}{c|}{1.00}       & 1.00        & \multicolumn{1}{c|}{1.00}       & 1.00        & \multicolumn{1}{c|}{1.00}       & 1.00        & \multicolumn{1}{c|}{1.00}       & 1.00        \\ \hline
			$L_D^{\prime}$                 & \multicolumn{1}{c|}{1.00}       & 1.00        & \multicolumn{1}{c|}{1.00}       & 1.00        & \multicolumn{1}{c|}{1.00}       & 1.00        & \multicolumn{1}{c|}{1.00}       & 1.00        \\ \hline
			%		$T_C^D$                 & \multicolumn{1}{c|}{1.00}       & 1.00        & \multicolumn{1}{c|}{1.00}       & 1.00        & \multicolumn{1}{c|}{1.00}       & 1.00        & \multicolumn{1}{c|}{1.00}       & 1.00        \\ \hline
			${T_C^D}^{\prime}$                 & \multicolumn{1}{c|}{1.00}       & 1.00        & \multicolumn{1}{c|}{1.00}       & 1.00        & \multicolumn{1}{c|}{1.00}       & 1.00        & \multicolumn{1}{c|}{1.00}       & 1.00        \\ \hline
			$S_R^{\prime}$                 & \multicolumn{1}{c|}{1.00}       & 1.00        & \multicolumn{1}{c|}{1.00}       & 1.00        & \multicolumn{1}{c|}{1.00}       & 1.00        & \multicolumn{1}{c|}{1.00}       & 1.00        \\ \hline
			$L_R$                 & \multicolumn{1}{c|}{1.00}       & 1.00        & \multicolumn{1}{c|}{1.00}       & 1.00        & \multicolumn{1}{c|}{1.00}       & 1.00        & \multicolumn{1}{c|}{1.00}       & 1.00        \\ \hline
			%		$T_C^R$                 & \multicolumn{1}{c|}{1.00}       & 1.00        & \multicolumn{1}{c|}{1.00}       & 1.00        & \multicolumn{1}{c|}{1.00}       & 1.00        & \multicolumn{1}{c|}{1.00}       & 1.00        \\ \hline
			${T_C^R}^{\prime}$                 & \multicolumn{1}{c|}{1.00}       & 1.00        & \multicolumn{1}{c|}{1.00}       & 1.00        & \multicolumn{1}{c|}{1.00}       & 1.00        & \multicolumn{1}{c|}{1.00}       & 1.00        \\ \hline
			$S_{\tau^{*}}^{\prime}$        & \multicolumn{1}{c|}{1.00}       & 1.00        & \multicolumn{1}{c|}{1.00}       & 1.00        & \multicolumn{1}{c|}{1.00}       & 1.00        & \multicolumn{1}{c|}{1.00}       & 1.00        \\ \hline
			$L_{\tau^{*}}$        & \multicolumn{1}{c|}{1.00}       & 1.00        & \multicolumn{1}{c|}{1.00}       & 1.00        & \multicolumn{1}{c|}{1.00}       & 1.00        & \multicolumn{1}{c|}{1.00}       & 1.00        \\ \hline
			%		${T_C^{\tau^{*}}}$        & \multicolumn{1}{c|}{1.00}       & 1.00        & \multicolumn{1}{c|}{1.00}       & 1.00        & \multicolumn{1}{c|}{1.00}       & 1.00        & \multicolumn{1}{c|}{1.00}       & 1.00        \\ \hline
			${T_C^{\tau^{*}}}^{\prime}$        & \multicolumn{1}{c|}{1.00}       & 1.00        & \multicolumn{1}{c|}{1.00}       & 1.00        & \multicolumn{1}{c|}{1.00}       & 1.00        & \multicolumn{1}{c|}{1.00}       & 1.00        \\ \hline
	\end{tabular}}
\end{table}

On the other hand, Table \ref{t3} summarizes the empirical power results of the involved tests under the alternative hypothesis with sparse settings \ref{sl1}-\ref{sn2} and distributions (i)-(ii). It suggests that under sparse alternatives, the proposed max-type and max-sum tests outperform the remaining ones in terms of empirical power.

\begin{table}[]
	\centering
	\caption{The empirical power of the tests under settings \ref{sl1}-\ref{sn2} with distributions (i)-(ii) at 5\% level.}\label{t3}
	\vspace{0.35cm}
	\renewcommand{\arraystretch}{0.98}
	\tabcolsep 3pt
	\scalebox{0.68}{
		\begin{tabular}{c|cc|cc|cc|cc}
			\hline
			& \multicolumn{4}{c|}{setting \ref{sl1}}   & \multicolumn{4}{c}{setting \ref{sn2}} \\ \hline
			\multicolumn{1}{c|}{} & \multicolumn{2}{c|}{normal distribution}                       & \multicolumn{2}{c|}{chi-square distribution}               & \multicolumn{2}{c|}{normal distribution}                      & \multicolumn{2}{c}{chi-square distribution}                 \\ \hline
			& \multicolumn{1}{c|}{n=100, p=50} & n=100, p=150 & \multicolumn{1}{c|}{n=100, p=50} & n=100, p=150 & \multicolumn{1}{c|}{n=100, p=50} & n=100, p=150 & \multicolumn{1}{c|}{n=100, p=50} & n=100, p=150 \\ \hline
			$T_n$         & \multicolumn{1}{c|}{0.44}       &   0.12      & \multicolumn{1}{c|}{ 0.39}       & 0.13
			& \multicolumn{1}{c|}{0.35}       & 0.12        & \multicolumn{1}{c|}{0.31}       & 0.12             \\ \hline
			$T_{dCov}$       & \multicolumn{1}{c|}{ 0.44 }       &  0.12      & \multicolumn{1}{c|}{0.39 }       & 0.13      & \multicolumn{1}{c|}{0.35}       & 0.11        & \multicolumn{1}{c|}{0.30}       & 0.12               \\ \hline
			$T_{mdCov}$    & \multicolumn{1}{c|}{0.23 }       &    0.08      & \multicolumn{1}{c|}{0.22 }       &  0.10             & \multicolumn{1}{c|}{0.22}       & 0.08        & \multicolumn{1}{c|}{0.20}       & 0.11          \\ \hline
			
			$T_{hCov_1}$      & \multicolumn{1}{c|}{0.44}       &  0.11        & \multicolumn{1}{c|}{0.39}       & 0.13        & \multicolumn{1}{c|}{0.35}       & 0.11        & \multicolumn{1}{c|}{0.30}       & 0.12             \\ \hline
			
			$T_{hCov_2}$         & \multicolumn{1}{c|}{ 0.42}       & 0.12        & \multicolumn{1}{c|}{ 0.37}       & 0.13     & \multicolumn{1}{c|}{0.33}       & 0.11        & \multicolumn{1}{c|}{0.29}       & 0.12           \\ \hline
			
			$D$      & \multicolumn{1}{c|}{ 0.23 }       &  0.09      & \multicolumn{1}{c|}{0.23 }       & 0.10        & \multicolumn{1}{c|}{0.25 }       &    0.10      & \multicolumn{1}{c|}{0.22}       & 0.11             \\ \hline
			$R$      & \multicolumn{1}{c|}{ 0.21 }       &   0.09      & \multicolumn{1}{c|}{0.21}       &  0.09         & \multicolumn{1}{c|}{0.20 }       & 0.08         & \multicolumn{1}{c|}{ 0.18 }       &  0.10              \\ \hline
			$\tau^*$      & \multicolumn{1}{c|}{0.21}       &  0.09       & \multicolumn{1}{c|}{ 0.22}       & 0.09        & \multicolumn{1}{c|}{0.22}       &  0.09       & \multicolumn{1}{c|}{0.19}       & 0.11             \\ \hline
			$S_{\rho}$      & \multicolumn{1}{c|}{0.24}       &  0.09
			& \multicolumn{1}{c|}{0.21}       &0.07   & \multicolumn{1}{c|}{0.23}       & 0.07        & \multicolumn{1}{c|}{0.19}       & 0.09             \\ \hline
			$L_{\rho}$      & \multicolumn{1}{c|}{0.91}       &0.84  & \multicolumn{1}{c|}{ 0.96}       & 0.87         & \multicolumn{1}{c|}{0.97}       & 0.87        & \multicolumn{1}{c|}{0.97}       & 0.96              \\ \hline
			
			$T_C^\rho$         & \multicolumn{1}{c|}{ 0.88}       &0.78       & \multicolumn{1}{c|}{ 0.93}       &   0.79           & \multicolumn{1}{c|}{0.95}       & 0.80        & \multicolumn{1}{c|}{0.94}       & 0.91          \\ \hline
			$S_\tau$       & \multicolumn{1}{c|}{ 0.25}       &  0.08        & \multicolumn{1}{c|}{ 0.23}       & 0.07         & \multicolumn{1}{c|}{0.26}       & 0.08        & \multicolumn{1}{c|}{0.22}       & 0.10              \\ \hline
			$L_\tau$      & \multicolumn{1}{c|}{ 0.93 }       &  0.88         & \multicolumn{1}{c|}{ 0.98}       &   0.91      & \multicolumn{1}{c|}{0.99}       & 0.95        & \multicolumn{1}{c|}{0.99}       & 0.99             \\ \hline
			
			$T_C^\tau$     & \multicolumn{1}{c|}{0.91  }       &  0.83         & \multicolumn{1}{c|}{ 0.96}       &    0.87             & \multicolumn{1}{c|}{0.98}       & 0.91        & \multicolumn{1}{c|}{0.98}       & 0.96            \\ \hline
			$S_D^{\prime}$            & \multicolumn{1}{c|}{0.82 }       &   0.29       & \multicolumn{1}{c|}{ 0.91}       &   0.35            & \multicolumn{1}{c|}{0.91}       & 0.37        & \multicolumn{1}{c|}{0.92}       & 0.47          \\ \hline
			$L_D^{\prime}$              & \multicolumn{1}{c|}{0.86 }       & 0.79         & \multicolumn{1}{c|}{0.96}       &0.89            & \multicolumn{1}{c|}{0.98}       & 0.94        & \multicolumn{1}{c|}{0.99}       & 0.98        \\ \hline
			
			${T_C^D}^{\prime}$              & \multicolumn{1}{c|}{  0.92}       &  0.78       & \multicolumn{1}{c|}{ 0.97}       & 0.88         & \multicolumn{1}{c|}{0.99}       & 0.91        & \multicolumn{1}{c|}{0.99}       & 0.97        \\ \hline
			$S_R^{\prime}$              & \multicolumn{1}{c|}{ 0.81  }       &  0.27             & \multicolumn{1}{c|}{0.86 }       &  0.29             & \multicolumn{1}{c|}{0.79}       & 0.24        & \multicolumn{1}{c|}{0.79}       & 0.28       \\ \hline
			$L_R$       & \multicolumn{1}{c|}{0.89}       &   0.81   & \multicolumn{1}{c|}{0.96}       &   0.86     & \multicolumn{1}{c|}{0.95}       & 0.83        & \multicolumn{1}{c|}{0.97}       & 0.95              \\ \hline
			
			${T_C^R}^{\prime}$              & \multicolumn{1}{c|}{ 0.92}       &    0.79       & \multicolumn{1}{c|}{ 0.97}       &   0.84        & \multicolumn{1}{c|}{0.96}       & 0.81        & \multicolumn{1}{c|}{0.97}       & 0.92        \\ \hline
			$S_{\tau^{*}}^{\prime}$       & \multicolumn{1}{c|}{ 0.82   }       & 0.28      & \multicolumn{1}{c|}{ 0.89}       &   0.31          & \multicolumn{1}{c|}{0.85}       & 0.28        & \multicolumn{1}{c|}{0.85}       & 0.35        \\ \hline
			$L_{\tau^{*}}$    & \multicolumn{1}{c|}{ 0.90}       & 0.82        & \multicolumn{1}{c|}{0.97 }       &   0.89    & \multicolumn{1}{c|}{0.97}       & 0.90        & \multicolumn{1}{c|}{0.98}       & 0.98              \\ \hline
			
			${T_C^{\tau^{*}}}^{\prime}$       & \multicolumn{1}{c|}{ 0.94}       &  0.82       & \multicolumn{1}{c|}{0.98}       & 0.88         & \multicolumn{1}{c|}{0.97}       & 0.86        & \multicolumn{1}{c|}{0.98}       & 0.96        \\ \hline
	\end{tabular}}
\end{table}

From Tables \ref{t2} and \ref{t3}, it can be seen that the sparsity of the dependency relationship between the random vectors $\bm X$ and $\bm Y$ has a significant impact on the power performance of the tests involved. Therefore, we consider an additional setting similar to that considered in \cite{zhouyeqing2024}, where the sparsity of dependency relationships varies.

\begin{enumerate}[label=(\alph*),resume]
	\item  Set $n=100$ and $p=q=50$. Let $\mathbf{\Sigma}=(\sigma_{st})_{1\leq s,t\leq p}$, where for each $1\leq s,t\leq p$, $\sigma_{s s}=s^{1/2}$, and $\sigma_{s t}=0.3 \sqrt{\sigma_{s s} \sigma_{t t}}$ when $s \neq t$ and $|s-t| \leq 3$, and $\sigma_{s t}=0$ when $s \neq t$ and $|s-t|>3$. $\{\bm X_k\}_{k=1}^n$ are iid from $N(\bm 0,\mathbf{\Sigma})$, and $Y_{k,j}=a X^2_{k,j} +\epsilon_{k,j}$ for each $1\leq k\leq n$ and $1\leq j \leq \lfloor(v/2)^{3}\rfloor$, where $\{\epsilon_{k,j}\}$ are mutually independent variables from $N(0,j^2)$, $a=6.8/\{(\log v )^{0.4}\sqrt{\log (pq) /n}\}$ and $\lfloor(v/2)^{3}\rfloor$ with $v\in \{2,3,4,5,6,7\}$ is the number of correlations in this setting. In addition, the remaining components $\{(Y_{k,\lfloor(v/2)^{3}\rfloor+1},\ldots,Y_{k,q})^\top\}_{k=1}^n$ are generated as follows: (1) $\{\tilde{\bm{Y}}_k\}_{k=1}^n$ are iid from $N(\bm0,\mathbf{\Sigma}^*)$ and independent of $\{\bm X_k\}_{k=1}^n$, where $\mathbf{\Sigma}^*=(\sigma_{st})_{1\leq s,t\leq q-\lfloor(v/2)^{3}\rfloor}$; (2) $(Y_{k,\lfloor(v/2)^{3}\rfloor+1},\ldots,Y_{k,q})^\top=\tilde {\bm{Y}}_k^2$. \label{ee}
\end{enumerate}

\begin{figure}[!t]
	\centering
	\includegraphics[width=6.2in]{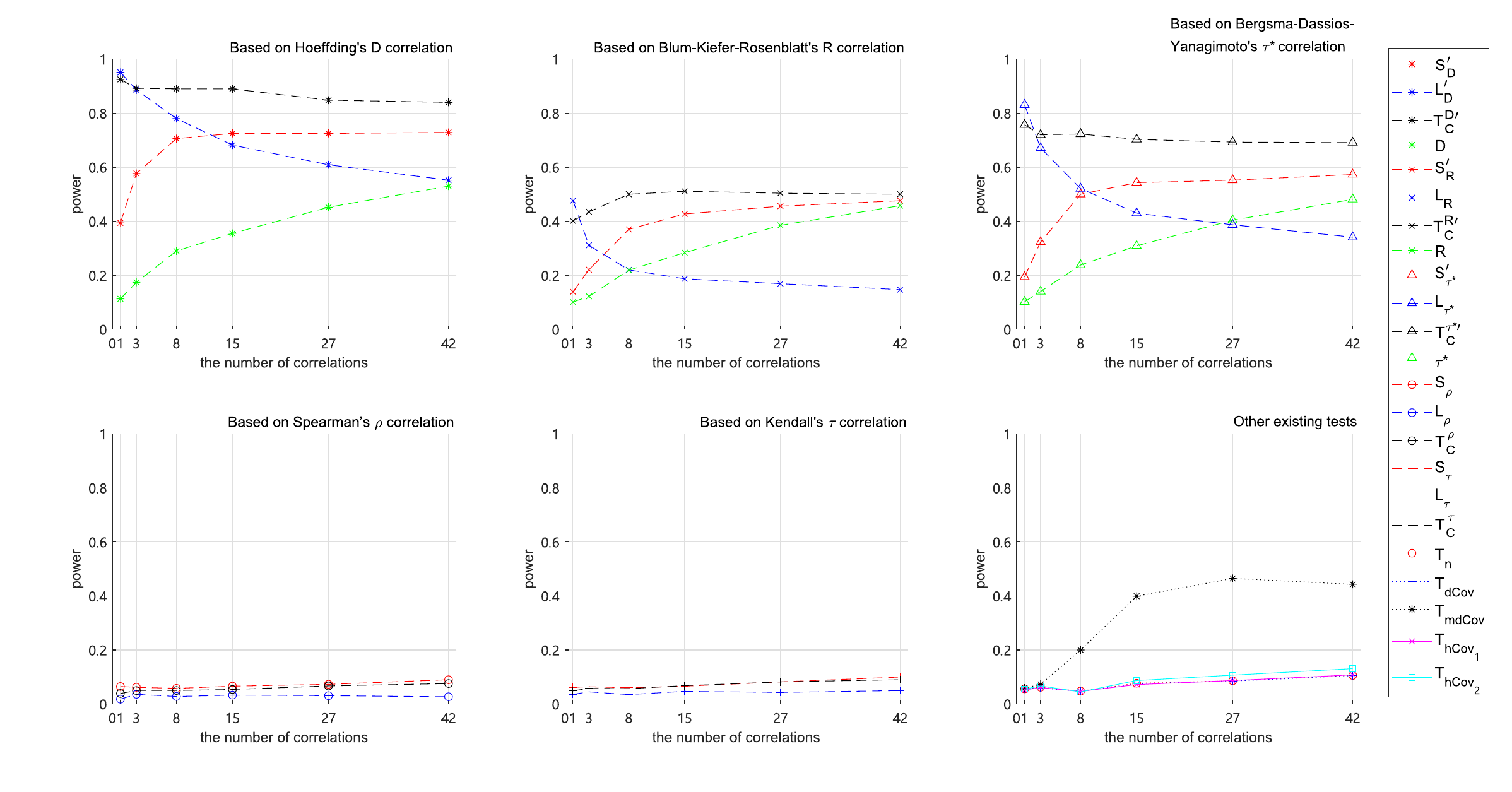}
	\caption{The empirical power curves of the tests at 5\% level.}\label{f1}
\end{figure}

Figure \ref{f1} presents the curves of empirical power of the involved tests, where the horizontal axis represents the number of correlations and the vertical axis represents the empirical power.
The results presented in Figure \ref{f1} can be summarized as follows: (1) the proposed tests based on Spearman's $\rho$ and Kendall's $\tau$ correlations, i.e. $S_\rho$, $L_{\rho}$, $T_C^\rho$, $S_\tau$, $L_{\tau}$, $T_C^\tau$, failed to detect the dependency between the two random vectors, because these two types of correlations are not suitable for characterizing non monotonic dependencies; (2)
the power curves of the max-type tests based on Hoeffding's D, Blum-Kiefer-Rosenblatt's R and Bergsma-Dassios-Yanagimoto's $\tau^*$ correlations, i.e. $L'_{D}$, $L_{R}$ and $L_{\tau^*}$, generally diminish with the increment of  the number of correlations, whereas the power curves of the sum-type tests, i.e. $S'_{D}$, $S'_{R}$, $S'_{\tau^*}$, $D$, $R$ and $\tau^*$, generally increase with the increment of the number of correlations; (3) the proposed max-sum tests ${T_C^{D}}^{\prime}$, ${T_C^{R}}^{\prime}$, ${T_C^{\tau^*}}^{\prime}$ based on Hoeffding's D, Blum-Kiefer-Rosenblatt's R and Bergsma-Dassios-Yanagimoto's $\tau^*$ correlations significantly outperform the existing tests in terms of empirical power, and each of them has more robust power performance on the number of correlations than the corresponding sum-type and max-type tests, which means that it performs well regardless of whether the number of correlations is large or small.

To further investigate the differences between the proposed tests and the sum-type tests proposed by \cite{zhouyeqing2024}, we consider an additional setting as follows.

\begin{enumerate}[label=(\alph*),resume]
	\item Set $n=100$ and $p=q\in \{50,150\}$. $\{\bm X_k\}_{k=1}^n$ are iid from $N(\bm 0,\mathbf{\Sigma})$, where $\mathbf{\Sigma}$ is the same as in setting \ref{ee}. For each $1\leq k \leq n$, $Y_{k,j}=2(-1)^{-j}X^2_{k,j} +\epsilon_{k,j}$ for $1\leq j \leq \lfloor p/9\rfloor$, where $\{\epsilon_{k,j}\}_{k=1}^n$ are iid from $N(0,j^2)$. The  remaining components $\{(Y_{k,\lfloor p/9\rfloor+1},\ldots,Y_{k,q})^\top\}_{k=1}^n$ are generated as follows: (1) $\{\tilde{\bm{Y}}_k\}_{k=1}^n$ are iid from $N(\bm0,\mathbf{\Sigma}^*)$ and independent of $\{\bm X_k\}_{k=1}^n$, where $\mathbf{\Sigma}^*=(\sigma_{st})_{1\leq s,t\leq q-\lfloor(v/2)^{3}\rfloor}$; (2) $(Y_{k,\lfloor p/9\rfloor+1},\ldots,Y_{k,q})^\top=\tilde {\bm{Y}}_k$.
	
	\label{drt}
\end{enumerate}

Table \ref{t4} presents the results of empirical power of the involved tests under setting \ref{drt}. In this table, we have obtained similar results as under the previous setting, except that the advantages of our proposed sum-type and max-sum tests based on Hoeffding's D, Blum-Kiefer-Rosenblatt's R and Bergsma-Dassios-Yanagimoto's $\tau^*$ correlations over the sum-type tests proposed in \cite{zhouyeqing2024} is significantly amplified. This may be because establishing statistics based on the sum of estimated correlations in this setting will cancel out positive and negative signals by summing them up.

\begin{table}[]
	\centering
	\caption{The empirical power of the tests under setting \ref{drt} at 5\% level.}\label{t4}
	\vspace{0.35cm}
	\renewcommand{\arraystretch}{0.98}
	\tabcolsep 3pt
	\scalebox{0.68}{
		\begin{tabular}{l|c|c|c|c|c|c|c|c|c|c|c|c|c|c|c|c|c}
			\hline
			%		$S_{\rho}$ & $L_{\rho}$ & $T_C^\rho$ & $S_\tau$ & $L_\tau$ & $T_C^\tau$
			& $T_n$ & $T_{dCov}$ & $T_{mdCov}$ & $T_{hCov_1}$ & $T_{hCov_2}$& $S'_D$ & $L'_D$ & ${T_C^D}^{\prime}$ & $S'_R$ & $L_R$ & ${T_C^R}^{\prime}$ & $S'_{\tau^{*}}$ & $L_{\tau^{*}}$ & ${T_C^{\tau^{*}}}^{\prime}$ & $D$   & $R$   & $\tau^{*}$  \\ \hline
			n=100, p=50   & 0.07 & 0.06      & 0.13      & 0.06        & 0.06   & 0.82  & 0.95 & 0.97 & 0.39 & 0.42 & 0.62  & 0.55           & 0.77          & 0.86           & 0.24 & 0.17 & 0.19          \\ \hline
			n=100, p=150	 & 0.04 & 0.04      & 0.10       & 0.04        & 0.04    & 0.34 & 0.72 & 0.76 & 0.17 & 0.08 & 0.18 & 0.20          & 0.30          & 0.39          & 0.16 & 0.14 & 0.15           \\ \hline
	\end{tabular}}
\end{table}

\section{Empirical application}\label{empirical study}

In this section, we study a dataset composed of the RNA microarrays from the eyes of 120 F2 rats, which was released and studied by \cite{scheetz2006regulation}. The original dataset contains more than 31,000 probes. In data preprocessing, the probes that were not expressed in eyes or lack sufficient variation were excluded. For the probes considered to be expressed, the maximum expression value observed in the 120 F2 rats needed to be greater than the 25th percentile of the entire set of RMA expression values. For the probes considered to be sufficiently variable, they must show at least a 2-fold change in expression levels across the 120 F2 animals. In total, there are 18,986 probes that met these two criteria. Probe 1389163\_at is one of the 18,986 probes, which is the probe for gene TRIM32. \cite{chiang2006homozygosity} found that mutations in TRIM32 can result in Bardet-Biedl syndrome (BBS).

Below, we will investigate whether this probe is independent of other probes. Let $\bm{X}=(X_1,\cdots,X_p)^{\top}$ with $p=1$ denote the expression value of probe 1389163\_at. Similar to \cite{zhouyeqing2024}, we selected 500 probes with the largest variance from the 18,985 probes, and let $\bm{Y}=(Y_1,\cdots,Y_q)^{\top}$ with $q=500$ denote the random vector composed of the expression values of these 500 probes.  We apply the testing procedures involved  in Section \ref{sec:simu} to test the null hypothesis $H_0$: $\bm{X}$ is independent of $\bm{Y}$. We found that all these testing procedures reject the null hypothesis. To further investigate the differences between these testing procedures, we reconstructed the data through resampling. Specifically, we randomly drew $n^{\prime} = 50, 65, 80$ samples from the original $n = 120$ samples, and then apply the testing procedures to the $n^{\prime}$ samples obtained by resampling. The results of rejection rate are presented in Table \ref{t5}, with all experiments repeated 200 times. Table \ref{t5} indicates that for each choice of $n^{\prime}$, the top five methods with the highest rejection rates are all max-sum methods.

\begin{figure}[!t]
	\centering\vspace{0.5cm}
	\includegraphics[width=5in]{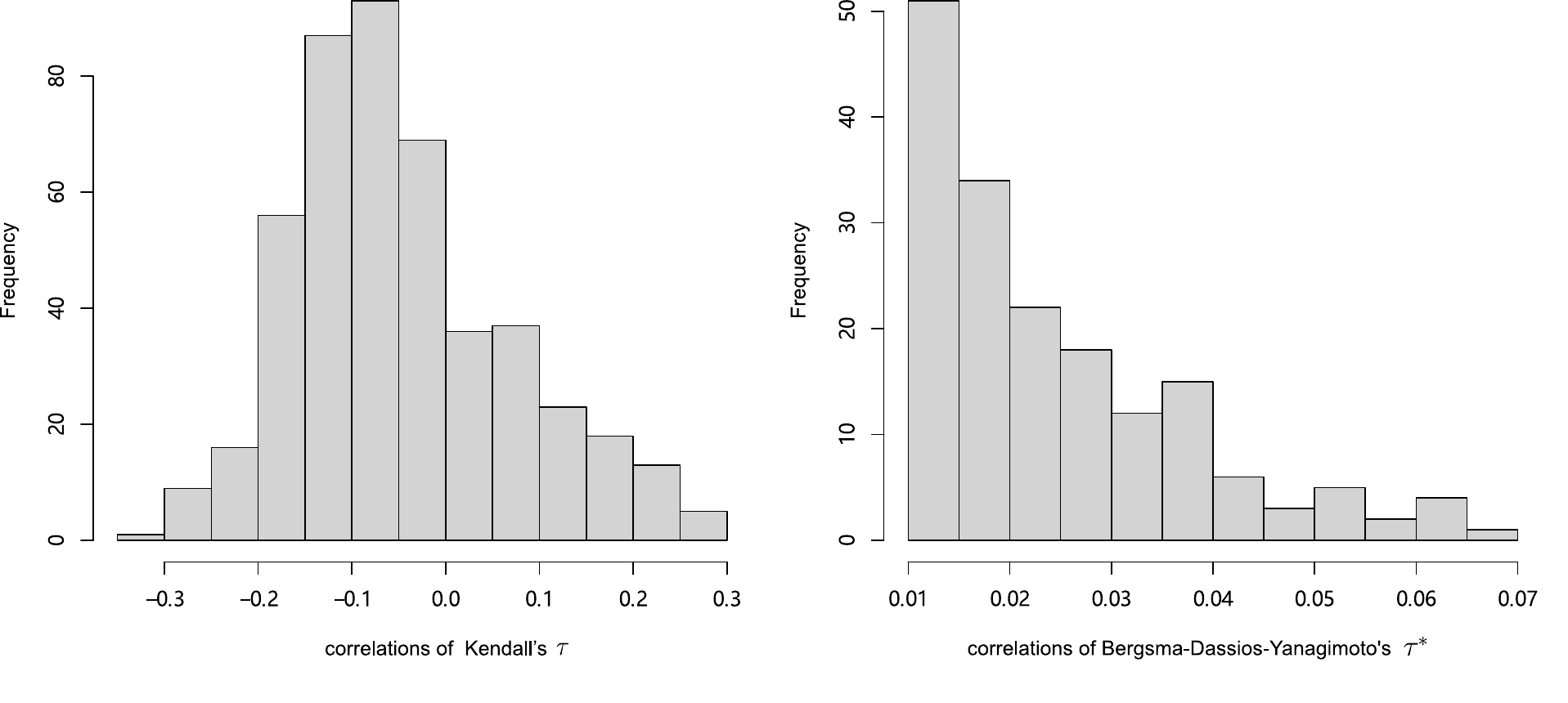}
	\caption{Histograms of the non-zero Kendall's $\tau$ and Bergsma-Dassios-Yanagimoto $\tau^*$ correlation coefficients.}\label{f2}	
\end{figure}

To demonstrate the reasonableness of rejecting the null hypothesis $H_0$, we will supplement with some descriptive analysis results.
Figure \ref{f2} presents the histograms of the non-zero Kendall's $\tau$ and Bergsma-Dassios-Yanagimoto $\tau^*$ correlation coefficients between $X_1$ and $Y_i$ for $i=1,\cdots,500$, respectively. Here, ``non-zero" means that the absolute value is less than 0.01. The histograms in Figure \ref{f2} indicate that there are a large number of non-zero correlations with small absolute values as well as a certain number of moderate-sized correlation coefficients. These results can serve as partial evidence of non-independence. In addition, \cite{zhouyeqing2024} concluded that there may exists nonlinear dependence between probe 1389163\_at and other probes. Therefore, we tend to consider rejecting the null hypothesis as the more reasonable outcome, suggesting that the max-sum testing methods proposed in this paper has an advantage in the aforementioned analysis.

\begin{table}[]
	\centering
	\caption{The results of rejection rate of the involved tests.}\label{t5}
	\vspace{0.5cm}
	\renewcommand{\arraystretch}{0.98}
	\tabcolsep 3pt
	\scalebox{0.68}{
		\begin{tabular}{lcccccccccccccccccc}
			\hline
			&$S_{\rho}$ & $L_{\rho}$ & $T_C^\rho$ & $S_\tau$ & $L_\tau$ & $T_C^\tau$
			& $S'_D$ & $L'_D$ & ${T_C^D}^{\prime}$ & $S'_R$ & $L_R$ & ${T_C^R}^{\prime}$ \\ \hline
			$n^{\prime}$=50   &0.540	&0.275	&{\bf 0.585}	&0.560	&0.410	&{\bf 0.680}	&0.525	&0.380	&{\bf 0.630}	&0.470	&0.355	&{\bf 0.605}
			\\
			$n^{\prime}$=65	&0.710	&0.610	&{\bf 0.840}	&0.725	&0.725	&{\bf 0.880}	&0.745	&0.645	&{\bf 0.850}	&0.740	&0.675	&{\bf 0.850}	
			\\
			$n^{\prime}$=80   &0.905	&0.870	&{\bf 0.970}	&0.910	&0.905	&{\bf 0.970}	&0.890	&0.845	&{\bf 0.955}	&0.880	&0.880	&{\bf 0.955}	
			%     \\
			%$n^{\prime}$=95   &0.980	&0.980	&1.000	&0.985	&0.995	&1.000	&0.985	&0.960	&1.000	&0.975	&0.985	&1.000	
			\\ \hline
			& $S'_{\tau^{*}}$ & $L_{\tau^{*}}$ & ${T_C^{\tau^{*}}}^{\prime}$ & $D$   & $R$   & $\tau^{*}$  	& $T_n$ & $T_{dCov}$ & $T_{mdCov}$ & $T_{hCov_1}$ & $T_{hCov_2}$ & \\ \hline
			$n^{\prime}$=50 	&0.475	&0.405	&{\bf 0.635}	&0.490	&0.510	&0.505	&0.435	&0.420	&0.415	&0.350	&0.350&
			\\ \hline
			$n^{\prime}$=65		&0.740	&0.710	&{\bf 0.870}	&0.680	&0.685	&0.685	&0.640	&0.635	&0.610	&0.520	&0.505&
			\\ \hline
			$n^{\prime}$=80   	&0.885	&0.885	&{\bf 0.955}	&0.835	&0.845	&0.845	&0.805	&0.800	&0.800	&0.730	&0.730&
			%\\ \hline
			%$n^{\prime}$=95   	&0.975	&0.985	&1.000	&0.985	&0.985	&0.985	&0.970	&0.970	&0.965	&0.920	&0.915 &
			\\ \hline
	\end{tabular}}
\end{table}

%
%The simulation results in the previous section reveal that the max-sum test based on Hoeffding's D correlation generally exhibits superior performance. Hence, we in this section we only use it to test the dependency between $\bm X$ and $\bm Y$. An examination of the results, where each p-value was derived from three years of continuous observations, revealed that the majority of the time, the p-value results were rejected. The global financial crisis, which spanned from 2007 to 2009, exerted a significant influence on both the Chinese and U.S. stock markets. This could potentially explain the high correlation observed between 2006 and 2009. In 2015, the initiation of the Shanghai-Hong Kong Stock Connect allowed substantial international capital to flow into the Chinese stock market, thereby significantly impacting it. This could account for the considerable number of correlations observed from 2014 to 2016. The correlation in 2018-2019 could be attributed to the global outbreak of the COVID-19 pandemic in 2020. In essence, the interconnection between the Chinese and U.S. stock markets could potentially be attributed to the influences of economic globalization.
%
%}
%\begin{figure}[!t]
%	\centering
%	\includegraphics[width=6.5in]{untitle.pdf}
%	\caption{P-value series of sliding annual independence test ${T_C^D}^{\prime}$ between return rates of two groups of stocks in CSI500 index and the S\&P 500 index.}\label{f2}
%\end{figure}

\section{Conclusion}\label{sec:dis}
We have studied the max-type and sum-type tests for testing the independence of two high-dimensional random vectors, based on three widely studied classes of rank correlations, including simple linear rank statistics, non-degenerate rank-based U-statistics and degenerate rank-based U-statistics. For each of the involved rank correlations, we have established the max-sum test by combining the max-type and sum-type test statistics based on it. This is premised on the establishment of asymptotic independence between the max-type and sum-type test statistics. The numerical study and empirical application demonstrate the significant advantages of the proposed tests.
\begin{appendices}

\section{Proofs of theorems}

\subsection*{Proof of Theorem \ref{th:LV}}
Without loss of generality, we assume that $\sum_{k=1}^n c_{n,k}=0$. Thus, $\mE_{H_0}\left(\tilde{V}_{i j}\right)=0$. Recall that $\left\{\upsilon_s\right\}_{s=1}^N=\left\{\tilde{V}_{i j} / \sigma_{L_V}\right\}_{ 1 \leq i\leq p,1\leq j \leq q}$. Define $z=(2 \log N-\log \log N+y)^{1 / 2}$. By Lemma C5 in \cite{distributionfree}, we have
$$
\mathbb{P}\left(\left|\upsilon_s\right| \geq z\right)=2\{1-\Phi(z)\}\{1+o(1)\} \sim \frac{1}{\sqrt{\pi}} \frac{e^{-y / 2}}{N},
$$
where $ a \sim b$ means that $a=b\{1+o(1)\}$.
Thus,
$$
\mathbb{P}\left(\max _{s \in C_N}\left|\upsilon_s\right|>z\right) \leq\left|C_N\right| \cdot P\left(\left|\upsilon_s\right| \geq z\right) \rightarrow 0,
$$
as $p \rightarrow \infty$. Set $D_N=\left\{1 \leq s \leq N: \left|B_{N, s}\right|<N^{\varsigma}\right\}$. Therefore, $|D_N| / N \rightarrow 1$ as $p \rightarrow \infty$. We see that
$$
\begin{aligned}
	\mathbb{P}\left(\max _{s \in D_N}\left|\upsilon_s\right|>z\right) & \leq \mathbb{P}\left(\max _{1 \leq s \leq N}\left|\upsilon_s\right|>z\right) \\
	& \leq \mathbb{P}\left(\max _{s \in D_N}\left|\upsilon_s\right|>z\right)+\mathbb{P}\left(\max _{s \in C_N}\left|\upsilon_s\right|>z\right).
\end{aligned}
$$
Therefore, to prove Theorem \ref{th:LV}, it is enough to show that
$$
\lim _{p \rightarrow \infty} \mathbb{P}\left(\max _{s \in D_N}\left|\upsilon_s\right|>z\right)=1-\exp \left(-\frac{1}{\sqrt{\pi}} e^{-x / 2}\right),
$$
as $p \rightarrow \infty$. Define
\begin{align}\label{alphat}
	\vartheta_t^{V}=\sum^{*} \mathbb{P}\left(\left|\upsilon_{i_1}\right|>z, \cdots,\left|\upsilon_{i_t}\right|>z\right),
\end{align}
for $1 \leq t \leq N$, where the sum runs over all $i_1<\cdots<i_t$ with $i_1 \in D_N, \cdots, i_t \in D_N$.

First, we will prove that
$$
\lim _{p \rightarrow \infty} \vartheta_t^{V}=\frac{1}{t !} \pi^{-t / 2} e^{-t y / 2},
$$
for each $t \geq1.$
Because $g\left(\cdot\right)$ is bounded by a constant $C_g$, thus all the assumptions in Theorem 1.1 in \cite{zaitsev} are satisfied. Thus, we have
$$
\begin{aligned}
	& \sum^{*} \mathbb{P}\left\{\left|Z_{i_1}\right|>z+\epsilon_n(\log N)^{-1}, \cdots,\left|Z_{i_t}\right|>z+\epsilon_n(\log N)^{-1}\right\} \\
	& -\binom{|D_N|}{t} c_1 t^{5 / 2} \exp \left\{-\frac{n^{1 / 2} \epsilon_n}{c_2 t^3(\log N)^{1 / 2}}\right\} \\
	\leq & \sum^* \mathbb{P}\left(\left|\upsilon_{i_1}\right|>z, \cdots,\left|\upsilon_{i_t}\right|>z\right)\\
	\leq & \sum^* \mathbb{P}\left\{\left|Z_{i_1}\right|>z-\epsilon_n(\log N)^{-1}, \cdots,\left|Z_{i_t}\right|>z-\epsilon_n(\log N)^{-1}\right\} \\
	& +\binom{|D_N|}{t} c_1 t^{5 / 2} \exp \left\{-\frac{n^{1 / 2} \epsilon_n}{c_2 t^3(\log N)^{1 / 2}}\right\},
\end{aligned}
$$
where $\left(Z_{i_1}, \cdots, Z_{i_t}\right)$ follows a multivariate normal distribution with mean zero and the same covariance matrix with $\left(\upsilon_{i_1}, \cdots, \upsilon_{i_t}\right)$. By the proof of Theorem 2 in \cite{feng2022dependent}, we have
$$
\begin{aligned}
	& \sum_*^* \mathbb{P}\left\{\left|Z_{i_1}\right|>z+\epsilon_n(\log N)^{-1}, \cdots,\left|Z_{i_t}\right|>z+\epsilon_n(\log N)^{-1}\right\} \rightarrow \frac{1}{t !} \pi^{-t / 2} e^{-t y / 2}, \\
	& \sum^* \mathbb{P}\left\{\left|Z_{i_1}\right|>z-\epsilon_n(\log N)^{-1}, \cdots,\left|Z_{i_t}\right|>z-\epsilon_n(\log N)^{-1}\right\} \rightarrow \frac{1}{t !} \pi^{-t / 2} e^{-t y / 2},
\end{aligned}
$$
with $\epsilon_n \rightarrow 0$ and $p \rightarrow \infty$. Additionally,
$$
\binom{|D_N|}{t} c_1 t^{5 / 2} \exp \left\{-\frac{n^{1 / 2} \epsilon_n}{c_2 t^3(\log N)^{1 / 2}}\right\} \leq C\binom{N}{t} \exp \left\{-\frac{n^{1 / 2} \epsilon_n}{c_2 t^3(\log N)^{1 / 2}}\right\} \rightarrow 0,
$$
for $\epsilon_n \rightarrow 0$ sufficiently slow. Thus, we have
$$
\sum^* \mathbb{P}\left(\left|\upsilon_{i_1}\right|>z, \cdots,\left|\upsilon_{i_t}\right|>z\right) \rightarrow \frac{1}{t !} \pi^{-t / 2} e^{-t y / 2} .
$$
Then, by Bonferroni inequality,
$$
\sum_{t=1}^{2 k}(-1)^{t-1} \vartheta_t^{V} \leq \mathbb{P}\left(\max _{s \in D_N}\left|\upsilon_s\right|>z\right) \leq \sum_{t=1}^{2 k+1}(-1)^{t-1} \vartheta_t^{V},
$$
for any $k \geq 1$. Let $p \rightarrow \infty$, we have
$$
\begin{aligned}
	\sum_{t=1}^{2 k}(-1)^{t-1} \frac{1}{t !}\left(\frac{1}{\sqrt{\pi}} e^{-x / 2}\right)^t & \leq \liminf _{p \rightarrow \infty} \mathbb{P}\left(\max _{s \in D_N}\left|\upsilon_s\right|>z\right) \\
	& \leq \limsup _{p \rightarrow \infty} \mathbb{P}\left(\max _{s \in D_N}\left|\upsilon_s\right|>z\right) \leq \sum_{t=1}^{2 k+1}(-1)^{t-1} \frac{1}{t !}\left(\frac{1}{\sqrt{\pi}} e^{-x / 2}\right)^t,
\end{aligned}
$$
for each $k \geq 1$. By letting $k \rightarrow \infty$ and using the Taylor expansion of the function $1-e^{-x}$, we complete the proof.   \hfill$\Box$
\subsection*{Proof of Theorem \ref{th:LVH1}}
By Lemma 2 in \cite{chen2022rank} or Lemma C6 in \cite{distributionfree}, there exists a constant $c$ such that, for any $t>0,$
$$
\mathbb{P}\left\{|\hat{V}_{i j}-\mE(\hat{V}_{i j})|>t\right\} \leq 2 e^{-nt^2 / c} .
$$
Then,
$$
\mathbb{P}\big\{\max _{1 \leq i\leq p,1\leq  j \leq q}|\hat{V}_{i j}-\mE(\hat{V}_{i j})|>t\big\} \leq N 2 e^{-nt^2 /c},
$$
which implies that, with probability at least $1-N^{-1}$,
$$
\max _{1 \leq i\leq p,1\leq  j \leq q}|\hat{V}_{i j}-\mE(\hat{V}_{i j})| \leq \sqrt{3 c \log N/n} .
$$
So, for large enough $n$, we have
$$
\begin{aligned}
	L_V^2 / \sigma_{L_V}^2 & =\max _{1 \leq i\leq p,1\leq  j \leq q} n\hat{V}_{i j}^2 \geq n\left\{\max _{1 \leq i\leq p,1\leq  j \leq q}|\mE(\hat{V}_{i j})|-\max _{1 \leq i\leq p,1\leq  j \leq q}|\hat{V}_{i j}-\mE(\hat{V}_{i j})|\right\}^2 \\
	& \geq(2+\varrho) \log N,
\end{aligned}
$$
for some small positive constant $\varrho$. Accordingly, for any given $q_\alpha$, with probability tending to one,
$$
L_V^2 / \sigma_{L_V}^2>2 \log N-\log \log N-q_\alpha.
$$
Then, we complete the proof.

\hfill$\Box$

\subsection*{Proof of Theorem \ref{R optimal}}
For each $1 \leq i \leq p$ and $1 \leq j \leq q$, let $e_i\in\mathbb{R}^{p}=(\underbrace{0, \ldots, 0}_{i-1}, 1,0, \ldots, 0)$ and $d_j\in\mathbb{R}^{q}=(\underbrace{0, \ldots, 0}_{j-1}, 1,0, \ldots, 0)$. Consider the Gaussian setting and a simple alternative set of parameters
$$
\mathcal{F}(\rho)=\left\{\mathbf{M}: \mathbf{M}=\left(\begin{array}{ccc}
	\mathbf{I}_{p} &\rho e_{i}d_{j}^{\top}  \\
	\rho d_{j}e_{i}^{\top} & \mathbf{I}_{q}  \\
\end{array}\right)\in\mathbb{R}^{d\times d},  1 \leq i \leq p,1 \leq j \leq q, \right\},
$$
where $d=p+q.$
Let $\mu_\rho$ be the uniform measure on $\mathcal{F}(\rho)$ and $\rho=c_0(\log N / n)^{1 / 2}$ for some small enough constant $c_0<1$. Let $\mathbb{P}_{\mathbf{\Sigma}}$ denote the probability measure of $N(\bm 0, \mathbf{\Sigma})$ and $\mathbb{P}_{\mu_\rho}=\int \mathbb{P}_{\mathbf{\Sigma}} \mathrm{d} \mu_\rho(\mathbf{\Sigma})$.  Let $\mathbb{P}_0$ denote the probability measure of $N\left(\bm 0, \mathbf{I}_d\right)$.
We then prove that $\mE_{\mathrm{p}_0}\{L_{\mu_\rho}^2\}=1+o(1)$, where
$$
L_{\mu_\rho}\doteq \frac{1}{N} \sum_{\mathbf{\Sigma} \in \mathcal{F}(\rho)}\left[\prod_{i=1}^n \frac{1}{|\mathbf{\Sigma}|^{1 / 2}} \exp \left\{-\frac{1}{2} \bm{Z}_{i}^{\mathrm{T}}\left(\mathbf{\Omega}-\mathbf{I}_d\right) \bm{Z}_{i}\right\}\right],
$$
$\mathbf{\Omega} = \mathbf{\Sigma}^{-1}$ and $\bm{Z}_{1}, \ldots, \bm{Z}_{n}$ are $d$-dimensional vectors to be specified later. We have
$$
\mE_{\mathrm{p}_0}\left\{L_{\mu_\rho}^2\right\}=\frac{1}{N^2} \sum_{\mathbf{\Sigma}_1, \mathbf{\Sigma}_2 \in \mathcal{F}(\rho)} \mE\left[\prod_{i=1}^n \frac{1}{\left|\mathbf{\Sigma}_1\right|^{1 / 2}} \frac{1}{\left|\mathbf{\Sigma}_2\right|^{1 / 2}} \exp \left\{-\frac{1}{2} \mathbf{Z}_{i}^{\mathrm{T}}\left(\mathbf{\Omega}_1+\mathbf{\Omega}_2-2 \mathbf{I}_d\right) \mathbf{Z}_{i}\right\}\right],
$$
where $\mathbf{\Omega}_i = \mathbf{\Sigma}_i^{-1}$ for $i=1,2$ and $\left\{\mathbf{Z}_{i}: 1 \leq i \leq n\right\}$ are independent and identically distributed from $N\left(\bm 0, \mathbf{I}_d\right)$. Let
$$
\mathbf{A}=\frac{\rho}{1-\rho^2}\left(\begin{array}{ccc}
	2 \rho & -1 & -1 \\
	-1 & \rho & 0 \\
	-1 & 0 & \rho
\end{array}\right), \quad \mathbf{B}=\frac{2 \rho}{1-\rho^2}\left(\begin{array}{cc}
	\rho & -1 \\
	-1 & \rho
\end{array}\right) .
$$
It is easy to derive that
$$
\begin{aligned}
	\mE_{\mathrm{p}_0}\left(L_{\mu_\rho}^2\right)= & \underbrace{\frac{1}{N^2} \sum_{\mathbf{\Sigma}_1\neq \mathbf{\Sigma}_2 \in \mathcal{F}(\rho)}\prod_{i=1}^n\mE\left[\frac{1}{1-\rho^2}\exp \left\{-\frac{1}{2} \mathbf{Z}_{i}^{\mathrm{T}}\left(\mathbf{\Omega}_1+\mathbf{\Omega}_2-2 \mathbf{I}_d\right) \mathbf{Z}_{i}\right\}\right]}_{E_1} \\
	& +\underbrace{\frac{1}{N^2}\sum_{\mathbf{\Sigma}\in \mathcal{F}(\rho)}\prod_{i=1}^n\mE\left[\frac{1}{1-\rho^2}\exp \left\{-\frac{1}{2} \mathbf{Z}_{i}^{\mathrm{T}}\left(2\mathbf{\Omega}-2 \mathbf{I}_d\right) \mathbf{Z}_{i}\right\}\right]}_{E_2}.
\end{aligned}
$$
After calculation, it is easy to obtain
$$
E_1=\frac{N(N-1)}{N^2}\frac{1}{\left(1-\rho^2\right)^n}\left(1-\rho^2\right)^n=1+o(1).
$$
For $E_2$, it is easy to calculate that $E_2=N^{-1}\left(1-\rho^2\right)^{-n}$. Recalling that $\rho=c_0(\log N / n)^{1 / 2}$ and $\log N / n=$ $o(1)$, we have
$$
E_2=N^{-1}\left(1-c_0^2 \log N / n\right)^{-n}=N^{-1} \exp \left(c_0^2 \log N\right)\{1+o(1)\}=o(1)
$$
as long as $c_0<1$. Hence, we can obtain that $\mE_{\mathrm{p}_0}\{L_{\mu_\rho}^2\}=1+o(1).$
This completes the proof.
\hfill$\Box$

\subsection*{Proof of Theorem \ref{th:LVH1 optimal}}
According to Theorem \ref{R optimal} and Assumption \ref{cond H1}, we can easily obtain
the result. \hfill$\Box$
\subsection*{Proof of Theorem \ref{th:SV}}
Without loss of generality, we assume that $\sum_{k=1}^n c_{n,k}=0$. Thus, $\mE_{H_0}\left(\tilde{V}_{i j}\right)=0$.
Let $G_{j}(\cdot)$ be the cumulative distribution function of $Y_{k,j},$ for $1\leq k\leq n$ and $1\leq j\leq q.$
According to Lemma C5 in \cite{distributionfree}, we have that
under $H_{0},$ $\tilde{V}_{ij}=\sum_{k=1}^{n}c_{n,k}g\left\{R_{n k}^{ij} /(n+1)\right\}$ is identically distributed to $\sum_{k=1}^{n}c_{n,k}g\left\{R_{k,j}^{2} /(n+1)\right\},$ where $R_{k,j}^{2}$ is the rank of $Y_{k,j}$ in $Y_{1,j},\cdots,Y_{n,j},$ and $\sum_{k=1}^{n}c_{n,k} g\left\{R_{k,j}^{2} /(n+1)\right\}-\frac{1}{n}\sum_{k=1}^{n}c_{n,k} g\left\{G_{j}(Y_{k,j})\right\}$ converges to zero in probability.
Because $\sum_{k=1}^{n}c_{n,k} g\left\{R_{k,j}^{2} /(n+1)\right\}$ is uniformly integrable, according to Lemma 2 in \cite{feng2022max}, for any $r\geq2$, $1\leq i\leq p,$ $1\leq j\leq q,$ we then have,
$$\mE_{H_{0}}\left\{\left(\tilde{V}_{ij}\right)^{r}\right\}=
\mE_{H_{0}}\big[\sum_{k=1}^{n}c_{nk                    }g\left\{G_{j}(X_{k,j})\right\}\big]^{r}=O(n^{-r/2}).$$
Hence, according to Corollary A.1 in \cite{mdependent}, we have for any $1\leq i\leq p,$
\begin{align*}
	&\mE|\sum_{j=1}^{q}\left\{\tilde{V}_{ij}^{2}-\mE_{H_0}(\tilde{V}_{ij}^{2})\right\}|^{2+\delta}\\
	\leq& C_{\delta}^{2+\delta}\mE(|\left\{\tilde{V}_{ij}^{2}-\mE_{H_0}(\tilde{V}_{ij}^{2})\right\}|^{2+\delta})(16m_{q}q)^{(2+\delta)/2}\\
	\leq& C_{\delta}^{2+\delta}O(n^{-(2+\delta)})(16m_{q}q)^{(2+\delta)/2},
\end{align*}
where $C_\delta $ is a positive constant depending only upon $\delta$.
Let $ \overline{\tilde{V}}_{ij}\doteq \tilde{V}_{ij}^{2}-\mE_{H_0}(\tilde{V}_{ij}^2)$ and $\overline{\tilde{V}}_{i.}\doteq\sum_{j=1}^{q}\overline{\tilde{V}}_{ij}.$
For all $a$ and all $k\geq m_{p}$, we have
\begin{align*}
	\var\Big(\sum_{i=a}^{a+k-1} \overline{\tilde{V}}_{i.}\Big)\leq& \sum_{i_{1}=a}^{a+k-1}\sum_{i_{2}=a}^{a+k-1} \mE(\overline{\tilde{V}}_{i_{1}.}\overline{\tilde{V}}_{i_{2}.})\\
	\leq &\sum_{i_{1}=a}^{a+k-1}m_{p}\max_{1\leq i\leq p}\mE(\overline{\tilde{V}}_{i.}^2)\\
	\leq & km_{p}\mE(\overline{\tilde{V}}_{1.}^2)= O(n^{-2}) km_{p}m_{q}q.
\end{align*}
When $m_p\leq k\leq p^{\alpha}$ and $\alpha=1/\{1+(1-\gamma)(1+2/\delta)\},$
we have $\var\Big(\sum_{i=a}^{a+k-1} \overline{\tilde{V}}_{i.}\Big)/k^{1+\gamma}\leq \max\{m_{p}^{-\gamma},p^{-\alpha\gamma}\} O(n^{-2}) m_{p}m_{q}q.$
Hence, according to Theorem 2.1 in \cite{mdependent} and Assumptions \ref{assump2} and \ref{assmp3}-\ref{cond5}, we can obtain the conclusion. \hfill$\Box$
\subsection*{Proof of Theorem \ref{th:CV}}
Define
\begin{align}\label{signdef3}
	&\Lambda_{N}=\{(i, j) ; 1 \leq i\leq p,1\leq j \leq q\}, \nonumber\\
	&E_{N}(x)=\left\{S_{V}/\sigma_{S_{V}} \leq x\right\} \quad \text { and } \quad
	J_{I}=\left\{\left|\tilde{V}_{ij}\right|\geq a_{N}\right\},
\end{align}
for any $I=(i, j) \in \Lambda_{N},$ where $$a_{N}=\sqrt{\sigma^2_{L_V}\left[2\log N -\log \log N  +y\right]}$$
and $\sigma^2_{L_V}=\var_{H_{0}}(\tilde{V}_{ij}).$
\begin{lemma}\label{lemmalinear}
	Assume $N=o\left(n^\epsilon\right)$ as $n \rightarrow \infty $. Under $H_0$, let
	$$
	H_{V}(t, N)=\sum_{I_{1},I_{2},\cdots,I_{t} \in \Lambda_{N}} \P_{H_{0}}\left(J_{I_{1}} J_{I_{2}} \cdots J_{I_{t}}\right).
	$$
	Then $\lim _{t \rightarrow \infty} \limsup _{p \rightarrow \infty} H_{V}(t, N)=0.$
\end{lemma}
\begin{lemma}\label{lemmalinear2}
	Under the assumptions of Lemma \ref{lemmalinear} and $H_{0}$,
	$$
	\sum_{I_{1}<I_{2}<\cdots<I_{t} \in \Lambda_{N}}\left\{\P_{H_{0}}\left(E_{N}(x) J_{I_{1}} J_{I_{2}} \cdots J_{I_{t}}\right)-\P_{H_{0}}\left(E_{N}(x)\right) \cdot \P_{H_{0}}\left(J_{I_{1}} J_{I_{2}} \cdots J_{I_{t}}\right)\right\} \rightarrow 0,
	$$
	as $n \rightarrow \infty$, for $t \geq 1.$
\end{lemma}
We need to prove that ${L}_{V}^2/\sigma^2_{L_V}-2 \log N+\log \log  N $ and $S_{V}/\sigma_{S_{V}}$ are asymptotically independent.
By Theorems \ref{th:LV} and \ref{th:SV} in this paper, the following holds:
\begin{align}
	&L^2_{{V}}/\sigma^2_{L_V}-2 \log N +\log \log  N  \rightarrow G(y)\,\text { in distribution;} \label{demax} \\
	&\frac{S_{V}}{\sigma_{S_{V}}} \rightarrow N(0,1)\text { in distribution. }\label{desum}
\end{align}
To show the asymptotic independence, it is enough to show the limit of
\begin{align}
	\lim _{n,p \rightarrow \infty} \P_{H_{0}}\left(\frac{S_{V}}{\sigma_{S_{V}}} \leq x, \frac{L_{V}}{\sigma_{L_{V}}}\leq {a}^{*}_{N}\right)=\Phi(x) \cdot G(y),
\end{align}
for any $x \in \mathbb{R}$ and $y \in \mathbb{R}$, where $\Phi(x)=(2 \pi)^{-1 / 2} \int_{-\infty}^{x} e^{-t^{2} / 2} d t$ and
$$
{a}^{*}_{N}=\sqrt{2 \log N -\log \log  N  +y},
$$
which makes sense for large $p$. Due to (\ref{demax}) and (\ref{desum}), the above is equivalent to
\begin{align}\label{indedengjia1}
	\lim _{n,p \rightarrow \infty} \P_{H_{0}}\left(\frac{S_{V}}{\sigma_{S_{V}}} \leq x, \frac{L_{{V}}}{\sigma_{L_{V}}}> {a}_{N}^{*}\right)=\Phi(x) \cdot\left\{1-G(y)\right\},
\end{align}
for any $x \in \mathbb{R}$ and $y \in \mathbb{R}$. Review notations $\Lambda_{N}, E_{N}(x)$ and $J_{I}$ for any $I=$ $(i, j) \in \Lambda_{N}$ in $(\ref{signdef3}) .$ Write
\begin{align}\label{union}
	\P_{H_{0}}\left(\frac{S_{V}}{\sigma_{S_{V}}} \leq x, \frac{L_{{V}}}{\sigma_{L_{V}}}> {a}_{N}^{*}\right)=\P_{H_{0}}\left(\bigcup_{I \in \Lambda_{N}} E_{N}(x) J_{I}\right).
\end{align}
Here the notation $E_{N}(x) J_{I}$ stands for $E_{N}(x) \cap J_{I}$. From the inclusion-exclusion principle,
\begin{align}\label{UAB2k+1}
	\P_{H_{0}}\left(\bigcup_{I \in \Lambda_{N}} E_{N}(x) J_{I}\right) \leq & \sum_{I_{1} \in \Lambda_{N}} \P_{H_{0}}\left(E_{N}(x) J_{I_{1}}\right)-\sum_{I_{1},I_{2} \in \Lambda_{N}} \P_{H_{0}}\left(E_{N}(x) J_{I_{1}} J_{I_{2}}\right)+\cdots+ \n\\
	&\sum_{I_{1},I_{2},\cdots,I_{2 l+1} \in \Lambda_{N}}\P_{H_{0}}\left(E_{N}(x) J_{I_{1}} J_{I_{2}} \cdots J_{I_{2 l+1}}\right)
\end{align}
and
\begin{align}\label{UAB2k}
	\P_{H_{0}}\left(\bigcup_{I \in \Lambda_{N}} E_{N}(x) J_{I}\right) \geq& \sum_{I_{1} \in \Lambda_{N}} \P_{H_{0}}\left(E_{N}(x) J_{I_{1}}\right)-\sum_{I_{1},I_{2} \in \Lambda_{N}} \P_{H_{0}}\left(E_{N}(x) J_{I_{1}} J_{I_{2}}\right)+\cdots-\n\\
	&\sum_{I_{1},I_{2},\cdots,I_{2 l} \in \Lambda_{N}} \P_{H_{0}}\left(E_{N}(x) J_{I_{1}} J_{I_{2}} \cdots J_{I_{2 l}}\right),
\end{align}
for any integer $l \geq 1$. Reviewing the definition
$$
H_{V}(t, N)=\sum_{I_{1},I_{2},\cdots,I_{t} \in \Lambda_{N}} \P_{H_{0}}\left(J_{I_{1}} J_{I_{2}} \cdots J_{I_{t}}\right),
$$
for $t \geq 1$ in Lemma $\ref{lemmalinear},$ we have from the lemma that
\begin{align}\label{lHpk}
	\lim _{t \rightarrow \infty} \limsup _{p \rightarrow \infty} H_{V}(t, N)=0.
\end{align}
Set
$$\zeta_{V}(t,N)
=\sum_{I_{1},I_{2},\cdots,I_{t} \in \Lambda_{N}}\left\{\P_{H_{0}}\left(E_{N}(x) J_{I_{1}} J_{I_{2}} \cdots J_{I_{t}}\right)-\P_{H_{0}}\left(E_{N}(x)\right) \cdot \P_{H_{0}}\left(J_{I_{1}} J_{I_{2}} \cdots J_{I_{t}}\right)\right\},
$$
for $t \geq 1 .$ By Lemma \ref{lemmalinear2},
\begin{align}\label{zeta}
	\lim _{p \rightarrow \infty} \zeta_{V}(t,N)=0,
\end{align}
for $t \geq 1$. The assertion (\ref{UAB2k+1}) implies that
\begin{align}\label{UzetaH2k+1}
	& \P_{H_{0}}\left(\bigcup_{I \in \Lambda_{N}} E_{N}(x) J_{I}\right) \n\\
	\leq & \P_{H_{0}}\left(E_{N}(x)\right)\left\{\sum_{I_{1} \in \Lambda_{N}} \P_{H_{0}}\left(J_{I_{1}}\right)-\sum_{I_{1},I_{2} \in \Lambda_{N}} \P_{H_{0}}\left(J_{I_{1}} J_{I_{2}}\right)+\cdots-\right.\n\\
	&\left.\sum_{I_{1},I_{2},\cdots,I_{2 l} \in \Lambda_{N}} \P_{H_{0}}\left(J_{I_{1}} J_{I_{2}} \cdots J_{I_{2 l}}\right)\right\}+\left\{\sum_{t=1}^{2 l} \zeta_{V}(t,N)\right\}+H( 2 l+1,N) \n\\
	\leq & \P_{H_{0}} \left(E_{N}(x)\right) \cdot \P_{H_{0}}\left(\bigcup_{I \in \Lambda_{N}} J_{I}\right)+\left\{\sum_{t=1}^{2 l} \zeta_{V}(t,N)\right\}+H_{V}( 2 l+1,N),
\end{align}
where the inclusion-exclusion formula is used again in the last inequality, that is
$$
\begin{array}{r}
	\P_{H_{0}}\left(\bigcup_{I \in \Lambda_{N}} J_{I}\right) \geq\left\{\sum_{I_{1} \in \Lambda_{N}} \P_{H_{0}}\left(J_{I_{1}}\right)-\sum_{I_{1},I_{2} \in \Lambda_{N}} \P_{H_{0}}\left(J_{I_{1}} J_{I_{2}}\right)+\cdots-\right. \\
	\left.\sum_{I_{1},I_{2},\cdots,I_{2 l} \in \Lambda_{N}} \P_{H_{0}}\left(J_{I_{1}} J_{I_{2}} \cdots J_{I_{2 l}}\right)\right\},
\end{array}
$$
for all $l \geq 1$. By the definition of ${a}_{N}^{*}$ and (\ref{demax}),
\begin{align*}
	\P_{H_{0}}\left(\bigcup_{I \in \Lambda_{N}} J_{I}\right)&=\P_{H_{0}}\left(\frac{L_{{V}}}{\sigma_{L_{V}}}>a_{N}^{*}\right)\n\\
	&=\P_{H_{0}}\left[\frac{L^2_{V}}{\sigma^2_{L_V}}-2 \log N +\log \log  N  >y\right] \n\\
	&\rightarrow 1-G(y),
\end{align*}
as $n,p\rightarrow \infty$. By (\ref{desum}), $\P_{H_{0}}\left(H_0\right) \rightarrow \Phi(x)$ as $n,p \rightarrow \infty$. From (\ref{union}), by fixing and sending $n,p \rightarrow \infty$, we get from (\ref{zeta}) that
$$
\begin{aligned}
	& \limsup _{n,p \rightarrow \infty} \P_{H_{0}}\left(\frac{S_{V}}{\sigma_{S_{V}}} \leq x, \frac{L_{{V}}}{\sigma_{L_{V}}}>{a}_{N}^{*}\right) \\
	\leq & \Phi(x) \cdot\left\{1-G(y)\right\}+\limsup _{p \rightarrow \infty} H_{V}( 2 l+1,N).
\end{aligned}
$$
Now, let $l\rightarrow \infty$ and use (\ref{lHpk}) to see
\begin{align}\label{limitright}
	\limsup _{n,p \rightarrow \infty} \P_{H_{0}}\left(\frac{S_{V}}{\sigma_{S_{V}}} \leq x, \frac{L_{{V}}}{\sigma_{L_{V}}}>{a}_{N}^{*}\right) \leq \Phi(x) \cdot\left\{1-G(y)\right\}.
\end{align}
By applying the same argument to (\ref{UAB2k}), we see that the counterpart of
(\ref{UzetaH2k+1}) becomes
\begin{align*}
	\P_{H_{0}}\left(\bigcup_{I \in \Lambda_{N}} H_0 J_{I}\right)
	\geq & \P_{H_{0}}\left(H_0\right)\left\{\sum_{I_{1} \in \Lambda_{N}} \P_{H_{0}}\left(J_{I_{1}}\right)-\sum_{I_{1},I_{2} \in \Lambda_{N}} \P_{H_{0}}\left(J_{I_{1}} J_{I_{2}}\right)+\cdots+\right.\n\\
	&\left.\sum_{I_{1},I_{2},\cdots,I_{2 l-1} \in \Lambda_{N}} \P_{H_{0}}\left(J_{I_{1}} J_{I_{2}} \cdots J_{I_{2 l-1}}\right)\right\}+\n\\
	&\left\{\sum_{n=1}^{2 l-1} \zeta_{V}(n, N)\right\}-H_{V}(2 l,N) \n\\
	\geq & \P_{H_{0}}\left(H_0\right) \cdot \P_{H_{0}}\left(\bigcup_{I \in \Lambda_{N}} J_{I}\right)+\left\{\sum_{n=1}^{2 l-1} \zeta_{V}(n, N)\right\}-H_{V}(2 l,N),
\end{align*}
where in the last step we use the inclusion-exclusion principle such that
\begin{align*}
	\P_{H_{0}}\left(\bigcup_{I \in \Lambda_{N}} J_{I}\right) \leq\left\{\sum_{I_{1} \in \Lambda_{N}} \P_{H_{0}}\left(J_{I_{1}}\right)-\sum_{I_{1},I_{2} \in \Lambda_{N}} \P_{H_{0}}\left(J_{I_{1}} J_{I_{2}}\right)+\cdots+\right. \\
	\left.\sum_{I_{1},I_{2},\cdots,I_{2 l-1} \in \Lambda_{N}} \P_{H_{0}}\left(J_{I_{1}} J_{I_{2}} \cdots J_{I_{2 l-1}}\right)\right\},
\end{align*}
for all $l \geq 1$. Review (\ref{union}) and repeat the earlier procedure to see
$$
\limsup _{n,p \rightarrow \infty} \P_{H_{0}}\left(\frac{S_{V}}{\sigma_{S_{V}}} \leq x, \frac{L_{{V}}}{\sigma_{L_{V}}}>{a}_{N}^{*}\right) \geq \Phi(x) \cdot\left\{1-G(y)\right\},
$$
by sending $n,p \rightarrow \infty$ and then sending $l \rightarrow \infty$. This and (\ref{limitright}) yield (\ref{indedengjia1}). The proof is completed.    \hfill$\Box$

\subsection*{Proof of Theorem \ref{th:LU}}
%
%\begin{lemma}
% Suppose that $U$ is a $U$-statistic with degree $m$ and bounded kernel $|h(\cdot)| \leq M$.
%We then have, for any $t>0$,
%$$
%P(|U-E U|>t) \leq 2 \exp \left\{-n t^2 /\left(2 m M^2\right)\right\}
%$$
%\end{lemma}
%\begin{lemma} Suppose that the boundedness assumption in Theorem 4 hold. We then have, in a region $x \in\left(0, o\left(n^{1 / 6}\right)\right)$
%$$
%P\left[\frac{U_{i j}-E\left(U_{i j}\right)}{\left\{\operatorname{var}\left(U_{i j}\right)\right\}^{1 / 2}}>x\right]=\{1-\Phi(x)\}\left\{1+O\left(\frac{1+x^3}{n^{1 / 2}}\right)\right\}
%$$
%\end{lemma}
First, we consider the following U-statistics with bounded and symmetric kernels, i.e.
$$
\tilde{U}_{i j}=\frac{(n-m)!m!}{n!} \sum_{1 \leq k_1<k_2, \cdots,<k_m \leq n} h\left\{\left(X_{k_1, i}, Y_{k_1, j}\right), \cdots,\left(X_{k_m, i}, Y_{k_m, j}\right)\right\}.
$$
Here we define $\left\{\tilde{u}_s\right\}_{s=1}^N=\big\{\tilde{U}_{i j}\big\}_{1 \leq i\leq p, 1 \leq j \leq q}$ and $\big\{\bm{Z}_{k, i j }\big\}=\big\{\left(X_{k, i}, Y_{k, j}\right)^{\top}\big\}_{1 \leq k \leq n}$. So we rewrite $\tilde{U}_{i j}$ in the following forms
\begin{align}
	\tilde{u}_s=\frac{(n-m)!m!}{n!} \sum_{1 \leq k_1<k_2, \cdots,<k_m \leq n} h\left(\bm{Z}_{k_1, i_s j_s}, \cdots, \bm{Z}_{k_m, i_s j_s}\right).\label{us}
\end{align}
We have $\mE\left(\tilde{u}_s\right)=0$ under the null hypothesis.
For any $1\leq i_{s}\leq p,1\leq j_{s}\leq q,$ the distribution of $\tilde{u}_{s}$ is the same.
To simplify notation, we omit the subscripts $i_{s}$ and $j_{s}.$
By the assumptions, we have
$$
\begin{aligned}
	\mu_r & \doteq \mE\left|h\left(\bm{Z}_{k_1}, \cdots, \bm{Z}_{k_m}\right)\right|^r<\infty, \\
	\sigma_\psi^2&=\operatorname{var}\left\{\mE\left(h\left(\bm{Z}_{k_1}, \cdots, \bm{Z}_{k_m}\right) \mid \bm{Z}_{k_1}\right)\right\}>0 .
\end{aligned}
$$
for any $r \geq 2$.
By Lemma 1 in \cite{malevich1979large}, we can rewrite $\tilde{u}_s$ as follow
$$
\tilde{u}_s=S_s+\eta_s, \quad S_s=\frac{m}{n} \sum_{k=1}^{n} \psi_s\left(\bm{Z}_{k,i_{s}j_{s}}\right), \quad \eta_s=\sum_{l=2}^m C_m^l {u}_{s, l},
$$
where for any $1\leq s \leq N,$ $\psi_s(\bm{x})\doteq\mE\left\{h\left(\bm{Z}_{k_1,i_{s}j_{s}}, \cdots, \bm{Z}_{k_m,i_{s}j_{s}}\right) \mid \bm{Z}_{k_1,i_{s}j_{s}}=\bm{x}\right\},$
for any $l=2, \cdots, m,$ ${u}_{s, l}$ is a $U$-statistics of the form (\ref{us}) with $\operatorname{kernel} \psi^{(l)}\left(\bm{x}_1, \cdots, \bm{x}_l\right)$ such that
$$
\begin{aligned}
	& \mathbb{P}\left\{{\mE}\left[\psi^{(l)}\left(\bm{Z}_1, \cdots, \bm{Z}_l\right) \mid \bm{Z}_1, \cdots, \bm{Z}_{l-1}\right]=0\right\}=1, \\
	&{\mE}\left|\psi^{(l)}\left(\bm{Z}_1, \cdots, \bm{Z}_l\right)\right|^r \leqq 2^{m r} \mu_r .
\end{aligned}
$$
By Lemma 2 in \cite{malevich1979large}, we have
$$
\mE\left|C_{n}^l {u}_{s, l}\right|^r \leq Cn^{l r / 2} .$$
Because $\tilde{U}_{ij}$ is a rank-based U-statistic with order of degeneracy 1, we have $\sigma_{L_U}=O(n^{-1/2})$ and $\sigma_{L_U}\geq C_1n^{-1/2}$ for large enough $n$ and some constant $C_1>0$ by Lemma 5.2.1A in \cite{serfling}.
Thus, by the Markov inequality, for some constant $C_1>0,$

\begin{align*}
	& \mathbb{P}\left\{\max _{1 \leq s \leq N}\left|\eta_s\right|/\sigma_{L_{U}}>(\log N)^{-1}\right\}\\
	\leq & N \mathbb{P}\left\{\left|\eta_s\right|>C_1(\log N)^{-1}n^{-1/2}\right\} \\
	\leq & C_1^{-2r}N(\log N)^{2 r}n^r \mE\left(\left|\eta_s\right|^{2 r}\right) \\
	\leq & C_1^{-2r}N(\log N)^{2 r}n^r \mE\left(\left|\sum_{l=2}^m C_m^l {u}_{s, l}\right|^{2 r}\right) \\
	= & C_1^{-2r}N(\log N)^{2 r}n^r m^{2 r} \mE\left(\left|\frac{1}{m} \sum_{l=2}^m C_m^l {u}_{s, l}\right|^{2 r}\right) \\
	\leq & C_1^{-2r}N(\log N)^{2 r}n^r m^{2 r-1}\left\{\sum_{l=2}^m\left(C_m^l\right)^{2 r} \mE\left(\left|{u}_{s, l}\right|^{2 r}\right)\right\} \\
	\leq & C_1^{-2r} N(\log N)^{2 r}n^r m^{2 r-1}\left\{\sum_{l=2}^m\left(C_m^l\right)^{2 r} C n^{-l r}\right\} \\
	\leq & C_1^{-2r} m^{2 r-1} C \left\{\sum_{l=2}^m\left(C_m^l\right)^{2 r}\right\} N(\log N)^{2 r}n^r n^{-2 r} \\
	= & C_0(m, r,C_{1}) N(\log N)^{2 r}n^{-r} \rightarrow 0,
\end{align*}
for some positive integer $r$ by $ N=o\left(n^\epsilon\right)$.
Thus, by
$$
\left|\max _{1 \leq s \leq N}\tilde{u}_s^2/\sigma_{L_U}^2-\max _{1 \leq s \leq N} S_s^2/\sigma_{L_U}^2\right| \leq 2 \max _{1 \leq s \leq N}\left|S_s\right|/\sigma_{L_U} \max _{1 \leq s \leq N}\left|\eta_s\right|/\sigma_{L_U}+\max _{1 \leq s \leq N} \eta_s^2/\sigma_{L_U}^2,
$$
we only need to show that
$$
\mathbb{P}\left(\max _{1 \leq s \leq N}S_s^2 / \sigma_{L_U}^2-2 \log N+\log \log N \leq y\right) \rightarrow \exp \left\{-\pi^{-1 / 2} \exp (-y / 2)\right\}.
$$
Here we define $\hat{u}_s=S_s / \sigma_{L_U}$ and $z=(2 \log N-\log \log N+y)^{1 / 2}$. Since $ \eta_s/ \sigma_{L_U}$ is negligible, the tail behavior of $\hat{u}_s$ is the same as that of $\tilde{u}_s / \sigma_{L_U}$. Therefore, by Lemma C5 in \cite{distributionfree}, we have
$$
\mathbb{P}\left(\left|\hat{u}_s\right| \geq z\right)=2\{1-\Phi(z)\}\{1+o(1)\} \sim \frac{1}{\sqrt{\pi}} \frac{e^{-y / 2}}{N}.
$$
Thus,
$$
\mathbb{P}\left(\max _{s \in C_N}\left|\hat{u}_s\right|>z\right) \leq\left|C_N\right| \cdot P\left(\left|\hat{u}_s\right| \geq z\right) \rightarrow 0,
$$
as $p \rightarrow \infty$. Set $D_N=\left\{1 \leq s \leq N : \left|B_{N, s}\right|<N^{\varsigma}\right\}$. Due to the assumptions, $|D_N| / N \rightarrow 1$ as $p \rightarrow \infty$
Easily,
$$
\begin{aligned}
	\mathbb{P}\left(\max _{s \in D_N}\left|\hat{u}_s\right|>z\right) & \leq \mathbb{P}\left(\max _{1 \leq s \leq N}\left|\hat{u}_s\right|>z\right) \\
	& \leq \mathbb{P}\left(\max _{s \in D_N}\left|\hat{u}_s\right|>z\right)+\mathbb{P}\left(\max _{s \in C_N}\left|\hat{u}_s\right|>z\right).
\end{aligned}
$$
Therefore, to prove Theorem \ref{th:LU}, it is enough to show
$$
\lim _{p \rightarrow \infty} \mathbb{P}\left(\max _{s \in D_N}\left|\hat{u}_s\right|>z\right)=1-\exp \left(-\frac{1}{\sqrt{\pi}} e^{-x / 2}\right),
$$
as $p \rightarrow \infty$. Define
$$
\vartheta_t^{U}=\sum^* \mathbb{P}\left(\left|\hat{u}_{i_1}\right|>z, \cdots,\left|\hat{u}_{i_t}\right|>z\right),
$$
for $1 \leq t \leq N$, where the sum runs over all $i_1<\cdots<i_t$ and $i_1 \in D_N, \cdots, i_t \in D_N$. First, we will prove next that
$$
\lim _{p \rightarrow \infty} \vartheta_t^{U}=\frac{1}{t !} \pi^{-t / 2} e^{-t y / 2},
$$
for each $t \geq 1$. Because $\psi_s\left(\bm{x}\right)$ is bounded, thus all the assumptions in Theorem 1.1 in \cite{zaitsev} are satisfied. Thus, we have
$$
\begin{aligned}
	& \sum^* \mathbb{P}\left(\left|Z_{i_1}\right|>z+\epsilon_n(\log N)^{-1 / 2}, \cdots,\left|Z_{i_t}\right|>z+\epsilon_n(\log N)^{-1 / 2}\right) \\
	& -\binom{|D_N|}{t} c_1 t^{5 / 2} \exp \left\{-\frac{n^{1 / 2} \epsilon_n}{c_2 t^3(\log N)^{1 / 2}}\right\} \\
	\leq & \sum_*^* \mathbb{P}\left\{\left|\hat{u}_{i_1}\right|>z, \cdots,\left|\hat{u}_{i_t}\right|>z\right\} \\
	\leq & \sum^* \mathbb{P}\left\{\left|Z_{i_1}\right|>z-\epsilon_n(\log N)^{-1 / 2}, \cdots,\left|Z_{i_t}\right|>z-\epsilon_n(\log N)^{-1 / 2}\right\} \\
	& +\binom{|D_N|}{t} c_1 t^{5 / 2} \exp \left\{-\frac{n^{1 / 2} \epsilon_n}{c_2 t^3(\log N)^{1 / 2}}\right\},
\end{aligned}
$$
where $\left(Z_{i_1}, \cdots, Z_{i_t}\right)$ follows a multivariate normal distribution with mean zero and the same covariance matrix with $\left(\hat{u}_{i_1}, \cdots, \hat{u}_{i_t}\right)$. By the proof of Theorem 2 in \cite{feng2022dependent}, we have
$$
\sum^* \mathbb{P}\left(\left|Z_{i_1}\right|>z+\epsilon_n(\log N)^{-1 / 2}, \cdots,\left|Z_{i_t}\right|>z+\epsilon_n(\log N)^{-1 / 2}\right) \rightarrow \frac{1}{t !} \pi^{-t / 2} e^{-t y / 2},
$$
$$
\sum^* \mathbb{P}\left(\left|Z_{i_1}\right|>z-\epsilon_n(\log N)^{-1 / 2}, \cdots,\left|Z_{i_t}\right|>z-\epsilon_n(\log N)^{-1 / 2}\right) \rightarrow \frac{1}{t !} \pi^{-t / 2} e^{-t y / 2},
$$
with $\epsilon_n \rightarrow 0$ and $p \rightarrow \infty$. Additionally,
$$
\binom{|D_N|}{t} c_1 t^{5 / 2} \exp \left\{-\frac{n^{1 / 2} \epsilon_n}{c_2 t^3(\log N)^{1 / 2}}\right\} \leq C\binom{N}{t} t^{5 / 2} \exp \left\{-\frac{n^{1 / 2} \epsilon_n}{c_2 t^3(\log N)^{1 / 2}}\right\} \rightarrow 0,
$$
as $\epsilon_n \rightarrow 0$ sufficiently slow. Thus, we have
$$
\sum^* \mathbb{P}\left\{\left|\hat{u}_{i_1}\right|>z, \cdots,\left|\hat{u}_{i_t}\right|>z\right\} \rightarrow \frac{1}{t !} \pi^{-t / 2} e^{-t y / 2}.
$$
Then, by Bonferroni inequality,
$$
\sum_{t=1}^{2 k}(-1)^{t-1} \vartheta_t^{U} \leq \mathbb{P}\left(\max _{s \in D_N}\left|\hat{u}_s\right|>z\right) \leq \sum_{t=1}^{2 k+1}(-1)^{t-1} \vartheta_t^{U},
$$
for any $k \geq 1$. let $p \rightarrow \infty$, we have
$$
\begin{aligned}
	\sum_{t=1}^{2 k}(-1)^{t-1} \frac{1}{t !}\left(\frac{1}{\sqrt{\pi}} e^{-x / 2}\right)^t & \leq \liminf _{p \rightarrow \infty} \mathbb{P}\left(\max _{s \in D_N}\left|\hat{u}_s\right|>z\right) \\
	& \leq \limsup _{p \rightarrow \infty} \mathbb{P}\left(\max _{s \in D_N}\left|\hat{u}_s\right|>z\right) \leq \sum_{t=1}^{2 k+1}(-1)^{t-1} \frac{1}{t !}\left(\frac{1}{\sqrt{\pi}} e^{-x / 2}\right)^t,
\end{aligned}
$$
for each $k \geq 1$. By letting $k \rightarrow \infty$ and using the Taylor expansion of the function $1-e^{-x}$, we obtain the result.     \hfill$\Box$
\subsection*{Proof of the Theorem \ref{th:LUH1}}
Because $\tilde{U}_{ij}$ is a U-statistic with degenerate degree $d=1,$
we have $$n\var_{H_0}(\tilde{U}_{ij})=m^2\operatorname{var}_{H_0}\left[\mE_{H_0}\left\{h\left(\bm{Z}_{1}, \ldots,\bm{Z}_{m}\right) \mid\bm{Z}_{1}\right\}\right]=a_{4}\{1+o(1)\}$$
by \cite{serfling}.
By Lemma 4 in \cite{chen2022rank} or Lemma C7 in \cite{distributionfree}, there exists a constant $c>0$, for any $t>0,$
$$
\mathbb{P}\left\{|\hat{U}_{i j}-\mE(\hat{U}_{i j})|>t\right\} \leq 2 e^{-nt^2 / c} .
$$
Taking the same procedure as in the proof of Theorem \ref{th:LVH1}, we can also obtain
the result. \hfill$\Box$

\subsection*{Proof of Theorem \ref{th:LUH1 optimal}}
According to Theorem \ref{R optimal} and Assumption \ref{assumpLUrate}, we can easily obtain
the result. \hfill$\Box$
\subsection*{Proof of Theorem \ref{th:SU}}
By Lemma 5.2.2B in \cite{serfling},
for any even $r>2$,
\begin{align}\label{Ustmoment}
	\mE_{H_{0}}\left[\left\{\tilde{U}_{ij}-\mE_{H_0}(\tilde{U}_{ij})\right\}^{r}\right]=O\left(n^{-r/ 2}\right).
\end{align}
We assume that $\mE_{H_0}(\tilde{U}_{ij})=0.$
Hence, we have for any $1\leq i\leq p,$
\begin{align*}
	&\mE|\sum_{j=1}^{q}\left\{\tilde{U}_{ij}^{2}-\mE_{H_0}(\tilde{U}_{ij}^{2})\right\}|^{2+\delta}\\
	\leq& C_{\delta}^{2+\delta}\mE(|\left\{\tilde{U}_{ij}^{2}-\mE_{H_0}(\tilde{U}_{ij}^{2})\right\}|^{2+\delta})(4m_{q}q)^{(2+\delta)/2}\\
	\leq& C_{\delta}^{2+\delta}O(n^{-(2+\delta)})(4m_{q}q)^{(2+\delta)/2},
\end{align*}
where $C_q $ is a positive constant depending only upon $\delta$.
Let $ \overline{\tilde{U}}_{ij}\doteq \tilde{U}_{ij}^{2}-\mE_{H_0}(\tilde{U}_{ij}^2)$ and $\overline{\tilde{U}}_{i.}\doteq\sum_{j=1}^{q}\overline{\tilde{U}}_{ij}.$
For all $a$ and all $k\geq m_{p}$, we have
\begin{align*}
	\var\Big(\sum_{i=a}^{a+k-1} \overline{\tilde{U}}_{i.}\Big)\leq& \sum_{i_{1}=a}^{a+k-1}\sum_{i_{2}=a}^{a+k-1} \mE(\overline{\tilde{U}}_{i_{1}.}\overline{\tilde{U}}_{i_{2}.})\\
	\leq &\sum_{i_{1}=a}^{a+k-1}m_{p}\max_{1\leq i\leq p}\mE(\overline{\tilde{U}}_{i_{1}.}^2)\\
	\leq & km_{p}\mE(\overline{\tilde{U}}_{i_{1}.}^2)= O(n^{-2}) km_{p}m_{q}q.
\end{align*}
When $m_p\leq k\leq p^{\alpha}$ and $\alpha=1/\{1+(1-\gamma)(1+2/\delta)\},$
we have $\var\Big(\sum_{i=a}^{a+k-1} \overline{\tilde{U}}_{i.}\Big)/k^{1+\gamma}\leq \max\{m_{p}^{-\gamma},p^{-\alpha\gamma}\} O(n^{-2}) m_{p}m_{q}q.$
Hence, according to Theorem 2.1 in \cite{mdependent} and Assumptions \ref{assmp3} and \ref{condi:sigmaSU}, we can obtain the conclusion. \hfill$\Box$
\subsection*{Proof of Theorem \ref{th:CU}}
Taking the similar procedure as in the proof of Theorem \ref{th:CV}, we can obtain the conclusion. \hfill$\Box$

\subsection*{Proof of Theorem \ref{th:LQ}}
\begin{lemma}\label{lemma5.8} Define $\tilde{v}_s=\sum_{v=1}^M \lambda_v\left\{n^{-1 / 2} \sum_{k=1}^{n} \phi_v\left(\boldsymbol{S}_{k, i_{s}j_{s}}\right)\right\}^2=\sum_{v=1}^M \lambda_v\left\{n^{-1 / 2} \sum_{k=1}^{n} Q_{v,ks}\right\}^2$ where $s \in\{1, \cdots, N\}$. Let
	$$
	\vartheta_t^{Q}=\sum^* \mathbb{P}\left(\tilde{v}_{i_1}>z, \cdots, \tilde{v}_{i_t}>z\right),
	$$
	for $1 \leq t \leq N$, where the sum runs over all $i_1<\cdots<i_t$ and $i_1 \in D_N, \cdots, i_t \in D_N$. Then,
	$$
	\lim _{p \rightarrow \infty} \vartheta_t^{Q}=\frac{1}{t !}\left(\frac{\kappa}{\Gamma\left(\mu_1 / 2\right)} e^{-\frac{y}{2}}\right)^{-t},
	$$
	for each $t \geq 1.$
\end{lemma}
We proceed in two steps, proving first the case $m=2$ and then generalizing to $m \geq 2$. For notational convenience we introduce the constants $b_1=\|h\|_{\infty}<$ $\infty$ and $b_2=\sup _v\left\|\phi_v\right\|_{\infty}<\infty$. Similar to the proof of Theorem \ref{th:LU}. we define $\left\{\tilde{u}_s\right\}_{s=1}^N=\big\{ \tilde{U}_{i j}\big\}_{1 \leq i\leq p,1\leq j \leq q}.$ Recall that $h\left(\boldsymbol{S}_{1, ij}, \cdots, \boldsymbol{S}_{m, i j}\right)=h\left(\bm{Z}_{1, ij}, \cdots, \bm{Z}_{m, i j}\right),$
where $\bm{Z}_{k,ij}=(X_{k,i},Y_{k,j})$.
So we rewrite $\tilde{U}_{i j}$ in the following forms
\begin{align*}
	\tilde{u}_s=&\frac{(n-m)!m!}{n!} \sum_{1 \leq k_1<k_2, \cdots,<k_m \leq n} h\left(\bm{Z}_{k_1, i_s j_s}, \cdots, \bm{Z}_{k_m, i_s j_s}\right)\n\\
	=&\frac{(n-m)!m!}{n!} \sum_{1 \leq k_1<k_2, \cdots,<k_m \leq n} h\left(\boldsymbol{S}_{k_1, i_s j_s}, \cdots, \boldsymbol{S}_{k_m, i_s j_s}\right).
\end{align*}

Step I. Suppose $m=2$. In this case, we naturally have $$h\left(\boldsymbol{S}_{k_1, i_s j_s},  \boldsymbol{S}_{k_2, i_s j_s}\right)=h_2\left(\boldsymbol{S}_{k_1, i_s j_s},\boldsymbol{S}_{k_2, i_s j_s}\right).
%=\sum_{v=1}^{\infty}\lambda_{v}\psi_{v}(\boldsymbol{S}_{k_1, i_s j_s})\psi_{v}(\boldsymbol{S}_{k_2, i_s j_s})
$$  We start with the scenario that there are infinitely many nonzero eigenvalues.
For a large enough integer $M=\lfloor n^{(1-3 \theta) / 5}\rfloor,$ we define the ``truncated" U-statistic
$$
\tilde{u}_{M, s}=\binom{n}{2}^{-1} \sum_{1 \leq k<l \leq n} h_{2, M}\left(\boldsymbol{S}_{k, i_s j_s}, \boldsymbol{S}_{l, i_s j_s}\right),
$$
where $h_{2, M}\left(\boldsymbol{S}_{k, i_s j_s}, \boldsymbol{S}_{l, i_s j_s}\right)=\sum_{v=1}^{M}\lambda_{v}\phi_{v}(\boldsymbol{S}_{k, i_s j_s})\phi_{v}(\boldsymbol{S}_{l, i_s j_s}).$
%For simpler presentation, define
Recall that $Q_{v, ks}=\phi_v\left(\boldsymbol{S}_{k, i_s j_s}\right)$ for all $v=1,2, \ldots,$ $s=1,\ldots,N$ and $k =1,2,\ldots,n$. In view of the expansions of $h_{2, M}(\cdot),$ $\tilde{u}_{M, s}$ can be written as
$$
\begin{aligned}
	\tilde{u}_{M, s} & =\frac{1}{n-1}\bigg\{\sum_{v=1}^M \lambda_v\left(n^{-1 / 2} \sum_{k=1}^{n} Q_{v, ks}\right)^2-\sum_{v=1}^M \lambda_v\left(\frac{\sum_{k=1}^{n} Q_{v, ks}^2}{n}\right)\bigg\}.
	%\tilde{u}_s & =\frac{1}{n-1}\left\{\sum_{v=1}^{\infty} \lambda_v\left(n^{-1 / 2} \sum_{k=1}^{n} Q_{v, ks}\right)^2-\sum_{v=1}^{\infty} \lambda_v\left(\frac{\sum_{k=1}^{n} Q_{v, ks}^2}{n}\right)\right\}
\end{aligned}
$$
By the definition of $\theta$, there exist a positive absolute constant $C_\theta$ such that $\sum_{v=M+1}^{\infty} \lambda_v \leq C_\theta n^{-\theta}$ for all sufficiently large $n$. Thus, for any $\epsilon>0$, we have
$$
\mathbb{P}\left\{\max _{1 \leq s \leq N}\left(n-1\right)\left|\tilde{u}_{M, s}-\tilde{u}_s\right| \geq \epsilon\right\}  \leq N \mathbb{P}\left\{\left(n-1\right)\left|\tilde{u}_{M, s}-\tilde{u}_s\right| \geq \epsilon\right\}.
$$
Here, for a single pair $(i_{s},j_{s}),$ because $X_{i_s}$ is independent with $Y_{j_s}$ under null hypothesis, $\tilde{u}_{s}$ can be written as
$$
\begin{aligned}
	\tilde{u}_s & =\frac{1}{n-1}\left\{\sum_{v=1}^{\infty} \lambda_v\left(n^{-1 / 2} \sum_{k=1}^{n} Q_{v, ks}\right)^2-\sum_{v=1}^{\infty} \lambda_v\left(\frac{\sum_{k=1}^{n} Q_{v, ks}^2}{n}\right)\right\}.
\end{aligned}
$$
Hence, we have
\begin{align*}
	\mathbb{P}\left(\max _{1 \leq s \leq N}\left(n-1\right)\left|\tilde{u}_{M, s}-\tilde{u}_s\right| \geq \epsilon\right)  \leq &N \mathbb{P}\left\{\left(n-1\right)\left|\tilde{u}_{M, s}-\tilde{u}_s\right| \geq \epsilon\right\}\\
	\leq& 2 N e^{1 / 12} \exp \left(-\frac{\epsilon}{12 b_2^2 \sum_{v=M+1}^{\infty} \lambda_v}\right) \\
	\leq &2 N e^{1 / 12} \exp \left(-\frac{\epsilon n^\theta}{12 b_2^2 C_\theta}\right) \rightarrow 0,
\end{align*}
by $\log N=o\left(n^\theta\right)$. Here the second inequality are followed by (A.9) in \cite{drton2020}.
Thus, by
$$
\left|\max _{1 \leq s \leq N}\left(n-1\right) \tilde{u}_s-\max _{1 \leq s \leq N}\left(n-1\right) \tilde{u}_{M, s}\right| \leq \max _{1 \leq s \leq N}\left(n-1\right)\left|\tilde{u}_s-\tilde{u}_{M, s}\right| \rightarrow 0,
$$
we only need to show that
$$
\begin{aligned}
	& \mathbb{P}\left\{\max _{1 \leq s \leq N}\left(n-1\right) \tilde{u}_{M, s}-2 \lambda_1 \log N-\lambda_1\left(\mu_1-2\right) \log \log N+\Lambda \leq \lambda_1 y\right\} \\
	& \rightarrow \exp \left\{-\frac{\kappa}{\Gamma\left(\mu_1 / 2\right)} \exp \left(-\frac{y}{2}\right)\right\}.
\end{aligned}
$$
Define
$$
\hat{u}_{M, s}=\frac{1}{n-1}\left\{\sum_{v=1}^M \lambda_v\left(n^{-1 / 2} \sum_{i=1}^{n} Q_{v, ks}\right)^2-\sum_{v=1}^M \lambda_v\right\}.
$$
Thus, for any $\epsilon>0$, we have
$$
\begin{aligned}
	\mathbb{P}\left(\max _{1 \leq s\leq  N}\left(n-1\right)\left|\tilde{u}_{M, s}-\hat{u}_{M, s}\right| \geq \epsilon\right) & \leq N \mathbb{P}\left\{\left(n-1\right)\left|\tilde{u}_{M, s}-\hat{u}_{M, s}\right| \geq \epsilon\right\}\\
	& \leq N \mathbb{P}\left\{\left|\sum_{v=1}^M \lambda_v \frac{\sum_{i=1}^{n}\left(Q_{v, ks}^2-1\right)}{n}\right| \geq \epsilon\right\} \\
	& \leq 2 N \exp \left\{-\frac{n \epsilon^2}{48 \Lambda^2\left(b_2^2+1\right)^2}\right\} \rightarrow 0,
\end{aligned}
$$
by $\log N=o\left(n^\theta\right)$. Here the second inequality are followed by (A.10) in \cite{drton2020}.
Thus, by
$$
\left|\max _{1 \leq s \leq N}\left(n-1\right) \hat{u}_{M, s}-\max _{1 \leq s \leq N}\left(n-1\right) \tilde{u}_{M, s}\right| \leq \max _{1 \leq s \leq N}\left(n-1\right)\left|\hat{u}_{M, s}-\tilde{u}_{M_, s}\right| \rightarrow 0,
$$
we only need to show that
$$
\begin{aligned}
	& \mathbb{P}\left\{\max _{1 \leq s \leq N}\left(n-1\right) \hat{u}_{M, s}-2 \lambda_1 \log N-\lambda_1\left(\mu_1-2\right) \log \log N+\Lambda \leq \lambda_1 y\right\} \\
	& \rightarrow \exp \left\{-\frac{\kappa}{\Gamma\left(\mu_1 / 2\right)} \exp \left(-\frac{y}{2}\right)\right\}.
\end{aligned}
$$
Define $\tilde{v}_s=\sum_{w=1}^M \lambda_w\left(n^{-1 / 2} \sum_{k=1}^{n} Q_{w, ks}\right)^2=\left(n-1\right) \hat{u}_{M, s}+\sum_{w=1}^M \lambda_w$. By the definition of $\Lambda$, we have $\Lambda-\sum_{w=1}^M \lambda_w=\sum_{w=M+1}^{\infty} \lambda_w\leq C_{\theta}n^{-\theta}$, thus, we only need to show that
$$
\begin{aligned}
	& \mathbb{P}\left\{\max _{1 \leq s \leq N} \tilde{v}_s-2 \lambda_1 \log N-\lambda_1\left(\mu_1-2\right) \log \log N \leq \lambda_1 y\right\} \\
	& \rightarrow \exp \left\{-\frac{\kappa}{\Gamma\left(\mu_1 / 2\right)} \exp \left(-\frac{y}{2}\right)\right\}.
\end{aligned}
$$
Define $z=2 \lambda_1 \log N+\lambda_1\left(\mu_1-2\right) \log \log N+\lambda_1 y+o(1/\log N)$.
By Theorem 4.1 in \cite{drton2020} and $\log N=o\left(n^\theta\right)$, we have
$$
\mathbb{P}\left(\tilde{v}_1 \geq z\right)=\frac{\kappa}{\Gamma\left(\mu_1 / 2\right)}\left(\frac{z}{2 \lambda_1}\right)^{\mu_1 / 2-1} \exp \left(-\frac{z}{2 \lambda_1}\right)\{1+o(1)\} \sim \frac{\kappa}{\Gamma\left(\mu_1 / 2\right)} \frac{e^{-y / 2}}{N}.
$$
Thus,
$$
\mathbb{P}\left(\max _{s \in C_N} \tilde{v}_s>z\right) \leq\left|C_N\right| \cdot P\left(\tilde{v}_s \geq z\right) \rightarrow 0,
$$
as $p\rightarrow \infty$. Set $D_N=\left\{1 \leq s \leq N ;\left|B_{N, s}\right|<N^{\varsigma}\right\}$. By assumption, $|D_N| / N \rightarrow 1$ as $p \rightarrow \infty$ Easily,
$$
\begin{aligned}
	\mathbb{P}\left(\max _{s \in D_N} \tilde{v}_s>z\right) & \leq \mathbb{P}\left(\max _{1 \leq s \leq N} \tilde{v}_s>z\right) \\
	& \leq \mathbb{P}\left(\max _{s \in D_N} \tilde{v}_s>z\right)+\mathbb{P}\left(\max _{s \in C_N} \tilde{v}_s>z\right).
\end{aligned}
$$
Therefore, to prove Theorem \ref{th:LQ}, it is enough to show
$$
\lim _{p \rightarrow \infty} \mathbb{P}\left(\max _{s \in D_N} \tilde{v}_s>z\right)=1-\exp \left\{-\frac{\kappa}{\Gamma\left(\mu_1 / 2\right)} \exp \left(-\frac{y}{2}\right)\right\},
$$
as $p \rightarrow \infty$. Define
$$
\vartheta_t^{Q}=\sum^{\ast} P\left(\tilde{v}_{i_1}>z, \cdots, \tilde{v}_{{i}_t}>z\right),
$$
for $1 \leq t \leq N$, where the sum rums over all $i_1<\cdots<i_t$ and $i_1 \in D_N, \cdots, i_t \in D_N$. By Lemma \ref{lemma5.8}, we have
$$
\lim _{p \rightarrow \infty} \vartheta_t^{Q}=\frac{1}{t !}\left(\frac{\kappa}{\Gamma\left(\mu_1 / 2\right)} e^{-\frac{y}{2}}\right)^{t},
$$
for each $t \geq 1$. Then, by Bonferroni inequality.
$$
\sum_{t=1}^{2 k}(-1)^{t-1} \vartheta_t^{Q} \leq \mathbb{P}\left(\max _{s \in D_N} \tilde{v}_s>z\right) \leq \sum_{t=1}^{2 k+1}(-1)^{t-1} \vartheta_t^{Q},
$$
for any $k \geq 1$. let $p \rightarrow \infty$, we have
$$
\begin{aligned}
	\sum_{t=1}^{2 k}(-1)^{t-1} \frac{1}{t !}\left\{\frac{\kappa}{\Gamma\left(\mu_1 / 2\right)} e^{-\frac{y}{2}}\right\}^t & \leq \liminf _{p \rightarrow \infty} P\left(\max _{i \in D_N} \tilde{v}_i>z\right) \\
	& \leq \limsup _{p \rightarrow \infty} \mathbb{P}\left(\max _{i \in D_N} \tilde{v}_i>z\right) \leq \sum_{i=1}^{2 k+1}(-1)^{t-1} \frac{1}{t !}\left\{\frac{\kappa}{\Gamma\left(\mu_1 / 2\right)} e^{-\frac{y}{2}}\right\}^t,
\end{aligned}
$$
for each $k \geq 1$. By letting $k \rightarrow \infty$ and using the Taylor expansion of the function $1-e^{-x}$, so we obtain the result.
If there are only finitely many nonzero eigenvalues, $M$ is the number of nonzero eigenvalues. One can verify that the conclusion still holds.

Step II. For $m \geq 2$, by the Hoeffding decomposition, we have
$$
\tilde{u}_s=C_m^2 H_{n,s}^{(2)}+\sum_{\ell=3}^m C_m^t H_{n,s}^{(\ell)}
$$
where $H_{n,s}^{(\ell)}$ is the U-statistic based on the completely degenerate kernel $h^{(\ell)}$ from (\ref{2.8}). Here
$$
H_{n,s}^{(\ell)}=\binom{n}{l}^{-1} \sum_{1 \leq k_1<k_2<\cdots<k_{\ell} \leq n} h^{(\ell)}\left(\boldsymbol{S}_{k_{1}, i_s j_s}, \ldots, \boldsymbol{S}_{k_{\ell}, i_s j_s} \right),
$$

$$
h^{(\ell)}\left(\boldsymbol{S}_{k_{1}, i_s j_s}, \ldots, \boldsymbol{S}_{k_{\ell}, i_s j_s}\right)=h_{\ell}\left(\boldsymbol{S}_{k_{1}, i_s j_s}, \ldots, \boldsymbol{S}_{k_{\ell}, i_s j_s}\right)-\sum_{r=1}^{\ell-1}
\underset{\substack{q_{1}< \dots <q_{r},\\
		q_{1}, \dots q_{r}\in \{k_1,\dots,k_{\ell}\}}}{\sum}h^{(r)}\left(\boldsymbol{S}_{q_{1}, i_s j_s}, \ldots, \boldsymbol{S}_{q_{r}, i_s j_s}\right)
$$
and $h_{\ell}\left(\boldsymbol{S}_{k_{1}, i_s j_s}, \ldots, \boldsymbol{S}_{k_{\ell}, i_s j_s}\right)=\mE(\boldsymbol{S}_{k_{1}, i_s j_s}, \ldots, \boldsymbol{S}_{k_{\ell}, i_s j_s},\boldsymbol{S}_{k_{\ell+1}, i_s j_s},\boldsymbol{S}_{k_{m}, i_s j_s}\mid\boldsymbol{S}_{k_{1}, i_s j_s}, \ldots, \boldsymbol{S}_{k_{\ell}, i_s j_s} ).$
To prove the result, we only need to show that $\max _{1 \leq s \leq N}\left(n-1\right) H_{n,s}^{(\ell)}\left(\boldsymbol{S}_{k_{1}, i_s j_s}, \ldots, \boldsymbol{S}_{k_{\ell}, i_s j_s}\right)=o(1)$ for $\ell \geq 3.$

By Proposition 2.3(c) in \cite{arcones1993limit}, there exist positive constant $C_1, C_2$ such that for all $\epsilon>0$,
$$
\mathbb{P}\left\{n^{\ell / 2}\left|H_{n,s}^{(\ell)}\left(\boldsymbol{S}_{k_{1}, i_s j_s}, \ldots, \boldsymbol{S}_{k_{\ell}, i_s j_s}\right)\right| \geq \epsilon\right\} \leq C_1 \exp \left\{-C_2\left(\frac{\epsilon}{2^{\ell} b_1}\right)^{2 / \ell}\right\}.
$$
So, for any $\epsilon_1>0$,
$$
\begin{aligned}
	&\mathbb{P}\left\{\max _{1 \leq s \leq N}\left(n-1\right) H_{n,s}^{(\ell)}\left(\boldsymbol{S}_{k_{1}, i_s j_s}, \ldots, \boldsymbol{S}_{k_{\ell}, i_s j_s}\right)\geq \epsilon_1\right\} \\
	\leq &N \mathbb{P}\left\{\left(n-1\right) H_{n,s}^{(\ell)}\left(\boldsymbol{S}_{k_{1}, i_s j_s}, \ldots, \boldsymbol{S}_{k_{\ell}, i_s j_s}\right) \geq \epsilon_1\right\} \\
	\leq& C_1 N \exp \left\{-C_2\left(\frac{n^{\ell / 2-1} \epsilon_1}{2^{\ell} b_1}\right)^{2 / \ell}\right\} \rightarrow 0,
\end{aligned}
$$
by the assumption $\log N=o\left(n^\theta\right)$. Here we complete the proof.
\hfill$\Box$

\subsection*{Proof of Theorem \ref{th:LQH1}}
The proof is similar to the proof of Theorem 4.3 in \cite{drton2020}. So we omit it here.\hfill$\Box$
%\subsection*{Proof of Theorem \ref{th:LQH1 optimal}}
%According to Theorem 3 in \cite{feng2022testing} and Assumption (A), we can easily obtain
%the result.\hfill$\Box$

\subsection*{Proof of Theorem \ref{th:LQH1 optimal}}
According to Theorem \ref{R optimal} and Assumption \ref{LQoptimal}, we can easily obtain
the result. \hfill$\Box$

\subsection*{Proof of Theorem \ref{th:SQ}}
By Lemma 5.2.2B in \cite{serfling},
for any even $r>2$,
\begin{align*}
	\mE_{H_{0}}\left[\left\{{\tilde{U}}_{ij}-\mE_{H_0}({\tilde{U}}_{ij})\right\}^{r}\right]=O\left(n^{-r}\right).
\end{align*}
Taking the similar procedure as Theorem \ref{th:SU}, we can obtain the conclusion. \hfill$\Box$
\subsection*{Proof of Theorem \ref{th:CQ}}
Taking the similar procedure as Theorem \ref{th:CV}, we can obtain the conclusion. \hfill$\Box$

\section{Proof of some Lemmas}
\subsection*{Proof of Lemma \ref{lemmalinear}}
Take $\varphi_N=1 /(\log N)^2$. Recall $B_{N, i}=\left\{1 \leq j \leq p: \left|\sigma_{i j}^{V}\right| \geq \varphi_N\right\}$ for $1 \leq i \leq p.$ By Assumption, we know $\max _{1 \leq i \leq p} \sum_{j=1}^p (\sigma_{i j}^{V})^2 \leq(\log N)^C$ for all $p \geq 1$. Then
$$
\left|B_{N, i}\right| \cdot \frac{1}{(\log N)^4} \leq \sum_{j=1}^p (\sigma_{i j}^{V})^2 \leq(\log N)^C,
$$
for each $i=1, \cdots, p$. This shows that $\max _{1 \leq i \leq N}\left|B_{N, i}\right| \leq(\log N)^{C+4}$. Take $\varsigma=\varsigma_N=(C+ 5) (\log \log N) / \log N$ for $N \geq e^e$. Then $\varsigma_N \rightarrow 0$ and $(\log N)^{C+4}<N^\varsigma$. As a result,
$$
C_N=\left\{1 \leq i \leq N ;\left|B_{N, i}\right| \geq N^\varsigma\right\}=\emptyset.
$$
Hence, $D_N=\left\{1 \leq i \leq N : \left|B_{N, i}\right|<N^\varsigma\right\}=\{1,2, \cdots, N\}$. Recall (\ref{alphat}). By identifying " $H_{V}(t, N)$ " here as " $\vartheta_t^{V}$ " there for each $t \geq 1$, we obtain
$$
\lim _{p \rightarrow \infty} H_{V}(t, N)=\frac{1}{t !} \pi^{-t / 2} e^{-y t / 2}.
$$
Then, we can obtain the conclusion. \hfill$\Box$
\subsection*{Proof of Lemma \ref{lemmalinear2}}
For $I_{1},I_{2},\cdots,I_{t} \in \Lambda_{N},$ write $I_{l}=\left(i_{l}, j_{l}\right)$ for $l=1,2, \cdots, t .$ Set
\begin{align*}
	\Lambda_{N, t}=\big\{\left(i,j \right) ; &\max(i_{l}-m_{p},0)\leq i\leq \max(i_{l}+m_{p},p),\\
	&\max(j_{l}-m_{q},0)\leq j\leq \max(j_{l}+m_{q},q),1\leq l\leq t\big\},
\end{align*}
for $t \geq 1 .$ It is easy to check that $ %\left(m_{p}+1\right)\left(m_{q}+1\right)\leq
\left|\Lambda_{N, t}\right|\leq t\left(2m_{p}+1\right)\left(2m_{q}+1\right) .$
Recall
$$
S_{V}=\sum_{i=1}^p\sum_{j=1}^{q}\left\{\tilde{V}^2_{ij}-\mE_{H_{0}}(\tilde{V}^2_{ij})\right\},
$$
$$
E_{N}(x)=\left\{S_{V}/\sigma_{S_{V}}\leq x\right\}, \quad x \in \mathbb{R}
$$
and for $t \geq 1$,
$$
S_{V}^{t}=\sum_{(i, j) \in \Lambda_{N, t}}\left\{\tilde{V}^2_{ij}-\mE_{H_{0}}(\tilde{V}^2_{ij})\right\}.
$$
Observe that $J_{I_{1}} \cdots J_{I_{t}}$ is an event generated by random vectors $\left\{\tilde{V}_{ij}: (i, j) \in \Lambda_{N, t}\right\}.$ A crucial observation is that $S_{V}-S_{V}^{k}$ is independent of $J_{I_{1}} J_{I_{2}} \cdots J_{I_{t}} .$
For even $\tau$ and $\tau>2$, we have
\begin{align*}
	\P_{H_{0}}\left(\left|S_{V}^{t}\right|>\sigma_{S_{V}}\epsilon\right)
	\leq  \frac{\left|\Lambda_{N, t}\right|^{\tau}\max_{(i,j) \in \Lambda_{N, t}} \mE_{H_{0}}\left[\left\{\tilde{V}^2_{ij}-\mE_{H_{0}}(\tilde{V}^2_{ij})\right\}^{\tau}\right]}
	{\sigma^{\tau}_{S_{V}}\epsilon^{\tau}}.
\end{align*}
Then we can easily obtain $
\mE_{H_{0}}(\left\{\tilde{V}_{ij}^{2}-\mE_{H_{0}}(\tilde{V}^2_{ij})\right\}^{\tau})=O(n^{-\tau})
$.
Next, due to Assumptions \ref{assmp3}-\ref{cond5}, for some constant $C>0,$ we have
$$
\P_{H_{0}}\left(\left|S_{V}^{t}\right|>\sigma_{S_{V}}\epsilon\right) \leq C \cdot \left(\frac{t^2m_p m_q}{N}\right)^{\tau/2},
$$
for large $N.$

Fix $I_{1},I_{2},\cdots,I_{t} \in \Lambda_{N} .$ By the definition of $H_0$,
$$
\begin{aligned}
	& \P_{H_{0}}\left\{E_{N}(x) J_{I_{1}} J_{I_{2}} \cdots J_{I_{t}}\right\} \\
	\leq & \P_{H_{0}}\left\{E_{N}(x) J_{I_{1}} J_{I_{2}} \cdots J_{I_{t}}, \frac{1}{\sigma_{S_{V}}}\left|S_{V}^{t}\right|<\epsilon\right\}+C \left(\frac{t^2m_p m_q}{N}\right)^{\tau/2} \\
	\leq & \P_{H_{0}}\left\{\frac{1}{\sigma_{S_{V}}}\left(S_{V}-S_{V}^{t}\right) \leq x+\epsilon, J_{I_{1}} J_{I_{2}} \cdots J_{I_{t}}\right\}+C \left(\frac{t^2m_p m_q}{N}\right)^{\tau/2} \\
	=& \P_{H_{0}}\left\{\frac{1}{\sigma_{S_{V}}}\left(S_{V}-S_{V}^{t}\right) \leq x+\epsilon\right\} \cdot \P_{H_{0}}\left(J_{I_{1}} J_{I_{2}} \cdots J_{I_{t}}\right)+C \left(\frac{t^2m_p m_q}{N}\right)^{\tau/2},
\end{aligned}
$$
by the independence between $S_{V}-S_{V}^{t}$ and $J_{I_{1}} J_{I_{2}} \cdots J_{I_{t}} .$ Now
$$
\begin{aligned}
	& \P_{H_{0}}\left\{\frac{1}{\sigma_{S_{V}}}\left(S_{V}-S_{V}^{t}\right) \leq x+\epsilon\right\} \\
	\leq & \P_{H_{0}}\left\{\frac{1}{\sigma_{S_{V}}}\left(S_{V}-S_{V}^{t}\right) \leq x+\epsilon,\frac{1}{\sigma_{S_{V}}}\left|S_{V}^{t}\right|<\epsilon\right\}+C \left(\frac{t^2m_p m_q}{N}\right)^{\tau/2} \\
	\leq & \P_{H_{0}}\left\{\frac{1}{\sigma_{S_{V}}}S_{V} \leq x+2 \epsilon\right\}+C \left(\frac{t^2m_p m_q}{N}\right)^{\tau/2} \\
	\leq & \P_{H_{0}}\left\{E_{N}(x+2 \epsilon)\right\}+C \left(\frac{t^2m_p m_q}{N}\right)^{\tau/2}.
\end{aligned}
$$
Combing the two inequalities to get
\begin{align}\label{leq2V}
	& \P_{H_{0}}\left\{E_{N}(x) J_{I_{1}} J_{I_{2}} \cdots J_{I_{t}}\right\}\n \\
	\leq & \P_{H_{0}}\left\{E_{N}(x+2 \epsilon)\right\} \cdot \P_{H_{0}}\left(J_{I_{1}} J_{I_{2}} \cdots J_{I_{t}}\right)+2 C \left(\frac{t^2m_p m_q}{N}\right)^{\tau/2}.
\end{align}
Similarly,
$$
\begin{aligned}
	& \P_{H_{0}}\left\{\frac{1}{\sigma_{S_{V}}}\left(S_{V}-S_{V}^{t}\right) \leq x-\epsilon, J_{I_{1}} J_{I_{2}} \cdots J_{I_{t}}\right\} \\
	\leq & \P_{H_{0}}\left\{\frac{1}{\sigma_{S_{V}}}\left(S_{V}-S_{V}^{t}\right) \leq x-\epsilon, J_{I_{1}} J_{I_{2}} \cdots J_{I_{t}},\frac{1}{\sigma_{S_{V}}}\left|S_{V}^{t}\right|<\epsilon\right\}+C \left(\frac{t^2m_p m_q}{N}\right)^{\tau/2} \\
	\leq & \P_{H_{0}}\left\{\frac{1}{\sigma_{S_{V}}}S_{V} \leq x, J_{I_{1}} J_{I_{2}} \cdots J_{I_{t}}\right\}+C \left(\frac{t^2m_p m_q}{N}\right)^{\tau/2}.
\end{aligned}
$$
In other words, by independence,
\begin{align*}
	&\P_{H_{0}}\left\{E_{N}(x) J_{I_{1}} J_{I_{2}} \cdots J_{I_{t}}\right\} \\
	\geq &\P_{H_{0}}\left\{\frac{1}{\sigma_{S_{V}}}\left(S_{V}-S_{V}^{t}\right) \leq x-\epsilon\right\} \P_{H_{0}}\left(J_{I_{1}} J_{I_{2}} \cdots J_{I_{t}}\right)-C \left(\frac{t^2m_p m_q}{N}\right)^{\tau/2}.
\end{align*}
Furthermore,
\begin{align}
	&\P_{H_{0}}\left\{\frac{1}{\sigma_{S_{V}}}S_{V} \leq x-2 \epsilon\right\}\n\\
	\leq &\P_{H_{0}}\left\{\frac{1}{\sigma_{S_{V}}}S_{V} \leq x-2 \epsilon, \frac{1}{\sigma_{S_{V}}}\left|S_{V}^{t}\right|<\epsilon\right\}+C \left(\frac{t^2m_p m_q}{N}\right)^{\tau/2} \n\\
	\leq& \P_{H_{0}}\left\{\frac{1}{\sigma_{S_{V}}}\left(S_{V}-S_{V}^{t}\right) \leq x-\epsilon\right\}+C \left(\frac{t^2m_p m_q}{N}\right)^{\tau/2}.
\end{align}
The above two strings of inequalities imply
$$
\begin{aligned}
	& \P_{H_{0}}\left(E_{N}(x) J_{I_{1}} J_{I_{2}} \cdots J_{I_{t}}\right) \\
	\geq & \P_{H_{0}}\left\{\frac{1}{\sigma_{S_{V}}}S_{V} \leq x-2 \epsilon\right\} \cdot \P_{H_{0}}\left(J_{I_{1}} J_{I_{2}} \cdots J_{I_{t}}\right)-2 C \left(\frac{t^2m_p m_q}{N}\right)^{\tau/2},
\end{aligned}
$$
which joining with (\ref{leq2V}) yields
$$
\begin{aligned}
	&\left|\P_{H_{0}}\left\{E_{N}(x) J_{I_{1}} J_{I_{2}} \cdots J_{I_{t}}\right\}-\P_{H_{0}}\left\{E_{N}(x)\right\} \cdot \P_{H_{0}}\left(J_{I_{1}} J_{I_{2}} \cdots J_{I_{t}}\right)\right| \\
	\leq & \Delta_{N, \epsilon} \cdot \P_{H_{0}}\left(J_{I_{1}} J_{I_{2}} \cdots J_{I_{t}}\right)+4 C \left(\frac{t^2m_p m_q}{N}\right)^{\tau/2},
\end{aligned}
$$
where
$$
\Delta_{N, \epsilon}\doteq\left| \P_{H_{0}}\left\{E_{N}(x)\right\}-\P_{H_{0}}\left\{E_{N}(x+2 \epsilon)\right\}\right|+\left| \P_{H_{0}}\left\{E_{N}(x)\right\}-\P_{H_{0}}\left\{E_{N}(x-2 \epsilon) \right\}\right|.
$$
In particular,
\begin{align}\label{diff of phiV}
	\Delta_{N, \epsilon} \rightarrow\left|\Phi(x+2 \epsilon)-\Phi(x)|+|\Phi(x-2 \epsilon)-\Phi(x)\right|,
\end{align}
as $n,p \rightarrow \infty$ by Theorem \ref{th:SV}. As a consequence,
$$
\begin{aligned}
	\zeta_{V}(t,N)\doteq& \sum_{I_{1}<I_{2}<\cdots<I_{t} \in \Lambda_{N}}\left[\P_{H_{0}}\left\{E_{N}(x) J_{I_{1}} J_{I_{2}} \cdots J_{I_{t}}\right\}-\right.\\
	&\left.\P_{H_{0}}\left\{D_{N}(x)\right\} \cdot \P_{H_{0}}\left(J_{I_{1}} J_{I_{2}} \cdots J_{I_{t}}\right)\right] \\
	\leq & \sum_{I_{1},I_{2},\cdots,I_{t} \in \Lambda_{N}}\left\{\Delta_{N, \epsilon} \cdot \P_{H_{0}}\left(J_{I_{1}} J_{I_{2}} \cdots J_{I_{t}}\right)+4 C \left(\frac{t^2m_p m_q}{N}\right)^{\tau/2}\right\} \\
	\leq & \Delta_{N, \epsilon} \cdot H_{V}(t, N)+\left(4 C\right) \cdot\binom{N}{t} \cdot \left(\frac{t^2m_p m_q}{N}\right)^{\tau/2},
\end{aligned}
$$
where
$$
H_{V}(t, N)=\sum_{I_{1},I_{2},\cdots,I_{t} \in \Lambda_{N}} \P_{H_{0}}\left(J_{I_{1}} J_{I_{2}} \cdots J_{I_{t}}\right),
$$
as defined in Lemmas \ref{lemmalinear} and \ref{lemmalinear2}. We know $\lim \sup _{p \rightarrow \infty} H_{V}(t, N) \leq C / t !$, where $C$ is a universal constant. Picking $\tau=2\lceil 4t\frac{1+(1-\gamma)(1+2/\delta)}{(1-\gamma)(1+2/\delta)}\rceil,$ and using the trivial fact $\binom{r}{d} \leq r^{d}$ for any integers $1 \leq d \leq r,$ we have that
$$
\binom{N}{t} \cdot \left(\frac{t^2m_p m_q}{N}\right)^{\tau/2}\leq N^{t} \cdot \frac{t^{\tau}}{N^{\frac{(1-\gamma)(1+2/\delta)\tau/2}{1+(1-\gamma)(1+2/\delta)} }} \leq \frac{t^{\tau}}{N^{2t}}.
$$
Hence, from (\ref{diff of phiV})
$$
\begin{aligned}
	\limsup _{p \rightarrow \infty} \zeta_{V}(t,N) & \leq \frac{C}{t !} \cdot \limsup _{n,p \rightarrow \infty} \Delta_{p, \epsilon} \\
	&=\frac{C}{t !} \cdot\left\{|\Phi(x+2 \epsilon)-\Phi(x)|+|\Phi(x-2 \epsilon)-\Phi(x)|\right\},
\end{aligned}
$$
for any $\epsilon>0 .$ The desired result follows by sending $\epsilon \downarrow 0$.
\hfill$\Box$

\subsection*{Proof of Lemma \ref{lemma5.8}}
Before proving Lemma \ref{lemma5.8}, we introduce two lemmas.
\begin{lemma}\label{lemma5.6}
	For $N \geq 1$, let $\iota_N$ be positive integers with $\lim _{p \rightarrow \infty} \iota_N / N=1$. Let $L_{ iv}, i=1, \cdots, \iota_N, v=1, \cdots, M,$ be standard normal
	distributed random variables and $\operatorname{cov}\left(L_{i v}, L_{i u}\right)=0$ for $v \neq u$. Let $\boldsymbol{L}_i=\left(L_{i 1}, \cdots, L_{i M}\right)^{\top}$ and $\mathbf{\Xi}_{i j}=\operatorname{cov}\left(\boldsymbol{L}_i, \boldsymbol{L}_j\right)$. Assume $\left|\lambda_{\max }\left(\mathbf{\Xi}_{i j} \mathbf{\Xi}_{i j}^{\top}\right)\right| \leq \varphi_N^{2+2 c}$ for $c>0$ and all $1 \leq i<j \leq \iota_N$, where $\left\{\varphi_N ; N \geq 1\right\}$ are constants satisfying $0<\varphi_N=o(1 / \log N)$. Define $W_{i_k}=\sum_{v=1}^M \lambda_v L_{i_k v}^2$. Given $y \in \mathbb{R}$, set $z=2 \lambda_1 \log N+\lambda_1\left(\mu_1-2\right) \log \log N+\lambda_1 y+$ $o(1 / \log N)$. Then, for any fixed $m \geq 1$, we have
	\begin{align}\label{15}
		\left\{\frac{\Gamma\left(\mu_1 / 2\right)}{\kappa} \frac{N}{e^{-y / 2}}\right\}^m \cdot \mathbb{P}\left(W_{i_1}>z, \cdots, W_{i_m}>z\right) \rightarrow 1,
	\end{align}
	as $p \rightarrow \infty$ uniformly for all $1 \leq i_1<\cdots<i_m \leq \iota_N$.
\end{lemma}

\begin{lemma}\label{lemma5.7}
	Let $W_i=\sum_{v=1}^M \lambda_v L_{i v}^2$ where $L_{i v} \stackrel{\text { i.i.d }}{\sim} N(0,1)$ for $v=1, \cdots, M$. Define $\boldsymbol{L}_i=$ $\left(L_{i 1}, \cdots, L_{i M}\right)^{\top}$ and $\boldsymbol{L}_S=\left(\boldsymbol{L}_{i_1}^{\top}, \cdots, \boldsymbol{L}_{i_{m-1}}^{\top}\right)^{\top}$, where $S=\left\{i_1, \cdots, i_{m-1}\right\}$, and for any $i, j \in S$, $\lambda_{\max }\left(\mathbf{\Xi}_{i j} \mathbf{\Xi}_{i j}^{\top}\right) \leq \varphi_N^{2+2 c_0}$ for some constant $c_0>0$. Let $\mathbf{\Xi}_{i j}=\operatorname{cov}\left(\boldsymbol{L}_i, \boldsymbol{L}_j\right)$. Let $\boldsymbol{L}_{i_m}$ satisfy $\max _{1 \leq j \leq m-1}\left|\lambda_{\max }\left(\mathbf{\Xi}_{i_m j} \mathbf{\Xi}_{i_{m j}}^{\top}\right)\right|>\varphi_N^{2+2 c_0}$ and $\lambda_{\max }\left(\boldsymbol{\Sigma}_{i_m S} \boldsymbol{\Sigma}_{Si_m }\right) \leq \delta \in(0,1)$ where $\boldsymbol{\Sigma}_{i_m S}=$ $\operatorname{cov}\left(\boldsymbol{L}_{i_m}, \boldsymbol{L}_S\right)$ and $\boldsymbol{\Sigma}_{S i_m}=\boldsymbol{\Sigma}_{i_m S}^{\top}$. Then, we have
	$$
	\mathbb{P}\left(W_{i_m} \geq z, \min _{1 \leq l \leq m-1} W_{i_l} \geq z\right) \leq C(\log N)^c N^{-m+(3+\varepsilon)/4},
	$$
	where $\varepsilon=\max\{2\delta,(1-\delta)/(1-\delta/2)\} \in(0,1), c, C$ are some positive constants.
\end{lemma}

According to Theorem 1.1 in \cite{zaitsev}, we have
\begin{align*}
	& \sum^* \mathbb{P}\left\{\sum_{v=1}^M \lambda_v L_{i_1 v}^2>z+\epsilon_n(\log N)^{-1}, \cdots, \sum_{v=1}^M \lambda_v L_{i_t v}^2>z+\epsilon_n(\log N)^{-1}\right\} \\
	& -\binom{|D_N|}{t} c_1 t^{5 / 2} \exp \left\{-\frac{n^{1 / 2} \epsilon_n}{c_2 t^3(\log N)}\right\} \\
	\leq & \sum^* \mathbb{P}\left\{\tilde{v}_{i_1}>z, \cdots, \tilde{v}_{i_t}>z\right\}) \\
	\leq & \sum^* \mathbb{P}\left\{\sum_{v=1}^M \lambda_v L_{i_1 v}^2>z-\epsilon_n(\log N)^{-1}, \cdots, \sum_{v=1}^M \lambda_v L_{i_t v}^2>z-\epsilon_n(\log N)^{-1}\right\}\\
	&+\binom{|D_N|}{t}  c_1 t^{5 / 2} \exp \left\{-\frac{n^{1 / 2} \epsilon_n}{c_2 t^3(\log N)}\right\},
\end{align*}
where $\left(L_{i_1 v}, \cdots, L_{i_t v}\right)$ follows a multivariate normal distribution with mean zero and the same covariance matrix with $\left(n^{-1 / 2} \sum_{k=1}^{n} Q_{v,ki_{1}}, \cdots,n^{-1 / 2} \sum_{k=1}^{n}  Q_{v,ki_{t}}\right)$. By the assumption $\log N=o\left(n^\theta\right)$, there exists a small enough $\epsilon_n \rightarrow 0$ satisfying
$$
\binom{\left|D_{N}\right|}{t}c_1 t^{5 / 2} \exp \left\{-\frac{n^{1 / 2} \epsilon_n}{c_2 t^3(\log N)}\right\} \rightarrow 0.
$$
Define $W_i=\sum_{v=1}^M \lambda_v L_{i v}^2$ and $\boldsymbol{L}_i=\left(L_{i 1}, \cdots, L_{i M}\right)$. So we only need to show that
$$
\sum^* \mathbb{P}\left(W_{i_1}>z, \cdots, W_{i_t}>z\right) \rightarrow \frac{1}{t !}\left(\frac{\kappa}{\Gamma\left(\mu_1 / 2\right)} e^{-\frac{y}{2}}\right)^{-t}.
$$
Recalling $D_N=\left\{1 \leq s \leq N : \left|B_{N, s}\right|<N^\varsigma\right\}$, we write
$$
\left\{\left(i_1, \cdots, i_t\right) \in\left(D_N\right)^t : i_1<\cdots<i_t\right\}=F_t \cup G_t,
$$
where $\sigma_{i_r i_s}=\lambda_{\max }\left(\mathbf{\Xi}_{i_r i_s} \mathbf{\Xi}_{i_r i_s}^{\top}\right)$ and $\mathbf{\Xi}_{i_r i_s}=\operatorname{cov}\left(\boldsymbol{L}_{i_r}, \boldsymbol{L}_{i_s}\right)$. Let
\begin{align}
	&F_t=\big\{\left(i_1, \cdots, i_t\right) \in\left(D_N\right)^t : i_1<\cdots<i_t \text{ and } \left|\sigma_{i_r i_s}\right| \leq \varphi_N^{2+2 c} \text{ for all } 1 \leq r<s \leq t\big\}, \n\\
	&G_t=\big\{\left(i_1, \cdots, i_t\right) \in\left(D_N\right)^t : i_1<\cdots<i_t \text{ and } \left|\sigma_{i_r i_s}\right|>\varphi_N^{2+2 c} \text{ for a pair }\left(i_r, i_s\right) \n\\
	&\quad\quad\quad\quad\quad\quad\quad\quad\quad\quad\quad\quad\quad\quad\quad\quad\quad\quad\quad\quad\quad\quad\quad\quad\quad\quad\quad\text{ with } 1 \leq r<s \leq t\big\}.
\end{align}
Now, think $D_N$ as graph with $|D_N|$ vertices. Keep in mind that $|D_N| \leq N$ and $|D_N| / N \rightarrow 1$. Any two different vertices from them, say, $i$ and $j$ are connected if $\left|\sigma_{i j}\right|>\varphi_N^{2+2 c}$. In this case we also say there is an edge between them. By the definition $D_N$, each vertex in the graph has at most $N^{\varsigma}$ neighbors. Replacing ``$n$ ", ``$q$ " and ``$t$ " in Lemma 7.1 in \cite{feng2022dependent} with ``$|D_N|$ ", ``$N^{\varsigma}$" and ``$t$ ", respectively, we have that $\left|G_t\right| \leq N^{t+\varsigma-1}$ for each $2 \leq t \leq N$. Therefore $\binom{|D_N|}{t}
\geq\left|F_t\right| \geq\binom{|D_N|}{t}-N^{t+\varsigma-1}$. Since $D_N / N \rightarrow 1$ and $\varsigma=\varsigma_N \rightarrow 0$ as $p \rightarrow \infty$, we know
\begin{align}\label{21}
	\lim _{p \rightarrow \infty} \frac{\left|F_t\right|}{N^t}=\frac{1}{t !} .
\end{align}
Here
$$
\begin{aligned}
	\vartheta_t^{Q}= & \sum_{\left(i_1, \cdots, i_t\right) \in F_t} \mathbb{P}\left(W_{i_1}>z, \cdots, W_{i_t}>z\right)+ \\
	& \sum_{\left(i_1, \cdots, i_t\right) \in G_t} \mathbb{P}\left(W_{i_1}>z, \cdots, W_{i_t}>z\right) .
\end{aligned}
$$
From Lemma \ref{lemma5.6} and (\ref{21}) we have
$$
\sum_{\left(i_1, \cdots, i_t\right) \in F_t} \mathbb{P}\left(W_{i_1}>z, \cdots, W_{i_t}>z\right) \rightarrow \frac{1}{t !}\left\{\frac{\kappa}{\Gamma\left(\mu_1 / 2\right)} e^{-\frac{y}{2}}\right\}^t,
$$
as $p \rightarrow \infty$. As a consequence, it remains to show
\begin{align}\label{22}
	\sum_{\left(i_1, \cdots, i_t\right) \in G_t} \mathbb{P}\left(W_{i_1}>z, \cdots, W_{i_t}>z\right) \rightarrow 0,
\end{align}
as $p \rightarrow \infty$ for each $t \geq 2.$
Next, we will prove (\ref{22}). If $t=2$, the sum of probabilities in (\ref{22}) is bounded by $\left|G_2\right| \cdot \max _{1 \leq {s_1}<{s_2} \leq N} \mathbb{P}\left(W_{s_1}>z, W_{s_2}>z\right)$. By Lemma 7.1 in \cite{feng2022dependent}, $\left|G_2\right| \leq N^{\varsigma+1}$. Since $\left|\sigma_{{s_1}{s_2}}\right| \leq \delta$, by Lemma \ref{lemma5.7}
\begin{align}\label{23}
	\mathbb{P}\left(W_{s_1}>z, W_{s_2}>z\right) \leq \frac{(\log N)^C}{N^{(5-\varepsilon) / 4}},
\end{align}
where $\varepsilon=\max\{2\delta,(1-\delta)/(1-\delta/2)\}\in(0,1),$
uniformly for all $1 \leq {s_1}<{s_2}\leq N$ as $p$ is sufficiently large, where $C>0$ is a constant not depending on $N$. We then know (\ref{22}) holds. So the remaining job is to show (\ref{22}) for $t \geq 3$.
Let $N \geq 2$ and $\left(\sigma_{{s_1} {s_2}}\right)_{N \times N}$ be a non-negative definite matrix. For $\varphi_N>0$ and a set $A \subset\{1,2, \cdots, m\}$ with $2 \leq m \leq N$, define
$$
\wp(A)=\max \left\{|S| : S \subset A \text { and } \max _{i \in S, j \in S, i \neq j}\left|\sigma_{i j}\right| \leq \varphi_N^{2+2 c}\right\}.
$$
Easily, $\wp(A)$ takes possible values $0,2 \cdots,|A|$, where we regard $|\varnothing|=0$. If $\wp(A)=0$, then $\left|\sigma_{{s_1} {s_2}}\right|>\varphi_N^{2+2 c}$ for all ${s_1} \in A$ and ${s_2} \in A$.
Now we will look at $G_t$ closely. To do so, we classify $G_t$ into the following subsets
$$
G_{t, j}=\left\{\left(i_1, \cdots, i_t\right) \in G_t: \wp\left(\left\{i_1, \cdots, i_t\right\}\right)=j\right\},
$$
for $j=0,2, \cdots, t-1$. By the definition of $G_t$, we see $G_t=\cup G_{t, j}$ for $j=0,2, \cdots, t-1$. Since $t \geq 3$ is fixed, to show (\ref{22}), it suffices to prove
\begin{align}\label{24}
	\sum_{\left(i_1, \cdots, i_t\right) \in G_{t, j}} \mathbb{P}\left(W_{i_1}>z, \cdots, W_{i_t}>z\right) \rightarrow 0,
\end{align}
for any $j \in\{0,2, \cdots, t-1\}$.
Assume $\left(i_1, \cdots, i_t\right) \in G_{t, 0}$. This implies that $\left|\sigma_{i_r i_s}\right|>\varphi_N^{2+2 c}$ for all $1 \leq r<s \leq t$. Therefore, the subgraph $\left\{i_1, \cdots, i_t\right\} \in G_t$ is a clique. Taking $n=|D_N| \leq N, t=t$ and $q=N^{\varsigma}$. Then by Lemma 7.1 in \cite{feng2022dependent}, $\left|G_{t, 0}\right| \leq N^{1+\varsigma(t-1)} \leq N^{1+t \varsigma}$. Thus, the sum from (\ref{24}) is bounded by
$$
N^{1+t \varsigma} \cdot \max _{1 \leq {s_1}<{s_2} \leq N} \mathbb{P}\left(W_{s_1}>z, W_{s_2}>z\right) \leq N^{1+t \varsigma} \cdot \frac{(\log N)^C}{N^{(5-\varepsilon) / 4}} \rightarrow 0,
$$
as $p \rightarrow \infty$ by using (\ref{23}). So (\ref{24}) holds with $j=0$.
Now we assume $\left(i_1, \cdots, i_t\right) \in G_{t, j}$ with $j \in\{2, \cdots, t-1\}$. By definition, there exits $S \subset\left\{i_1, \cdots, i_t\right\}$ such that $$\max _{{s_1} \in S, {s_2} \in S, {s_1} \neq {s_2}}\left|\sigma_{{s_1} {s_2}}\right| \leq \varphi_N^{2+2 c}$$ and for each $k \in\left\{i_1, \cdots, i_t\right\} \backslash S$, there exists $i \in S$ satisfying $\left|\sigma_{i k}\right|>\varphi_N^{2+2 c}$. Looking at the last statement we see two possibilities: (i) for each $k \in\left\{i_1, \cdots, i_t\right\} \backslash S$, there exist at least two indices, say, $i \in S, j \in S$ with $i \neq j$ satisfying $\left|\sigma_{i k}\right|>\varphi_N^{2+2 c}$ and $\left|\sigma_{j k}\right|>\varphi_N^{2+2 c}$; (ii) there exists $k \in\left\{i_1, \cdots, i_t\right\} \backslash S$ such that $\left|\sigma_{i k}\right|>\varphi_N^{2+2 c}$ for an unique $i \in S$. However, for $\left(i_1, \cdots, i_t\right) \in G_{t, j}$, (i) and (ii) could happen at the same time for different $S$, say, (i) holds for $S_1$ and (ii) holds for $S_2$ simultaneously. Thus, to differentiate the two cases, we introduce following two definitions. Set
\begin{align}\label{26}
	H_{t, j}=&\big\{\left(i_1, \cdots, i_t\right) \in G_{t, j} : \ \text{ there exist } S \subset\left\{i_1, \cdots, i_t\right\}\text{ with }|S|=j\text{ and
	}\n\\
	&\quad\max _{i \in S, j \in S, i \neq j}\left|\sigma_{i j}\right| \leq \varphi_N^{2+2 c}\text{ such that for any }k \in\left\{i_1, \cdots, i_t\right\} \backslash S\text{ there exist}r \in S, s \in S,\n\\
	&\quad\quad\quad\quad r \neq s \text{ satisfying }\min \left\{\left|\sigma_{k r}\right|,\left|\sigma_{k s}\right|\right\}>\varphi_N^{2+2 c}\big\}.
\end{align}

Replacing ``$n$ ", ``$q$ " and ``$t$ " in Lemma 7.1 in \cite{feng2022dependent} with ``$|D_N|$", ``$N^{\varsigma}$" and ``$t$", respectively, we have that $\left|H_{t, j}\right| \leq t^t \cdot N^{j-1+(t-j+1) \varsigma}$ for each $t \geq 3$. Again, set
$$
\begin{aligned}
	H_{t, j}^{\prime}= & \left\{\left(i_1, \cdots, i_t\right) \in G_{t, j}: \text { for any } S \subset\left\{i_1, \cdots, i_t\right\} \text { with }|S|=j\right. \text { and } \\
	& \max _{i \in S, j \in S, i \neq j}\left|\sigma_{i j}\right| \leq \varphi_N^{2+2 c} \text { there exists } k \in\left\{i_1, \cdots, i_t\right\} \backslash S \text { such that }\left|\sigma_{k r}\right|>\varphi_N^{2+2 c}
\end{aligned}
$$
$$
\text { for a unique } r \in S\} \text {. }
$$
From Lemma 7.1 in \cite{feng2022dependent}, we see $\left|H_{t, j}^{\prime}\right| \leq t^t \cdot N^{j+(t-j) \varsigma}$. It is easy to see $G_{t, j}=H_{t, j} \cup H_{t, j}^{\prime}$. Therefore, to show (\ref{24}) , we only need to prove
\begin{align}\label{28}
	\sum_{\left(i_1, \cdots, i_t\right) \in H_{t, j}} \mathbb{P}\left(W_{i_1}>z, \cdots, W_{i_t}>z\right) \rightarrow 0
\end{align}
and
\begin{align}\label{29}
	\sum_{\left(i_1, \cdots, i_t\right) \in H_{t, j}^{\prime}} \mathbb{P}\left(W_{i_1}>z, \cdots, W_{i_t}>z\right) \rightarrow 0,
\end{align}
as $p \rightarrow \infty$ for $j=2, \cdots, t-1$. In fact, let $S$ be as in $(\ref{26})$, then by using Lemma \ref{lemma5.6}, the probability in (\ref{28}) is bounded by $\mathbb{P}\left(\cap_{l \in S}\left\{W_l>z\right\}\right) \leq C \cdot N^{-j}$ uniformly for all $S$ as $p$ is
sufficiently large, where $C$ is a constant not depending on $N$. Thus,
$$
\begin{aligned}
	\sum_{\left(i_1, \cdots, i_t\right) \in H_{t, j}} \mathbb{P}\left(W_{i_1}>z, \cdots, W_{i_t}>z\right) & \leq t^t \cdot N^{j-1+(t-j+1) \varsigma} \cdot\left(C \cdot N^{-j}\right) \\
	& \leq\left(C t^t\right) \cdot N^{-1+t \varsigma},
\end{aligned}
$$
as $p$ is sufficiently large. By assumption $\varsigma=\varsigma_N \rightarrow 0$, we then get (\ref{28}).
Now we show (\ref{29}). Recall the definition of $H_{t, j}^{\prime}$. For $\left(i_1, \cdots, i_t\right) \in H_{t, j}^{\prime}$, pick $S \subset$ $\left\{i_1, \cdots, i_t\right\}$ with $|S|=j,$ $\max _{i \in S, j \in S, i \neq j}\left|\sigma_{i j}\right| \leq \varphi_N^{2+2 c}$ and $k \in\left\{i_1, \cdots, i_t\right\} \backslash S$ such that $\varphi_N^{2+2 c}<$ $\left|\sigma_{k r}\right| \leq \delta$ for a unique $r \in S$. Then the probability from (\ref{29}) is bounded by
$$
\mathbb{P}\left(W_k>z, \bigcap_{l \in S}\left\{W_l>z\right\}\right),
$$
for $2 \leq j \leq t-1$. Taking $m=j+1$ in Lemma \ref{lemma5.7}, then the probability above is dominated by
$
O\left\{\frac{(\log N)^{c_1}}{N^{j+(1-\varepsilon) / 4}}\right\},
$
for some constant $c_1$ not depending on $N$. As stated earlier, $\left|H_{t, j}^{\prime}\right| \leq t^t \cdot N^{j+(t-j) \varsigma}$. Multiplying
the two quantities, since $\varsigma=\varsigma_N \rightarrow 0$, we see the sum from (\ref{29}) is of order $O\left\{N^{-(1-\varepsilon) / 8}\right\}$. Therefore (\ref{29}) holds. We then have proved (\ref{24}) for any $j \in\{0,2, \cdots, t-1\}$. The proof is completed.

\hfill$\Box$

\subsection*{Proof of Lemma \ref{lemma5.6}}
For $m=1,$ (\ref{15}) is followed by Equation (6) in \cite{zolotarev1961concerning}. Assume Equation (\ref{15}) holds with $m=k-1$. We will prove it also holds with $m=k$.

Define $\boldsymbol{L}_S=\left(\boldsymbol{L}_{i_1}^{\top}, \cdots, \boldsymbol{L}_{i_{m-1}}^{\top}\right)^{\top},$ for $S=\{i_1,\cdots,i_{m-1}\},$
$\boldsymbol{\Sigma}_{S S}=\operatorname{cov}\left(\boldsymbol{L}_{S}, \boldsymbol{L}_S\right),$
$\boldsymbol{\Sigma}_{i_m S}=\operatorname{cov}\left(\boldsymbol{L}_{i_m}, \boldsymbol{L}_S\right)$ and $\boldsymbol{\Sigma}_{S i_m}=\boldsymbol{\Sigma}_{i_m S}^{\top}$. So $\boldsymbol{L}_{i_m}=$ $\left(\boldsymbol{L}_{i_m}-\boldsymbol{\Sigma}_{i_m S} \boldsymbol{\Sigma}_{SS}^{-1}\boldsymbol{L}_S\right)+\boldsymbol{\Sigma}_{i_m S} \boldsymbol{\Sigma}_{SS}^{-1}\boldsymbol{L}_S \doteq U_Y+V_Y$. Thus, by the assumptional distribution of multivariate normal distributions, we have $U_Y$ is independent of $V_Y$.
Define $\mathbf{A}=\operatorname{diag}\left\{\lambda_1, \cdots, \lambda_M\right\}$. Thus, we have
$$
\begin{aligned}
	& \mathbb{P}\left(W_{i_1}>z, \cdots, W_{i_m}>z\right) \\
	= & \mathbb{P}\left(U_Y^{\top} \mathbf{A} U_Y+2 U_Y^{\top} \mathbf{A} V_Y+V_Y^{\top} \mathbf{A} V_Y \geq z, \min _{1 \leq l \leq m-1} \boldsymbol{L}_{i_l}^{\top} \mathbf{A} \boldsymbol{L}_{i_l} \geq z\right) \\
	= & \mathbb{P}\left(U_Y^{\top} \mathbf{A} U_Y+2 U_Y^{\top} \mathbf{A} V_Y+V_Y^{\top} \mathbf{A} V_Y \geq z, 2 U_Y^{\top} \mathbf{A} V_Y+V_Y^{\top} \mathbf{A} V_Y \leq C \varphi_N, \min _{1 \leq l \leq m-1} \boldsymbol{L}_{i_l}^{\top} \mathbf{A} \boldsymbol{L}_{i_l} \geq z\right) \\
	& +\mathbb{P}\left(U_Y^{\top} \mathbf{A} U_Y+2 U_Y^{\top} \mathbf{A} V_Y+V_Y^{\top} \mathbf{A} V_Y \geq z, 2 U_Y^{\top} \mathbf{A} V_Y+V_Y^{\top} \mathbf{A} V_Y>C \varphi_N, \min _{1 \leq l \leq m-1} \boldsymbol{L}_{i_l}^{\top} \mathbf{A} \boldsymbol{L}_{i_l} \geq z\right) \\
	\leq & \mathbb{P}\left(U_Y^{\top} \mathbf{A} U_Y \geq z-C \varphi_N, \min _{1 \leq l \leq m-1} \boldsymbol{L}_{i_l}^{\top} \mathbf{A} \boldsymbol{L}_{i_l} \geq z\right)+\mathbb{P}\left(2 U_Y^{\top} \mathbf{A} V_Y+V_Y^{\top} \mathbf{A} V_Y>C \varphi_N\right) \\
	\leq & \mathbb{P}\left(U_Y^{\top} \mathbf{A} U_Y \geq z-C \varphi_N\right) \mathbb{P}\left(\min _{1 \leq l \leq m-1} \boldsymbol{L}_{i_l}^{\top} \mathbf{A} \boldsymbol{L}_{i_l} \geq z\right)+\mathbb{P}\left(U_Y^{\top} \mathbf{A} V_Y>C \varphi_N / 4\right)\\
	&+\mathbb{P}\left(V_Y^{\top} \mathbf{A} V_Y>C \varphi_N / 2\right).
\end{aligned}
$$
Since $U_Y$ follows $N\left(\mathbf{0}, \mathbf{I}_M-\boldsymbol{\Sigma}_{i_m S} \boldsymbol{\Sigma}_{SS}^{-1}\boldsymbol{\Sigma}_{Si_m }\right)$, we have
$$
U_Y^{\top} \mathbf{A} U_Y \sim \xi_U^{\top}\left(\mathbf{I}_M-\boldsymbol{\Sigma}_{i_m S} \boldsymbol{\Sigma}_{SS}^{-1}\boldsymbol{\Sigma}_{Si_m }\right)^{1 / 2} \mathbf{A}\left(\mathbf{I}_M-\boldsymbol{\Sigma}_{i_m S} \boldsymbol{\Sigma}_{SS}^{-1}\boldsymbol{\Sigma}_{Si_m }\right)^{1 / 2} \xi_U,
$$
where $\xi_U \sim N\left(\mathbf{0}, \mathbf{I}_M\right)$. Define the eigenvalues of $\left(\mathbf{I}_M-\boldsymbol{\Sigma}_{i_m S} \boldsymbol{\Sigma}_{SS}^{-1}\boldsymbol{\Sigma}_{Si_m }\right)^{1 / 2} \mathbf{A}\left(\mathbf{I}_M-\boldsymbol{\Sigma}_{i_m S} \boldsymbol{\Sigma}_{SS}^{-1}\boldsymbol{\Sigma}_{Si_m }\right)^{1 / 2}$ are $\tilde{\lambda}_1 \geq \tilde{\lambda}_2 \geq \cdots \geq \tilde{\lambda}_M$ and $\tilde{\Lambda}, \tilde{\kappa}, \tilde{\mu}_1$ are the corresponding parameters as in Proposition 3.2 in \cite{drton2020}. Because $\lambda_{\text {max }}\left(\boldsymbol{\Sigma}_{i_m S} \boldsymbol{\Sigma}_{SS}^{-1}\boldsymbol{\Sigma}_{Si_m }\right) \leq (m-1)\{1+o(1)\}\varphi_N^{2+2 c}, \tilde{\lambda}_1=\lambda_1\left\{1+o\left(\varphi_N\right)\right\}$. So does $\tilde{\Lambda}, \tilde{\kappa}, \tilde{\mu}_1$. So by Equation (6) in \cite{zolotarev1961concerning}, we have
$$
\begin{aligned}
	\mathbb{P}\left(U_Y^{\top} \mathbf{A} U_Y \geq z-C \varphi_N\right) & \rightarrow \frac{\tilde{\kappa}}{\Gamma\left(\tilde{\mu}_1 / 2\right)}\left(\frac{z-C \varphi_N+\tilde{\Lambda}}{2 \tilde{\lambda}_1}\right)^{\tilde{\mu}_1 / 2-1} \exp \left(-\frac{z-C \varphi_N+\tilde{\Lambda}}{2 \tilde{\lambda}_1}\right) \\
	& \rightarrow \frac{\kappa}{\Gamma\left(\mu_1 / 2\right)} \frac{e^{-y / 2}}{N},
\end{aligned}
$$
by the assumption $\varphi_N=o(1 / \log N)$. So by the
$$
\mathbb{P}\left(U_Y^{\top} \mathbf{A} U_Y \geq z-C \varphi_N\right) \mathbb{P}\left(\min _{1 \leq l \leq m-1} \boldsymbol{L}_{i_l}^{\top} \mathbf{A} \boldsymbol{L}_{i_l} \geq z\right) \rightarrow\left\{\frac{\kappa}{\Gamma\left(\mu_1 / 2\right)} \frac{e^{-y / 2}}{N}\right\}^m.
$$
Next, we will show that $ \mathbb{P}\left(U_Y^{\top} \mathbf{A} V_Y>C \varphi_N / 4\right)+ \mathbb{P}\left(V_Y^{\top} \mathbf{A} V_Y>C \varphi_N / 2\right)=o\left(N^{-m}\right)$. Similarly, $V_Y^{\top} \mathbf{A} V_Y$ follows
$$
\xi_V^{\top}  \boldsymbol{\Sigma}_{SS}^{-1/2}\boldsymbol{\Sigma}_{S i_m} \mathbf{A} \boldsymbol{\Sigma}_{i_m S} \boldsymbol{\Sigma}_{SS}^{-1/2}\xi_V,
$$
where  $\boldsymbol{\Sigma}_{SS}^{-1/2}$ is the principal square root
matrix of $\boldsymbol{\Sigma}_{SS}^{-1}$ and $\xi_V \sim N\left(\mathbf{0}, \mathbf{I}_{(m-1) M}\right)$. Define that the eigenvalues of $\boldsymbol{\Sigma}_{SS}^{-1/2}\boldsymbol{\Sigma}_{S i_m} \mathbf{A} \boldsymbol{\Sigma}_{i_m S}\boldsymbol{\Sigma}_{SS}^{-1/2}$ are $\zeta_1, \cdots, \zeta_M$. So $V_Y^{\top} \mathbf{A} V_Y$ follows
$$
\sum_{k=1}^{(m-1) M} \zeta_k \xi_k^2,
$$
where $\xi_k$ are all independently distributed as $N(0,1)$. Thus, for small enough constant $\varpi$,
$$
\begin{aligned}
	\mE\left\{\exp \left(\varpi \varphi_N^{-1-c} V_Y^{\top} \mathbf{A} V_Y\right)\right\} & =\prod_{k=1}^M \mE e^{\varpi \varphi_N^{-1-c} \zeta_k \xi_k^2}=\exp \left\{-\frac{1}{2} \sum_{k=1}^{(m-1) M} \log \left[1-2 \varpi \varphi_N^{-1-c} \zeta_k\right]\right\} \\
	& \leq \exp \left\{2 \varpi \varphi_N^{-1-c} \sum_{k=1}^{(m-1) M} \zeta_k\right\}.
\end{aligned}
$$
In addition,
$$
\operatorname{tr}\left(\boldsymbol{\Sigma}_{SS}^{-1/2}\boldsymbol{\Sigma}_{S i_m} \mathbf{A} \boldsymbol{\Sigma}_{i_m S}\boldsymbol{\Sigma}_{SS}^{-1/2}\right) \leq \lambda_{m a x}\left(\boldsymbol{\Sigma}_{SS}^{-1/2}\boldsymbol{\Sigma}_{S i_m} \boldsymbol{\Sigma}_{i_m S}\boldsymbol{\Sigma}_{SS}^{-1/2}\right) \operatorname{tr}(\mathbf{A})
\leq(m-1)\{1+o(1)\} \varphi_N^{2+2 c} \Lambda.
$$
So $\mE\left\{\exp \left(\varpi \varphi_N^{-1-c} V_Y^{\top} \mathbf{A} V_Y\right)\right\} \leq \exp \left(2 \varpi \varphi_N^{1+c} \Lambda\right) \leq C_2$ for some constant $C_2>0$. By the Markov inequality, we have
$$
\mathbb{P}\left(V_Y^{\top} \mathbf{A} V_Y>C \varphi_N / 2\right)= \mathbb{P}\left(\varpi \varphi_N^{-1-c} V_Y^{\top} \mathbf{A} V_Y>C \varpi \varphi_N^{-c} / 2\right)
$$
$$
\begin{aligned}
	& \leq \exp \left(-C \varpi \varphi_N^{-c} / 2\right) \mE\left\{\exp \left(\varpi \varphi_N^{-1-c} V_Y^{\top} \mathbf{A} V_Y\right)\right\} \\
	& \leq C_2 \exp \left(-C \varpi \varphi_N^{-c} / 2\right)=o\left(N^{-m}\right),
\end{aligned}
$$
for large enough constant $C$. Similarly, $U_Y^{\top} \mathbf{A} V_Y $ follows
$$
\sum_{k=1}^M \rho_k \xi_k \eta_k,
$$
where $\eta_k$ are all independently distributed as $N(0,1)$ and $\rho_1, \cdots, \rho_M$ are the singular values of $\left(\mathbf{I}_M-\boldsymbol{\Sigma}_{i_m S} \boldsymbol{\Sigma}_{SS}^{-1}\boldsymbol{\Sigma}_{Si_m }\right)^{1 / 2} \mathbf{A} \boldsymbol{\Sigma}_{i_m S}\boldsymbol{\Sigma}_{SS}^{-1/2}$. And then
$$
\begin{aligned}
	& \mathbb{P}\left(U_Y^{\top} \mathbf{A} V_Y>C \varphi_N / 4\right) \\
	= & \mathbb{P}\left(\varpi \varphi_N^{-1-c} U_Y^{\top} \mathbf{A} V_Y>C \varpi \varphi_N^{-c} / 4\right) \\
	\leq & \exp \left(-C \varpi \varphi_N^{-c} / 4\right) \mE\left\{\exp \left(\varpi \varphi_N^{-1-c} U_Y^{\top} \mathbf{A} V_Y\right)\right\} \\
	\leq & \exp \left(-C \varpi \varphi_N^{-c} / 4\right) \exp \left\{-\frac{1}{2} \sum_{k=1}^M \log \left(1-\varpi^2 \varphi_N^{-2-2 c} \rho_k^2\right)\right\}\\
	\leq& \exp \left(-C \varpi \varphi_N^{-c} / 4\right) \exp \left(\varpi^2 \varphi_N^{-2-2 c} \sum_{k=1}^M \rho_k^2\right) \\
	\leq &\exp \left(-C \varpi \varphi_N^{-c} / 4\right) \exp \left(\varpi^2 \varphi_N^{-2-2 c} \operatorname{tr}\left[\left\{\left(\mathbf{I}_M-\boldsymbol{\Sigma}_{i_m S} \boldsymbol{\Sigma}_{SS}^{-1}\boldsymbol{\Sigma}_{Si_m }\right)^{1 / 2} \mathbf{A} \boldsymbol{\Sigma}_{i_m S}\boldsymbol{\Sigma}_{SS}^{-1/2}\right\}^2\right]\right) \\
	\leq& \exp \left(-C \varpi \varphi_N^{-c} / 4\right) \exp \left\{\varpi^2 \varphi_N^{-2-2 c} \lambda_{\max }\left(\boldsymbol{\Sigma}_{S i_m}\boldsymbol{\Sigma}_{SS}^{-1} \boldsymbol{\Sigma}_{i_m S}\right) \operatorname{tr}\left(\mathbf{A}^2\right)\right\} \\
	\leq & \exp \left(-C \varpi \varphi_N^{-c} / 4\right) \exp \left[\varpi^2 (m-1)\{1+o(1)\} \operatorname{tr}\left(\mathbf{A}^2\right)\right]=o\left(N^{-m}\right),
\end{aligned}
$$
for large enough constant $C>0$. Thus, we have
\begin{align}\label{8.20}
	\mathbb{P}\left(W_{i_1}>z, \cdots, W_{i_m}>z\right) \leq\left\{\frac{\kappa}{\Gamma\left(\mu_1 / 2\right)} \frac{e^{-y / 2}}{N}\right\}^m+o\left(N^{-m}\right) .
\end{align}
Further more,
$$
\begin{aligned}
	& \mathbb{P}\left(W_{i_1}>z, \cdots, W_{i_m}>z\right) \\
	= & \mathbb{P}\left(U_Y^{\top} \mathbf{A} U_Y+2 U_Y^{\top} \mathbf{A} V_Y+V_Y^{\top} \mathbf{A} V_Y \geq z, \min _{1 \leq l \leq m-1} \boldsymbol{L}_{i_l}^{\top} \mathbf{A} \boldsymbol{L}_{i_l} \geq z\right) \\
	\geq & \mathbb{P}\left(U_Y^{\top} \mathbf{A} U_Y \geq z+C \varphi_N, \min _{1<l<m-1} \boldsymbol{L}_{i_l}^{\top} \mathbf{A} \boldsymbol{L}_{i_l} \geq z\right)- \mathbb{P}\left(2 U_Y^{\top} \mathbf{A} V_Y+V_Y^{\top} \mathbf{A} V_Y \leq-C \varphi_N\right)\\
	\geq & \mathbb{P}\left(U_Y^{\top} \mathbf{A} U_Y \geq z+C \varphi_N\right) \mathbb{P}\left(\min _{1 \leq l \leq m-1} \boldsymbol{L}_{i_l}^{\top} \mathbf{A} \boldsymbol{L}_{i_l} \geq z\right)- \mathbb{P}\left(2 U_Y^{\top} \mathbf{A} V_Y+V_Y^{\top} \mathbf{A} V_Y \leq-C \varphi_N\right),
\end{aligned}
$$
where the first inequality is based on the fact that:
$$
\begin{aligned}
	& \mathbb{P}\left(\min _{1 \leq l \leq m-1} \boldsymbol{L}_{i_l}^{\top} \mathbf{A} \boldsymbol{L}_{i_l} \geq z\right)- \mathbb{P}\left(U_Y^{\top} \mathbf{A} U_Y+2 U_Y^{\top} \mathbf{A} V_Y+V_Y^{\top} \mathbf{A} V_Y \geq z, \min _{1 \leq l \leq m-1} \boldsymbol{L}_{i_l}^{\top} \mathbf{A} \boldsymbol{L}_{i_l} \geq z\right) \\
	= & \mathbb{P}\left(U_Y^{\top} \mathbf{A} U_Y+2 U_Y^{\top} \mathbf{A} V_Y+V_Y^{\top} \mathbf{A} V_Y<z, \min _{1 \leq l \leq m-1} \boldsymbol{L}_{i_l}^{\top} \mathbf{A} \boldsymbol{L}_{i_l} \geq z\right) \\
	= & \mathbb{P}\left(U_Y^{\top} \mathbf{A} U_Y+2 U_Y^{\top} \mathbf{A} V_Y+V_Y^{\top} \mathbf{A} V_Y<z, 2 U_Y^{\top} \mathbf{A} V_Y+V_Y^{\top} \mathbf{A} V_Y<-C \varphi_N, \min _{1 \leq l \leq m-1} \boldsymbol{L}_{i_l}^{\top} \mathbf{A} \boldsymbol{L}_{i_l} \geq z\right) \\
	& + \mathbb{P}\left(U_Y^{\top} \mathbf{A} U_Y+2 U_Y^{\top} \mathbf{A} V_Y+V_Y^{\top} \mathbf{A} V_Y<z, 2 U_Y^{\top} \mathbf{A} V_Y+V_Y^{\top} \mathbf{A} V_Y \geq-C \varphi_N, \min _{1 \leq l \leq m-1} \boldsymbol{L}_{i_l}^{\top} \mathbf{A} \boldsymbol{L}_{i_l} \geq z\right) \\
	\leq & \mathbb{P}\left(2 U_Y^{\top} \mathbf{A} V_Y+V_Y^{\top} \mathbf{A} V_Y<-C \varphi_N\right)+ \mathbb{P}\left(U_Y^{\top} \mathbf{A} U_Y<z+C \varphi_N, \min _{1 \leq l \leq m-1} \boldsymbol{L}_{i_l}^{\top} \mathbf{A} \boldsymbol{L}_{i_l} \geq z\right) .
\end{aligned}
$$
Obviously,
$$
\mathbb{P}\left(U_Y^{\top} \mathbf{A} U_Y \geq z+C \varphi_N\right) \mathbb{P}\left(\min _{1 \leq l \leq m-1} \boldsymbol{L}_{i_l}^{\top} \mathbf{A} \boldsymbol{L}_{i_l} \geq z\right) \rightarrow\left(\frac{\kappa}{\Gamma\left(\mu_1 / 2\right)} \frac{e^{-y / 2}}{N}\right)^m,
$$
by $\varphi_N=o(1 / \log N)$. Next,
$$
\begin{aligned}
	& \mathbb{P}\left(2 U_Y^{\top} \mathbf{A} V_Y+V_Y^{\top} \mathbf{A} V_Y \leq-C \varphi_N\right) \\
	\leq & \mathbb{P}\left(U_Y^{\top} \mathbf{A} V_Y \leq-C \varphi_N / 2\right)= \mathbb{P}\left(U_Y^{\top} \mathbf{A} V_Y \geq C \varphi_N / 2\right)=o\left(N^{-m}\right).
\end{aligned}
$$
So
\begin{align}\label{8.21}
	\mathbb{P}\left(W_{i_1}>z, \cdots, W_{i_m}>z\right) \geq\left\{\frac{\kappa}{\Gamma\left(\mu_1 / 2\right)} \frac{e^{-y / 2}}{N}\right\}^m+o\left(N^{-m}\right) .
\end{align}
Then, we obtain the result by (\ref{8.20}) and (\ref{8.21}).
\hfill$\Box$
\subsection*{Proof of Lemma \ref{lemma5.7}}
Define $\boldsymbol{L}_{i_m}=\left(\boldsymbol{L}_{i_m}-\boldsymbol{\Sigma}_{i_m S} \boldsymbol{\Sigma}_{SS}^{-1}\boldsymbol{L}_S\right)+\boldsymbol{\Sigma}_{i_m S} \boldsymbol{\Sigma}_{SS}^{-1}\boldsymbol{L}_S \doteq U_Y+V_Y$. Thus, by the assumptional distribution of multivariate normal distributions, we have $\boldsymbol{L}_{i_m}-\boldsymbol{\Sigma}_{i_m S} \boldsymbol{\Sigma}_{SS}^{-1}\boldsymbol{L}_S$ that follows $N\left(\mathbf{0}, \mathbf{I}_M-\right.$ $\left.\boldsymbol{\Sigma}_{i_m S} \boldsymbol{\Sigma}_{SS}^{-1}\boldsymbol{\Sigma}_{Si_m }\right)$ is independent of $\boldsymbol{L}_S$. Define $\mathbf{A}=\operatorname{diag}\left\{\lambda_1, \cdots, \lambda_M\right\}$. Thus, we have
$$
\begin{aligned}
	& \mathbb{P}\left(W_{i_m} \geq z, \min _{1 \leq l \leq m-1} W_{i_l} \geq z\right) \\
	= & \mathbb{P}\left(\boldsymbol{L}_{i_m}^{\top} \mathbf{A} \boldsymbol{L}_{i_m} \geq z, \min _{1 \leq l \leq m-1} \boldsymbol{L}_{i_l}^{\top} \mathbf{A} \boldsymbol{L}_{i_l} \geq z\right) \\
	= & \mathbb{P}\left(U_Y^{\top} \mathbf{A} U_Y+2 U_Y^{\top} \mathbf{A} V_Y+V_Y^{\top} \mathbf{A} V_Y \geq z, \min _{1 \leq l \leq m-1} \boldsymbol{L}_{i_l}^{\top} \mathbf{A} \boldsymbol{L}_{i_l} \geq z\right) \\
	\leq & \mathbb{P}\left(2 U_Y^{\top} \mathbf{A} U_Y+2 V_Y^{\top} \mathbf{A} V_Y \geq z, \min _{1 \leq l \leq m-1} \boldsymbol{L}_{i_l}^{\top} \mathbf{A} \boldsymbol{L}_{i_l} \geq z\right) \\
	\leq & \mathbb{P}\left(U_Y^{\top} \mathbf{A} U_Y \geq \frac{1}{4} z, \min _{1 \leq l \leq m-1} \boldsymbol{L}_{i_l}^{\top} \mathbf{A} \boldsymbol{L}_{i_l} \geq z\right)+ \mathbb{P}\left(V_Y^{\top} \mathbf{A} V_Y \geq \frac{1}{4} z, \min _{1 \leq l \leq m-1} \boldsymbol{L}_{i_l}^{\top} \mathbf{A} \boldsymbol{L}_{i_l} \geq z\right) \\
	= & \mathbb{P}\left(U_Y^{\top} \mathbf{A} U_Y \geq \frac{1}{4} z\right) \mathbb{P}\left(\min _{1 \leq l \leq m-1} \boldsymbol{L}_{i_l}^{\top} \mathbf{A} \boldsymbol{L}_{i_l} \geq z\right)+ \mathbb{P}\left(V_Y^{\top} \mathbf{A} V_Y \geq \frac{1}{4} z, \min _{1 \leq l \leq m-1} \boldsymbol{L}_{i_l}^{\top} \mathbf{A} \boldsymbol{L}_{i_l} \geq z\right).
\end{aligned}
$$
By Lemma \ref{lemma5.6}, we have
$$
\mathbb{P}\left(\min _{1 \leq l \leq m-1} \boldsymbol{L}_{i_l}^{\top} \mathbf{A} \boldsymbol{L}_{i_l} \geq z\right) \leq(1+2 \epsilon)^{m-1}\left\{\frac{\kappa}{\Gamma\left(\mu_1 / 2\right)} \frac{e^{-y / 2}}{N}\right\}^{m-1} \leq C N^{1-m}.
$$
By $U_Y \sim N\left(\mathbf{0}, \mathbf{I}_M-\boldsymbol{\Sigma}_{i_m S} \boldsymbol{\Sigma}_{SS}^{-1}\boldsymbol{\Sigma}_{Si_m }\right)$, we have
$$
U_Y^{\top} \mathbf{A} U_Y \sim \xi_U^{\top}\left(\mathbf{I}_M-\boldsymbol{\Sigma}_{i_m S} \boldsymbol{\Sigma}_{SS}^{-1}\boldsymbol{\Sigma}_{Si_m }\right)^{1 / 2} \mathbf{A}\left(\mathbf{I}_M-\boldsymbol{\Sigma}_{i_m S} \boldsymbol{\Sigma}_{SS}^{-1}\boldsymbol{\Sigma}_{Si_m }\right)^{1 / 2} \xi_U,
$$
where $\xi_U \sim N\left(\mathbf{0}, \mathbf{I}_M\right)$. So
$$
\begin{aligned}
	\tilde{\lambda}_1 & \doteq \lambda_{\max } \left\{\left(\mathbf{I}_M-\boldsymbol{\Sigma}_{i_m S} \boldsymbol{\Sigma}_{SS}^{-1}\boldsymbol{\Sigma}_{Si_m }\right)^{1 / 2} \mathbf{A}\left(\mathbf{I}_M-\boldsymbol{\Sigma}_{i_m S} \boldsymbol{\Sigma}_{SS}^{-1}\boldsymbol{\Sigma}_{Si_m }\right)^{1 / 2}\right\}  \\
	& \geq \lambda_{\max }(\mathbf{A}) \lambda_{\min }\left(\mathbf{I}_M-\boldsymbol{\Sigma}_{i_m S} \boldsymbol{\Sigma}_{SS}^{-1}\boldsymbol{\Sigma}_{Si_m }\right) \geq \lambda_1[1-\{1+o(1)\}\delta]>\lambda_1(1-2\delta),
\end{aligned}
$$
for sufficiently large $p$ by the assumption. By Equation (6) in \cite{zolotarev1961concerning}, we have
$$
\begin{aligned}
	\mathbb{P}\left(U_Y^{\top} \mathbf{A} U_Y \geq \frac{1}{4} z\right) & \leq(1+\epsilon) \frac{\tilde{\kappa}}{\Gamma\left(\tilde{\mu}_1 / 2\right)}\left(\frac{z / 4+\tilde{\Lambda}}{2 \tilde{\lambda}_1}\right)^{\tilde{\mu}_1 / 2-1} \exp \left(-\frac{z / 4+\tilde{\Lambda}}{2 \tilde{\lambda}_1}\right) \\
	& \leq C\left\{\frac{z / 4+\tilde{\Lambda}}{2 \lambda_1(1-2\delta)}\right\}^{\tilde{\mu}_1 / 2-1} \exp \left\{-\frac{z / 4+\tilde{\Lambda}}{2 \lambda_1(1-2\delta)}\right\} \\
	& \leq C(\log N)^c N^{-(1-2\delta) / 4},
\end{aligned}
$$
where $c=\left(\tilde{\mu}_1-\mu_1\right) / 2$. Thus, we have
$$
\mathbb{P}\left(U_Y^{\top} \mathbf{A} U_Y \geq \frac{1}{4} z\right) \mathbb{P}\left(\min _{1 \leq l \leq m-1} \boldsymbol{L}_{i_l}^{\top} \mathbf{A} \boldsymbol{L}_{i_l} \geq z\right) \leq C(\log N)^c N^{-m+(3+2\delta) / 4}.
$$
Define $\tilde{\mathbf{A}}=\operatorname{diag}\{\mathbf{A}, \cdots, \mathbf{A}\} \in \mathbb{R}^{(m-1) M \times(m-1) M}$. Next, we consider
$$
\begin{aligned}
	& \mathbb{P}\left(V_Y^{\top} \mathbf{A} V_Y \geq \frac{1}{4} z, \min _{1 \leq l \leq m-1} \boldsymbol{L}_{i_l}^{\top} \mathbf{A} \boldsymbol{L}_{i_l} \geq z\right) \\
	= & \mathbb{P}\left(\boldsymbol{L}_S^{\top}\boldsymbol{\Sigma}_{SS}^{-1} \boldsymbol{\Sigma}_{S i_m} \mathbf{A} \boldsymbol{\Sigma}_{i_m S} \boldsymbol{\Sigma}_{SS}^{-1}\boldsymbol{L}_S \geq \frac{1}{4} z ,\min _{1 \leq l \leq m-1} \boldsymbol{L}_{i_l}^{\top} \mathbf{A} \boldsymbol{L}_{i_l} \geq z\right) \\
	\leq & \mathbb{P}\left(\boldsymbol{L}_S^{\top}\boldsymbol{\Sigma}_{SS}^{-1} \boldsymbol{\Sigma}_{S i_m} \mathbf{A} \boldsymbol{\Sigma}_{i_m S} \boldsymbol{\Sigma}_{SS}^{-1}\boldsymbol{L}_S \geq \frac{1}{4} z, \boldsymbol{L}_S^{\top} \tilde{\mathbf{A}} \boldsymbol{L}_S \geq(m-1) z\right) \\
	\leq & \mathbb{P}\left[\boldsymbol{L}_S^{\top}\boldsymbol{\Sigma}_{SS}^{-1}\left\{(1-\epsilon) \tilde{\mathbf{A}}+\epsilon \boldsymbol{\Sigma}_{S i_m} \mathbf{A} \boldsymbol{\Sigma}_{i_m S}\right\}\boldsymbol{\Sigma}_{SS}^{-1} \boldsymbol{L}_S \geq\left\{(m-1)(1-\epsilon)+\frac{1}{4} \epsilon\right\} z\right] \\
	= & \mathbb{P}\left[\xi_S^{\top} \boldsymbol{\Sigma}_{SS}^{-1 / 2}\left\{(1-\epsilon) \tilde{\mathbf{A}}+\epsilon \boldsymbol{\Sigma}_{S i_m} \mathbf{A} \boldsymbol{\Sigma}_{i_m S}\right\} \boldsymbol{\Sigma}_{SS}^{-1 / 2} \xi_S \geq\left\{(1-\epsilon) m-1+\frac{5}{4} \epsilon\right\} z\right],
\end{aligned}
$$
where $\xi_S \sim N\left\{\mathbf{0}, \mathbf{I}_{(m-1) M}\right\}$. We have
$$
\begin{aligned}
	\breve{\lambda}_1 & \doteq \lambda_{\max }\left\{\boldsymbol{\Sigma}_{SS}^{-1 / 2}\left((1-\epsilon) \tilde{\mathbf{A}}+\epsilon \boldsymbol{\Sigma}_{S i_m} \mathbf{A} \boldsymbol{\Sigma}_{i_m S}\right) \boldsymbol{\Sigma}_{SS}^{-1 / 2}\right\} \\
	& \geq1/\left\{1+(m-1) \varphi_N\right\} \lambda_{\max }\left\{(1-\epsilon) \tilde{\mathbf{A}}+\epsilon \boldsymbol{\Sigma}_{S i_m} \mathbf{A} \boldsymbol{\Sigma}_{i_m S}\right\} \\
	& \geq1/\left\{1+(m-1) \varphi_N\right\}\left[\lambda_{\max }\{(1-\epsilon) \tilde{\mathbf{A}}\}-\lambda_{\max }\left(\epsilon \boldsymbol{\Sigma}_{S i_m} \mathbf{A} \boldsymbol{\Sigma}_{i_m S}\right)\right] \\
	& \geq \lambda_1/\left\{1+(m-1) \varphi_N\right\}\{1-(1+\delta) \epsilon\} \geq(1-\delta / 2) \lambda_1,
\end{aligned}
$$
for a small enough positive real number $\epsilon$.
By Equation (6) in \cite{zolotarev1961concerning}, we have
$$
\begin{aligned}
	& \mathbb{P}\left[\xi_S^{\top} \boldsymbol{\Sigma}_{SS}^{-1 / 2}\left\{(1-\epsilon) \tilde{\mathbf{A}}+\epsilon \boldsymbol{\Sigma}_{S i_m} \mathbf{A} \boldsymbol{\Sigma}_{i_m S}\right\} \boldsymbol{\Sigma}_{SS}^{-1 / 2} \xi_S \geq\left\{(1-\epsilon) m-1+\frac{5}{4} \epsilon\right\} z\right] \\
	\leq & (1+\epsilon) \frac{\breve{\kappa}}{\Gamma\left(\breve{\mu}_1 / 2\right)}\left[\frac{\left\{(1-\epsilon) m-1+\frac{5}{4} \epsilon\right\} z+\breve{\Lambda}}{2 \breve{\lambda}_1}\right]^{\breve{\mu}_{1 / 2-1}} \exp \left[-\frac{\left\{(1-\epsilon) m-1+\frac{5}{4} \epsilon\right\} z+\breve{\Lambda}}{2 \breve{\lambda}_1}\right] \\
	\leq & C\left[\frac{\left\{(1-\epsilon) m-1+\frac{5}{4} \epsilon\right\} z+\breve{\Lambda}}{2 \lambda_1(1-\delta / 2)}\right]^{\breve{\mu}_1 / 2-1} \exp \left[-\frac{\left\{(1-\epsilon) m-1+\frac{5}{4} \epsilon\right\} z+\breve{\Lambda}}{2 \lambda_1(1-\delta / 2)}\right] \\
	\leq & C(\log N)^c N^{-\frac{(1-\epsilon) m-1+\frac{5}{4} \epsilon}{1-\delta / 2}} \leq C(\log N)^c N^{-m+(1-\frac{5\delta}{8}) / (1-\delta/2)},
\end{aligned}
$$
by setting $\frac{(1-\epsilon) m-1+\frac{5}{4} \epsilon}{1-\delta / 2} \geq m-\frac{(1-\frac{5\delta}{8}) }{(1-\delta/2)}$.
\hfill$\Box$

\end{appendices}

\bibliographystyle{agsm}
%\bibliography{refer1}

@book{farcomeni2005multiple,
  title={Multiple testing procedures under dependence, with applications},
  author={Farcomeni, Alessio},
  year={2009},
  publisher={VDM Verlag}
}

@article{li2022simultaneous,
  title={Simultaneous detection of signal regions using quadratic scan statistics with applications to whole genome association studies},
  author={Li, Zilin and Liu, Yaowu and Lin, Xihong},
  journal={Journal of the American Statistical Association},
  volume={117},
  number={538},
  pages={823--834},
  year={2022}
}

@article{MOON2013143,
title = {Tests for m-dependence based on sample splitting methods},
journal = {Journal of Econometrics},
volume = {173},
number = {2},
pages = {143-159},
year = {2013},
author = {Moon, Seongman and Velasco, Carlos}
}

@article{chiang2006homozygosity,
  title={Homozygosity mapping with SNP arrays identifies TRIM32, an E3 ubiquitin ligase, as a Bardet--Biedl syndrome gene (BBS11)},
  author={Chiang, Annie P and Beck, John S and Yen, Hsan-Jan and Tayeh, Marwan K and Scheetz, Todd E and Swiderski, Ruth E and Nishimura, Darryl Y and Braun, Terry A and Kim, Kwang-Youn A and Huang, Jian and others},
  journal={Proceedings of the National Academy of Sciences},
  volume={103},
  number={16},
  pages={6287--6292},
  year={2006}
}

@article{scheetz2006regulation,
  title={Regulation of Gene Expression in the Mammalian Eye and Its Relevance to Eye Disease},
  author={Scheetz, Todd E and Kim, Kwang-Youn A and Swiderski, Ruth E and Philp, Alisdair R and Braun, Terry A and Knudtson, Kevin L and Dorrance, Anne M and DiBona, Gerald F and Huang, Jian and Casavant, Thomas L and others},
  journal={Proceedings of the National Academy of Sciences},
  volume={103},
  number={39},
  pages={14429--14434},
  year={2006}
}

@article{Gretton2010,
  author  = {Gretton, Arthur and Gy{\"o}rfi, L{\'a}szl{\'o}},
  title   = {Consistent Nonparametric Tests of Independence},
  journal = {Journal of Machine Learning Research},
  year    = {2010},
  volume  = {11},
  number  = {46},
  pages   = {1391--1423}
}

@article{szekely2013distance,
  title={The Distance Correlation T-Test of Independence in High Dimension},
  author={Sz{\'e}kely, G{\'a}bor J and Rizzo, Maria L},
  journal={Journal of Multivariate Analysis},
  volume={117},
  pages={193--213},
  year={2013},
  publisher={Elsevier}
}

@article{josse2016measuring,
  title={Measuring Multivariate Association and Beyond},
  author={Josse, Julie and Holmes, Susan},
  journal={Statistics Surveys},
  volume={10},
  pages={132--167},
  year={2016}
}

@article{guerrero1998measures,
  title={Measures of Dependence for the Multivariate T Distribution with Applications to the Stock Market},
  author={Guerrero-Cusumano, Jos{\'e}-Luis},
  journal={Communications in Statistics-Theory and Methods},
  volume={27},
  number={12},
  pages={2985--3006},
  year={1998},
  publisher={Taylor \& Francis}
}

@article{Spearman1904,
author = {Spearman, C.},
title = {The proof and measurement of association between two things},
volume = {15},
number = {1},
pages = {72--101},
journal = {The American Journal of Psychology},
year = {1904}
}

@article{zhouyeqing2024,
author = {Zhou, Yeqing and Xu, Kai and Zhu, Liping and Li, Runze},
title = {Rank-based indices for testing independence between two high-dimensional vectors},
journal = {The Annals of Statistics},
volume = {52},
number = {1},
pages = {184--206},
year = {2024}}

@article{yu2022jasa,
	title = {Fisher's combined probability test for high-dimensional covariance matrices},
	journal = {Journal of the American Statistical Association},
	volume = {119},
	number = {545},
	pages = {511--524},
	year = {2024},
	author ={Xiufan Yu and Danning Li and Lingzhou Xue}
}

@article{wang2023,
	title = {Computationally efficient and data-adaptive changepoint inference in high dimension},
	journal = {Journal of the Royal Statistical Society Series B: Statistical Methodology},
	volume = {85},
	number = {3},
	pages = {936--958},
	year = {2023},
	author ={Guanghui Wang and Long Feng}}

@article{xu2016adaptive,
	title={An adaptive two-sample test for high-dimensional means},
	author={Xu, Gongjun and Lin, Lifeng and Wei, Peng and Pan, Wei},
	journal={Biometrika},
	volume={103},
	number={3},
	pages={609--624},
	year={2016},
	publisher={Oxford University Press}
}

@article{wang2024,
author = {Hongfei Wang and Binghui Liu and Long Feng and Yanyuan Ma},
title = {Rank-based max-sum tests for mutual independence of high-dimensional random vectors},
volume = {238},
journal = {Journal of Econometrics},
pages = {1--21},
year = {2024}
}

@article{hoeiffding1948n1,
  title={A class of statistics with asymptotically normal distributions},
  author={Hoeffding, Wassily},
  journal={The Annals of Mathematical Statistics},
  volume={19},
  pages={293--325},
  year={1948}
}

@article{heller2013consistent,
  title={A consistent multivariate test of association based on ranks of distances},
  author={Heller, Ruth and Heller, Yair and Gorfine, Malka},
  journal={Biometrika},
  volume={100},
  number={2},
  pages={503--510},
  year={2013}
}

@article{Cai2023,
author = {Cai, Zhanrui and  Lei, Jing and  Roeder, Kathryn},
title = {Asymptotic Distribution-Free Independence Test for High-Dimension Data},
journal = {Journal of the American Statistical Association, In press},
year={2024}
}

@article{chakraborty2021new,
  title={A new framework for distance and kernel-based metrics in high dimensions},
  author={Chakraborty, Shubhadeep and Zhang, Xianyang},
  journal={Electronic Journal of Statistics},
  volume={15},
  number={2},
  pages={5455--5522},
  year={2021}
}

@article{zhu2017,
    author = {Zhu, Liping and Xu, Kai and Li, Runze and Zhong, Wei},
    title = {Projection correlation between two random vectors},
    journal = {Biometrika},
    volume = {104},
    number = {4},
    pages = {829--843},
    year = {2017}
}

@article{Kim2018,
author = {Kim, Ilmun and Balakrishnan, Sivaraman and Wasserman, Larry},
title = {Robust multivariate nonparametric tests via projection averaging},
volume = {48},
journal = {The Annals of Statistics},
number = {6},
pages = {3417--3441},
year = {2020}
}

@article{he2021,
	title = {Asymptotically independent U-statistics in high-dimensional testing},
	journal = {The Annals of Statistics},
	volume = {49},
	number = {1},
	pages = {151--181},
	year = {2021},
	author = {Yinqiu He and Gongjun Xu and Chong Wu and Wei Pan}
}

@article{Qiutao,
title = {Independence tests with random subspace of two random vectors in high dimension},
author = {Qiu, Tao and Xu, Wangli and Zhu, Lixing},
journal = {Journal of Multivariate Analysis},
volume = {195},
pages = {105--160},
year = {2023}
}

@article{arcones1993limit,
author = {Arcones, Miguel A and Gin{\'e}, Evarist},
title = {Limit Theorems for $U$-Processes},
volume = {21},
journal = {The Annals of Probability},
number = {3},
pages = {1494--1542},
year = {1993}
}

@book{serfling,
  title={Approximation theorems of mathematical statistics},
  author={Serfling, Robert J},
  year={1980},
  publisher={John Wiley \& Sons}
}

@article{malevich1979large,
  title={Large deviation probabilities for U-statistics},
  author={Malevich, TL and Abdalimov, B},
  journal={Theory of Probability \& Its Applications},
  volume={24},
  number={1},
  pages={215--219},
  year={1979}
}

@article{feng2022max,
  title={Max-sum tests for cross-sectional independence of high-dimensional panel data},
  author={Feng, Long and Jiang, Tiefeng and Liu, Binghui and Xiong, Wei},
  journal={The Annals of Statistics},
  volume={50},
  number={2},
  pages={1124--1143},
  year={2022}
}

@article{chen2022rank,
  title={Rank Based Tests for High Dimensional White Noise},
  author={Chen, Dachuan and Song, Fengyi and Feng, Long},
  journal={Statistica Sinica, In press},
  year={2024}
}

@article{drton2020,
author = {Drton, Mathias and Han, Fang and Shi, Hongjian},
title = {High-dimensional consistent independence testing with maxima of rank correlations},
volume = {48},
journal = {The Annals of Statistics},
number = {6},
pages = {3206--3227},
year = {2020}
}

@article{testsum,
author = {Leung, Dennis and Drton, Mathias},
title = {Testing independence in high dimensions with sums of rank correlations},
volume = {46},
journal = {The Annals of Statistics},
number = {1},
pages = {280--307},
year = {2018}
}

@article{distributionfree,
    author = {Han, Fang and Chen, Shizhe and Liu, Han},
    title = {Distribution-free tests of independence in high dimensions},
    journal = {Biometrika},
    volume = {104},
    number = {4},
    pages = {813--828},
    year = {2017}
}

@article{DebandSen2023,
author = {Deb, Nabarun and Sen, Bodhisattva},
title = {Multivariate Rank-Based Distribution-Free Nonparametric Testing Using Measure Transportation},
journal = {Journal of the American Statistical Association},
volume = {118},
number = {541},
pages = {192--207},
year = {2023}
}

@article{Shi2022,
author = {Shi, Hongjian and Drton, Mathias and Han, Fang},
title = {Distribution-Free Consistent Independence Tests via Center-Outward Ranks and Signs},
journal = {Journal of the American Statistical Association},
volume = {117},
number = {537},
pages = {395--410},
year = {2022}
}

@article{Sejdinovic2013,
author = {Sejdinovic, Dino and Sriperumbudur, Bharath and Gretton, Arthur and Fukumizu, Kenji},
title = {Equivalence of distance-based and RKHS-based statistics in hypothesis testing},
volume = {41},
journal = {The Annals of Statistics},
number = {5},
pages = {2263--2291},
year = {2013}
}

@article{Szekely2007,
author = {Sz{\'e}kely, G{\'a}bor J and Rizzo, Maria L and Bakirov, Nail K},
title = {Measuring and testing dependence by correlation of distances},
volume = {35},
journal = {The Annals of Statistics},
number = {6},
pages = {2769--2794},
year = {2007}
}

@article{lu2009financial,
  title={Financial time series forecasting using independent component analysis and support vector regression},
  author={Lu, Chi-Jie and Lee, Tian-Shyug and Chiu, Chih-Chou},
  journal={Decision support systems},
  volume={47},
  number={2},
  pages={115--125},
  year={2009}
}

@article{grover1985probabilistic,
  title={A probabilistic model for testing hypothesized hierarchical market structures},
  author={Grover, Rajiv and Dillon, William R},
  journal={Marketing Science},
  volume={4},
  number={4},
  pages={312--335},
  year={1985},
  publisher={INFORMS}
}

@article{liu2010versatile,
  title={A versatile gene-based test for genome-wide association studies},
  author={Liu, Jimmy Z and Mcrae, Allan F and Nyholt, Dale R and Medland, Sarah E and Wray, Naomi R and Brown, Kevin M and Hayward, Nicholas K and Montgomery, Grant W and Visscher, Peter M and Martin, Nicholas G and others},
  journal={The American Journal of Human Genetics},
  volume={87},
  number={1},
  pages={139--145},
  year={2010},
  publisher={Elsevier}
}

@book{imbens2015causal,
  title={Causal inference in statistics, social, and biomedical sciences},
  author={Imbens, Guido W and Rubin, Donald B},
  year={2015},
  publisher={Cambridge University Press}
}

@book{maathuis2018handbook,
  title={Handbook of graphical models},
  author={Maathuis, Marloes and Drton, Mathias and Lauritzen, Steffen and Wainwright, Martin},
  year={2018},
  publisher={CRC Press}
}

@book{fan2020statistical,
  title={Statistical foundations of data science},
  author={Fan, Jianqing and Li, Runze and Zhang, Cun-Hui and Zou, Hui},
  year={2020},
  publisher={CRC press}
}

@article{Gretton2005,
  author  = {Gretton, Arthur and Herbrich, Ralf and Smola, Alexander and Bousquet, Olivier and Sch{\"o}lkopf, Bernhard},
  title   = {Kernel Methods for Measuring Independence},
  journal = {Journal of Machine Learning Research},
  year    = {2005},
  volume  = {6},
  number  = {70},
  pages   = {2075--2129}
}

@article{Bergsma2014,
author = {Bergsma, Wicher and Dassios, Angelos},
title = {A consistent test of independence based on a sign covariance related to Kendall's tau},
volume = {20},
journal = {Bernoulli},
number = {2},
pages = {1006--1028},
year = {2014}
}

@article{Blum1961,
author = {Blum, Julius R and Kiefer, Jack and Rosenblatt, Murray},
title = {Distribution Free Tests of Independence Based on the Sample Distribution Function},
volume = {32},
journal = {The Annals of Mathematical Statistics},
number = {2},
pages = {485--498},
year = {1961}
}

@article{Hoeffding1948,
author = {Hoeffding, Wassily},
title = {A Non-Parametric Test of Independence},
volume = {19},
journal = {The Annals of Mathematical Statistics},
number = {4},
pages = {546--557},
year = {1948}
}

@article{kendall1938new,
  title={A new measure of rank correlation},
  author={Kendall, Maurice G},
  journal={Biometrika},
  volume={30},
  number={1/2},
  pages={81--93},
  year={1938},
  publisher={JSTOR}
}

@article{Zhuchangbo,
author = {Zhu, Changbo and Zhang, Xianyang and Yao, Shun and Shao, Xiaofeng},
title = {Distance-based and RKHS-based dependence metrics in high dimension},
volume = {48},
journal = {The Annals of Statistics},
number = {6},
pages = {3366--3394},
year = {2020}
}

@article{Gaolan,
author = {Gao, Lan and Fan, Yingying and Lv, Jinchi and Shao, Qi-Man},
title = {Asymptotic distributions of high-dimensional distance correlation inference},
volume = {49},
journal = {The Annals of Statistics},
number = {4},
pages = {1999--2020},
year = {2021}
}

@article{zolotarev1961concerning,
  title={Concerning a certain probability problem},
  author={Zolotarev, Vladimir Mikhailovich},
  journal={Theory of Probability \& Its Applications},
  volume={6},
  number={2},
  pages={201--204},
  year={1961},
  publisher={SIAM}
}

@article{mdependent,
title = {A more general central limit theorem for m-dependent random variables with unbounded m},
journal = {Statistics \& Probability Letters},
author = {Romano, Joseph P and Wolf, Michael},
volume = {47},
number = {2},
pages = {115--124},
year = {2000}
}

@article{zaitsev,
  title={On the Gaussian approximation of convolutions under multidimensional analogues of SN Bernstein's inequality conditions},
  author={Zaitsev, A Yu},
  journal={Probability theory and related fields},
  volume={74},
  number={4},
  pages={535--566},
  year={1987},
  publisher={Springer}
}

@article{feng2022dependent,
  title={Asymptotic independence of the sum and maximum of dependent random variables with applications to high-dimensional tests},
  author={Feng, Long and Jiang, Tiefeng and Li, Xiaoyun and Liu, Binghui},
  journal={Statistica Sinica, In press},
  year={2024}
}

\end{document}